\documentclass[12pt]{article}
\pdfoutput=1
\usepackage[margin=1.0in]{geometry}
\usepackage{textcomp}
\usepackage{amsmath,amssymb,graphicx}
\usepackage{hyperref}
\usepackage{slashed}
\usepackage{wasysym}
\usepackage{hhline}
\usepackage[section,above,below]{placeins}
\usepackage[normalem]{ulem}

\usepackage{multirow}
\usepackage[table]{xcolor}
\usepackage{xspace}
\usepackage{feynman}
\usepackage[utf8]{inputenc}

\usepackage[greek,english]{babel}

\usepackage{subcaption}
\usepackage{booktabs}

\usepackage[font=small,labelfont=bf]{caption}
\usepackage{cite}
\usepackage{enumerate}

\newcommand{\pt}{\ensuremath{p_{\mathrm T}}\xspace}

\newcommand{\be}{\begin{equation}}
\newcommand{\ee}{\end{equation}}
\newcommand{\beq}{\begin{equation}}
\newcommand{\eeq}{\end{equation}}

\newcommand{\beqn}{\begin{eqnarray}}
\newcommand{\eeqn}{\end{eqnarray}}
\newcommand{\bea}{\begin{eqnarray}}
\newcommand{\eea}{\end{eqnarray}}

\usepackage[table]{xcolor}
\usepackage{bm}

\definecolor{gray}{cmyk}{0,0,0,0.05}
\newcolumntype{a}{>{\columncolor{gray}} l}

\numberwithin{equation}{section}

\usepackage{soul}
\usepackage{lineno}

\begin{document}
\captionsetup[table]{skip=0pt}
\captionsetup[table]{belowskip=0pt}

\begin{center}

\begin{flushright}
IFIC/22-13
\end{flushright}
~
\vskip 1cm

{\Large \bf 
Discovery prospects for long-lived 
\\[0.3cm] 
multiply charged particles at the LHC
}

\vskip 1.5cm

Mohammad Mahdi Altakach$^{(a)}$,
Priyanka Lamba$^{(a)}$,
Rafa\l{} Mase\l{}ek$^{(a)}$,
\\[0.3cm]
Vasiliki A.~Mitsou$^{(b)}$
\,and\,
Kazuki Sakurai$^{(a)}$

\vskip 1.0cm

$^{(a)}${\em
\small
Institute of Theoretical Physics, Faculty of Physics,\\University of Warsaw, ul.~Pasteura 5, PL-02-093 Warsaw, Poland\\[0.1cm]
}

\vspace{3mm}

$^{(b)}${\em
\small
Instituto de F\'isica Corpuscular (IFIC), CSIC -- Universitat de Val\`encia, \\ 
C/ Catedr\'atico Jos\'e Beltr\'an 2, E-46980 Paterna (Valencia), Spain\\[0.1cm]
}

\end{center}

\vskip 1.5cm
\begin{abstract}
In this work, we aim to provide a comprehensive and largely model independent investigation on prospects to detect long-lived multiply charged particles at the LHC. We consider particles with spin 0 and $\frac{1}{2}$, with electric charges in range $1 \le |Q/e| \le 8$, which are singlet or triplet under $SU(3)_C$. 
Such particles might be produced as particle-antiparticle pairs and propagate through detectors, or form a positronium(quarkonium)-like bound state. We consider both possibilities and estimate lower mass bounds on new particles, that can be provided by ATLAS, CMS and MoEDAL experiments at the end of Run 3 and HL-LHC data taking periods. 
We find out that the sensitivities of ATLAS and CMS are generally stronger than those of MoEDAL at Run 3, while 
they may be competitive at HL-LHC
for $3 \lesssim |Q/e| \lesssim 7$ for all types of long-lived particles we consider.

\end{abstract}

\vskip 2.0cm

\section{Introduction}
\label{sec:intro}

One of the motivations behind the design, construction and operation of the Large Hadron Collider (LHC)~\cite{Evans:2008zzb} is the discovery of new particles beyond the Standard Model, or the (partial) exclusion of their potential existence. Priorities during the LHC Run~1 and Run~2 included scans for peaks in invariant-mass spectra, searches for excesses of missing transverse energy in events with various final states, and precision tests of Standard Model (SM) predictions. Apart from the discovery of a SM-like Higgs boson, no search has led to evidence of a new, still unobserved particle.

As conventional searches kept on showing no deviation from the SM predictions, the interest of the experimental --- and theoretical --- community started shifting towards less orthodox final states, such as those by \emph{long-lived particles (LLPs)}~\cite{Alimena:2019zri,Lee:2018pag,Mitsou:2021tti}. In theory, long lifetimes may be due to a symmetry (leading to stable particles), narrow mass splittings, small couplings, or a heavy mediator. Here, we focus on multiply charged particles that either are pair produced and leave a high ionising trace in the detectors, or they form a bound state which decays into two photons.
In this article we are interested in the lifetime that is long enough such that the particles can be treated as stable within the detectors.
For shorter lifetimes of the order of mm and cm, 
other exotic signatures, such as displaced vertices and disappearing tracks, are available. However, these are beyond the scope of the paper.

ATLAS and CMS experiments
have been actively searching for
multiply charged LLPs 
\cite{ATLAS:2015hau,ATLAS:2018imb,CMS:2013czn,CMS:2016kce} with electric charge $Q$ ranging 
$2 \le |Q/e| \le 7$
with $e$ being the electric charge of the proton.
Their searches exploit 
the fact that multi-charged LLPs 
give rise to highly ionising tracks with anomalously large energy loss, $dE/dx$.
So far, their analyses
focus on 
colourless fermions 
and include only Drell-Yan production channels.
As pointed out in Refs.~\cite{Barrie:2017eyd,Jager:2018ecz},
however, production channels involving photons in the initial state may become important for $|Q| > 1e$ and should be included in the analysis.
In this paper we recast the large $dE/dx$ search \cite{CMS:2016kce}
for
various types of LLPs with spin 0 and $\frac{1}{2}$
and with and without colour charges.
Our analysis includes production channels with photonic initial states, 
and take two important effects into account.
One is underestimation of the transverse momentum, \pt, of charged tracks for $|Q| > 1e$. 
It is caused by the estimation of track \pt's based on the track curvatures 
($r \propto Q/\pt$) assuming $Q = 1e$.
This leads to a loss of efficiency 
in \pt cuts.
The second effect comes from the fact that
produced charged particles 
lose kinetic energy
while traveling inside the detector via strong electromagnetic interactions with detector materials.  For large $|Q|$, this effect may be so strong that
for a large fraction of events
charged LLPs arrive at the muon system too late after the next bunch crossing. Since such events are rejected,
this negatively affects the overall sensitivity.

Pair-produced charged particles  
can also form a bound state, ${\cal B}$.
Those bound states are generally short-lived and decay into the Standard Model particles.
The best way to look for such bound states is to focus on the $\gamma \gamma$ decay mode and search for a bump in diphoton invariant mass distributions.  
We compute the event rate of the $pp \to {\cal B} \to \gamma \gamma$ process for various types of
charged LLPs and estimate the current limit as well as the projected sensitivities achievable at Run 3 and HL-LHC.

The MoEDAL experiment, which is designed to search for highly ionising  avatars~\cite{Acharya:2014nyr}, has published results on isolated magnetic charges, namely magnetic monopoles~\cite{MoEDAL:2016jlb,MoEDAL:2016lxh,MoEDAL:2017vhz}. Rather recently, it also constrained high electric charges either indirectly through dyons~\cite{MoEDAL:2020pyb}, which possess both electric and magnetic charge, or as highly electrically charged LLPs~\cite{MoEDAL:2021mpi}.     
Its capability to probe heavy highly electrically charged LLPs has been demonstrated quantitatively in previous studies~\cite{Felea:2020cvf,Acharya:2020uwc, Hirsch:2021wge}, which 
have focused on $|Q| \le 4e$. These studies 
indicated that MoEDAL's sensitivity 
is generally less significant than those obtained from large $dE/dx$ analyses by ATLAS and CMS, apart from some exceptional cases \cite{Felea:2020cvf}. 
This is largely because 
the data collected at MoEDAL is roughly 10 times smaller than those at ATLAS and CMS.
Also, MoEDAL can register only
slow-moving charged particles with
$\beta < 0.15 \cdot |Q/e|$,
whilst those particles are often produced with relativistic velocities 
and the above condition is not satisfied
in majority of events
for $|Q| \lesssim 4e$.

The situation may, however, be different for $|Q| \gtrsim 4e$.
While the signal efficiency of large $dE/dx$ analyses drops for larger $|Q|$
due to the two aforementioned obstacles (underestimation of \pt and late arrival time at the muon system due to the velocity loss),
MoEDAL's detection rate only increases with $|Q|$.
In this paper we estimate, for the first time, 
MoEDAL's sensitivity to multiply charged LLPs with $|Q| > 4e$
including photonic initial states 
and compare the sensitivities obtained from ATLAS and CMS analyses at Run~3 and HL-LHC.

The paper is structured as follows. 
In the next section we define the relevant theoretical models for this study.
In section~\ref{sec:open}
the open production modes 
for charged LLPs are studied. 
Our prescription to estimate
the effect of hadronisation
for coloured LLPs is also outlined.
In section~\ref{sec:open_ATLAS},
we discuss the large $dE/dx$ analysis 
and the methodology to recast 
the CMS analysis.
In section~\ref{sec:open_MoEDAL},
we briefly describe the MoEDAL detector 
and outline the estimation of
the MoEDAL's sensitivity to the charged LLPs.
The calculation of the bound state production and decay are discussed in section \ref{sec:closed}, where
the sensitivity from the diphoton resonance searches is also studied.
The results of our numerical studies are 
summarised in section \ref{sec:results}.
Section \ref{sec:concl} is devoted to conclusions.


\section{Models}
\label{sec:models}

In this study four types of electrically charged long-lived particles are considered:
(spin 0 and $\frac{1}{2}$)
$\times$
(colour-singlet and triplet).
In order to write a simple gauge invariant Lagrangian, we 
introduce 
the corresponding quantum field, 
$\phi$, for scalars 
and, 
$\psi$, for fermions, 
and postulate that
they are $SU(2)_L$ singlet and have the hypercharge $Y = Q/e$,
where $Q$ is the electric charge measured in the proton charge unit, $e$.
With these fields, we extend the Standard Model Lagrangian, ${\cal L}_{\rm SM}$,
as
\begin{equation}
{\cal L} \,=\, {\cal L}_{\rm SM} \,+\, |D_\mu \phi|^2 \,-\, m^2 |\phi|^2 \,+\, \cdots \,.
\label{eq:Ls}
\end{equation}
for scalars and
\begin{equation}
{\cal L} \,=\, {\cal L}_{\rm SM} \,+\, i \overline \psi \slashed{D} \psi \,-\, m \overline \psi \psi  \,+\, \cdots \,.
\label{eq:Lf}
\end{equation}
for fermions with
$\slashed D = \gamma^\mu D_\mu$.
The covariant derivative is
$D_\mu = \partial_\mu - i g_Y Q B_\mu$
for colour-singlet fields, while it is
$D_\mu = \partial_\mu - i g_Y Q B_\mu - i g_c T^a G^a_\mu$
for colour-triplet fields
with $B_\mu$ and $G_\mu^a$ being the $U(1)_Y$
and $SU(3)_C$ gauge fields, respectively, 
and $T^a$ being the generator of $SU(3)_C$.
After the electroweak symmetry breaking,
the interaction with the $B_\mu$ field
should be rewritten 
in terms of the photon ($A_\mu$)
and $Z$-boson ($Z_\mu$) fields as
\beq
g_Y Q B_\mu \,=\, e Q A_\mu - e Q \tan \theta_W Z_\mu,
\eeq
where $\theta_W$ is the weak mixing angle.

We denote the scalar (fermion) particle $\phi^{+Q}$ ($\psi^{+Q}$)
and
its antiparticle
$\phi^{-Q}$ ($\psi^{-Q}$).
We also write 
$\xi^{+Q}$ ($\xi^{-Q}$)
to indicate the charged particle (antiparticle) when not specifying the spin. 
Without the terms 
denoted by $\cdots$ in Eqs.~\eqref{eq:Ls} and \eqref{eq:Lf},
these charged particles are absolutely stable.
Since the presence of stable charged particles is problematic in cosmology,
we postulate extra terms responsible for decays 
in $\cdots$ in the Lagrangian.
In favour of model-independence of the analysis,
we do not specify these terms as well as the details of the decay.
We instead  
treat the lifetime of $\xi^{\pm Q}$ as a free parameter 
when studying the sensitivity at the MoEDAL experiment.
In estimating the sensitivities of ATLAS and CMS, we assume the charged particles are meta-stable, i.e.~their typical decay length  is much larger than the size of the detectors.

The models described above are implemented 
in {\tt MadGraph\,5} \cite{madgraph5}
with the help of the {\tt FeynRules}~\cite{feynrules} package.
{\tt MadGraph\,5} is used for later analyses in calculating tree-level production cross sections and generating Monte Carlo events. 

\section{Open production mode}
\label{sec:open}


In the models described in the previous section,
long-lived charged particles may be produced at the LHC
in a particle-antiparticle pair 
and travel through the detectors before decaying.
We call this type of production process the {\it open production mode}.
In this case, the production modes are similar to those for other exotic particles such as the magnetic monopoles~\cite{Baines:2018ltl}.

\begin{figure}[t!]
\centering
\includegraphics[width=0.36\textwidth]{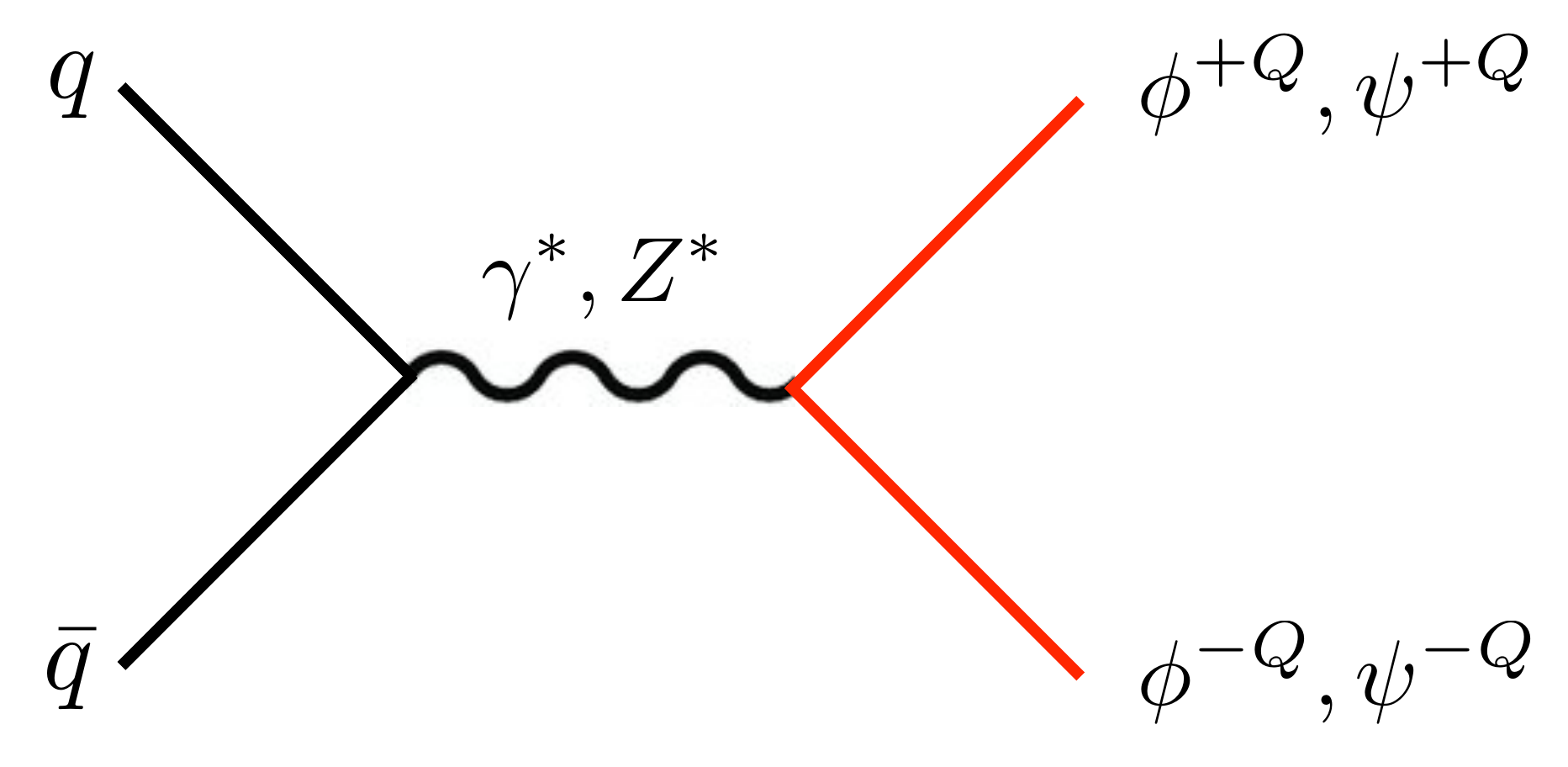}
\hspace{0.2cm}
\includegraphics[width=0.3\textwidth]{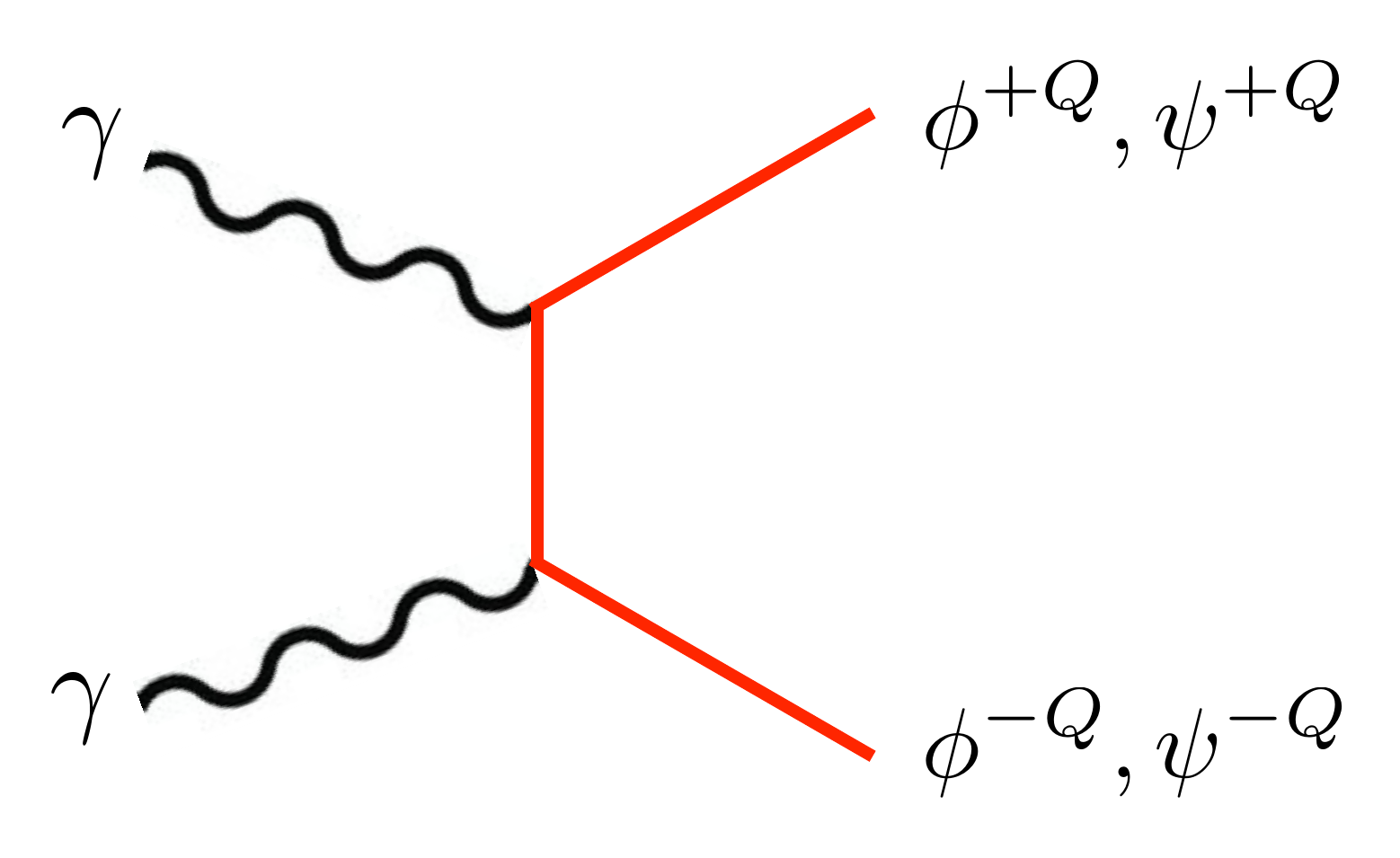}
\hspace{0.1cm}
\includegraphics[width=0.26\textwidth]{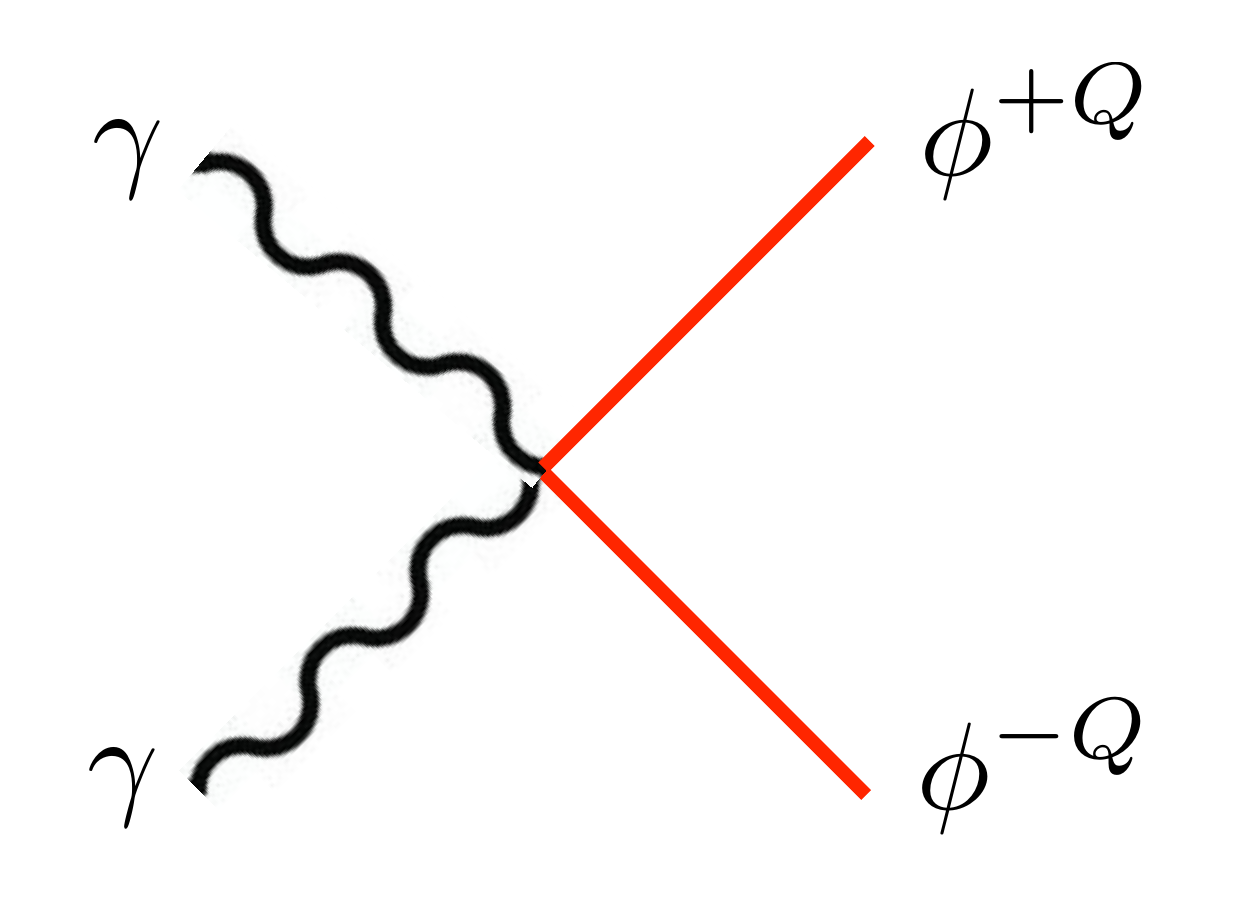}
\caption{\label{fig:diagrams}Diagrams for colour-singlet open production modes}
\end{figure}

\subsection{Colour-singlet particles}

For colour-singlet particles, 
the following processes are possible (see Fig.~\ref{fig:diagrams}):
%
\begin{itemize}
\item
Drell-Yan: $q \bar q$ initial state with $s$-channel $\gamma/Z$ exchange 
%
\item
Photon fusion: $\gamma \gamma$ initial state with a $t$-channel interaction
(and a 4-point interaction for scalars)
\end{itemize}
%
%
As can be seen in the diagrams of Fig.~\ref{fig:diagrams}, the production rate is proportional to $Q^2$ 
for the Drell-Yan $q \bar q$ initiated process, while
the $\gamma \gamma$ initiated process has the production rate proportional to $Q^4$. 
Although the latter process is generally suppressed 
by the parton distribution function (PDF) of photons,
it can be dominant for large values of $|Q|$.\footnote{One might worry that initial states involving $Z$-bosons also give sizeable contribution since the cross section also grows with $Q^4$.
However, this contribution turns out to be small ($\lesssim 15\%$) with respect to that from the $\gamma \gamma$ initial state.
See Appendix \ref{ap:a0} for further information.
}

\begin{figure}[t!]
\centering
\includegraphics[width=0.48\textwidth]{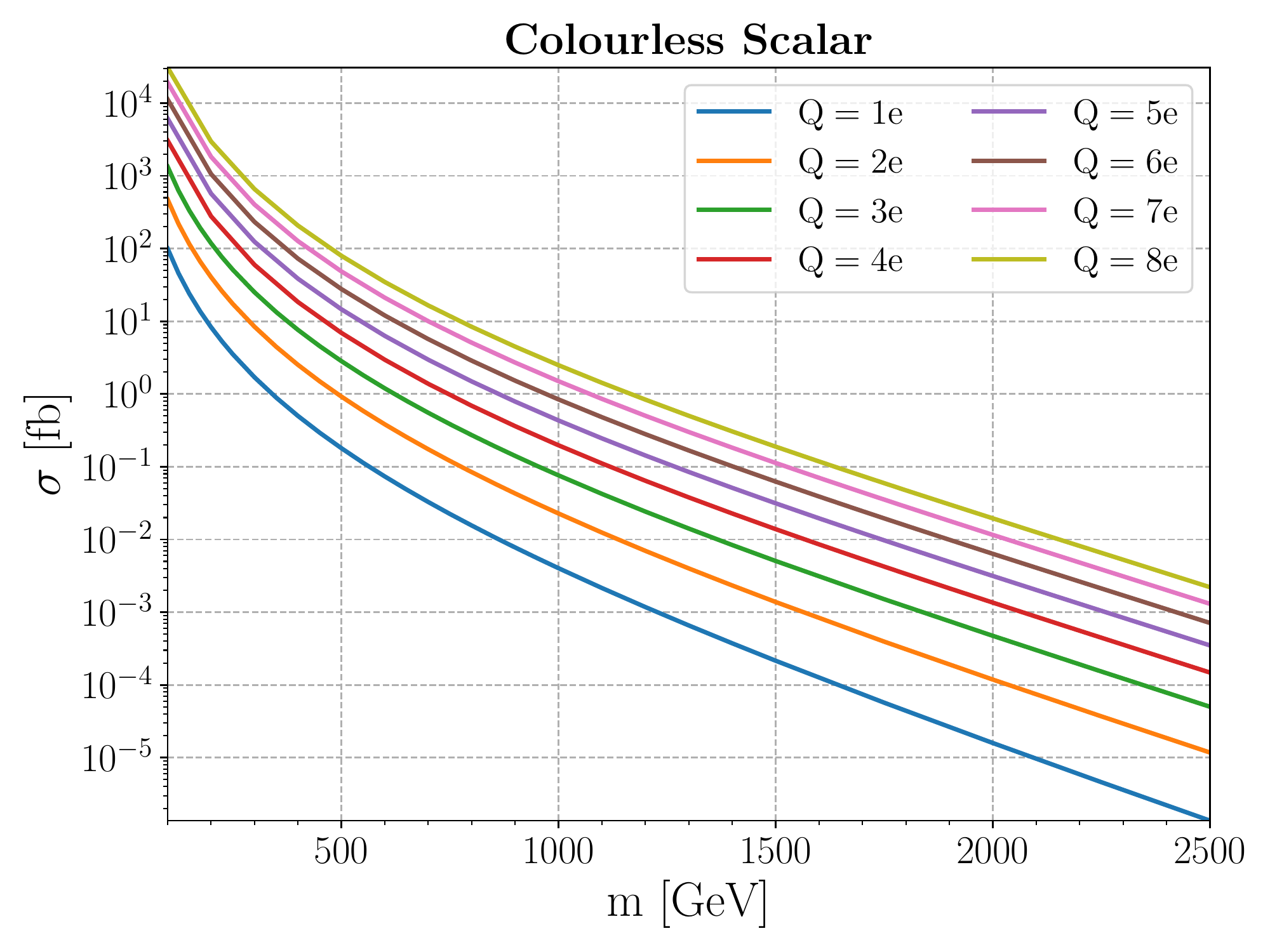}
\includegraphics[width=0.48\textwidth]{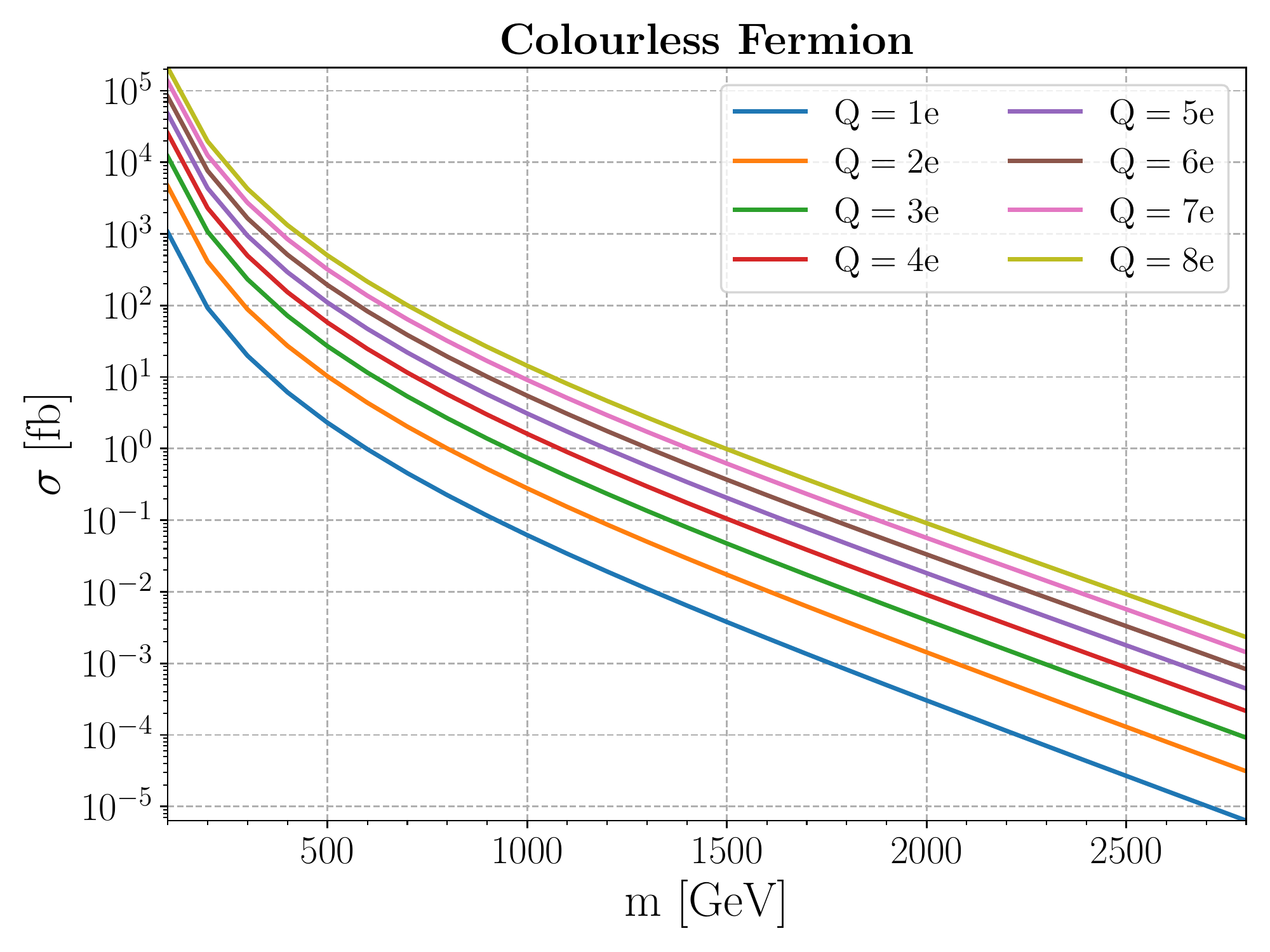}
\caption{\label{fig:xs_cless}  Pair-production cross section for colourless 
scalars (left) and fermions (right). Lines with different colours correspond to particles with electric charges from 1 to 8 times the elementary charge. 
}
\end{figure}

In Fig.~\ref{fig:xs_cless}
we show the leading order cross sections 
of open production modes 
for colour-singlet 
scalars (left) and fermions (right).
The cross sections are calculated with {\tt MadGraph\,5} 
\cite{Alwall:2007st,madgraph5,Alwall:2014hca}
using
{\texttt LUXqed}
PDF set 
(\texttt{LUXqed17\_plus\_PDF4LHC15\_nnlo\_100})
\cite{Manohar:2016nzj,Manohar:2017eqh}, 
which includes improved calculation for the photon PDF.
We see that around $m = 100$ GeV
the cross section varies more than two orders of magnitude from $|Q/e| = 1$ to 8.
The impact of $Q$ on the cross section is stronger for larger masses and for scalars since the photon fusion channel is relatively more important than the Drell-Yan channel for those cases. 
For scalars with $m = 2$ TeV, the cross section with $Q=8e$ is more than three orders of magnitude larger than that with $Q=1e$.    
For $Q \gg 1e$, we expect
that the electromagnetic higher order correction is sizeable.  
However, the next-to-leading order cross sections
for multicharged particles have not been well studied and including this effect is beyond the scope of this 
study.

\begin{figure}[b!]
\centering
\includegraphics[width=0.49\textwidth]{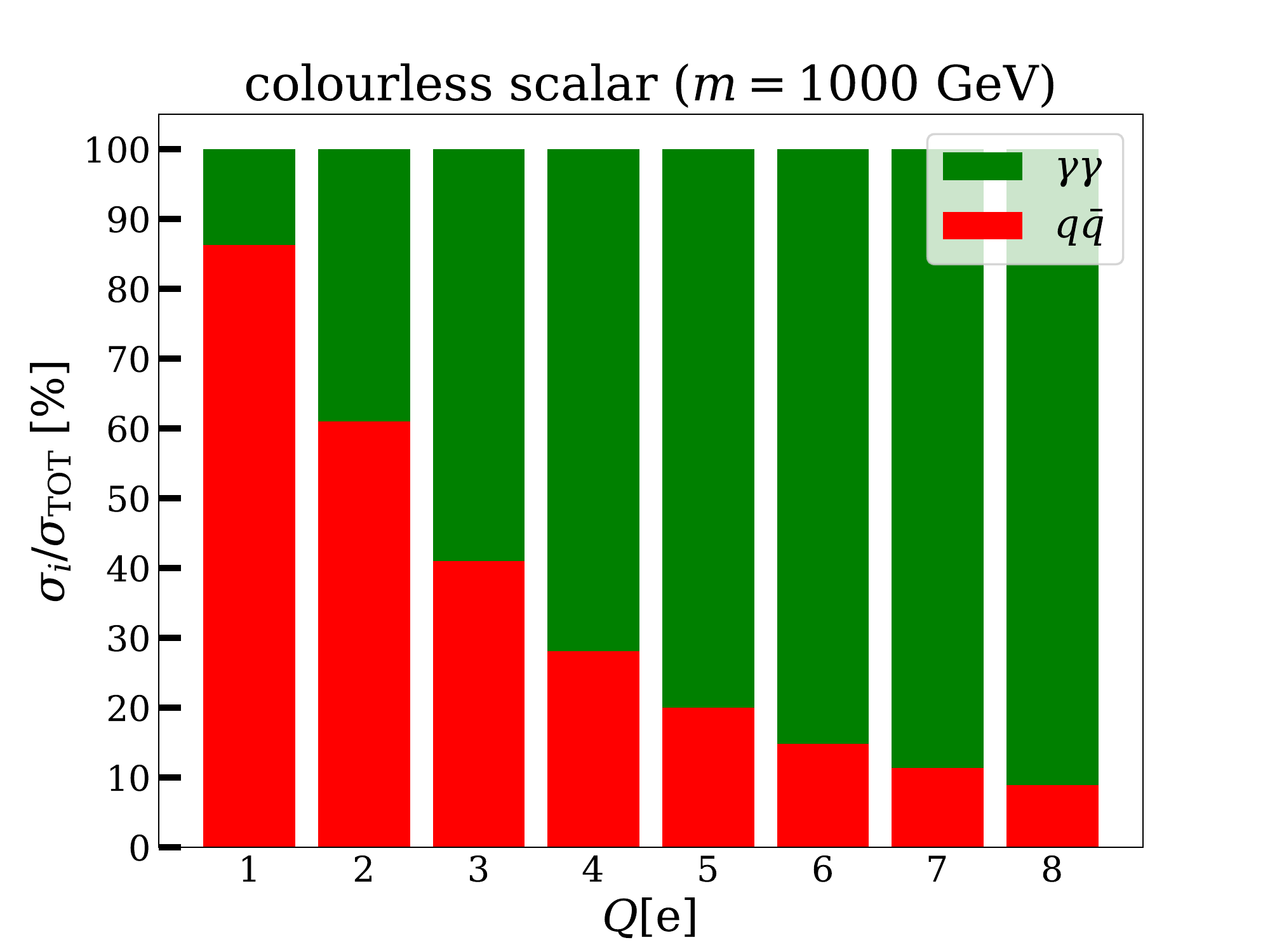}
\includegraphics[width=0.49\textwidth]{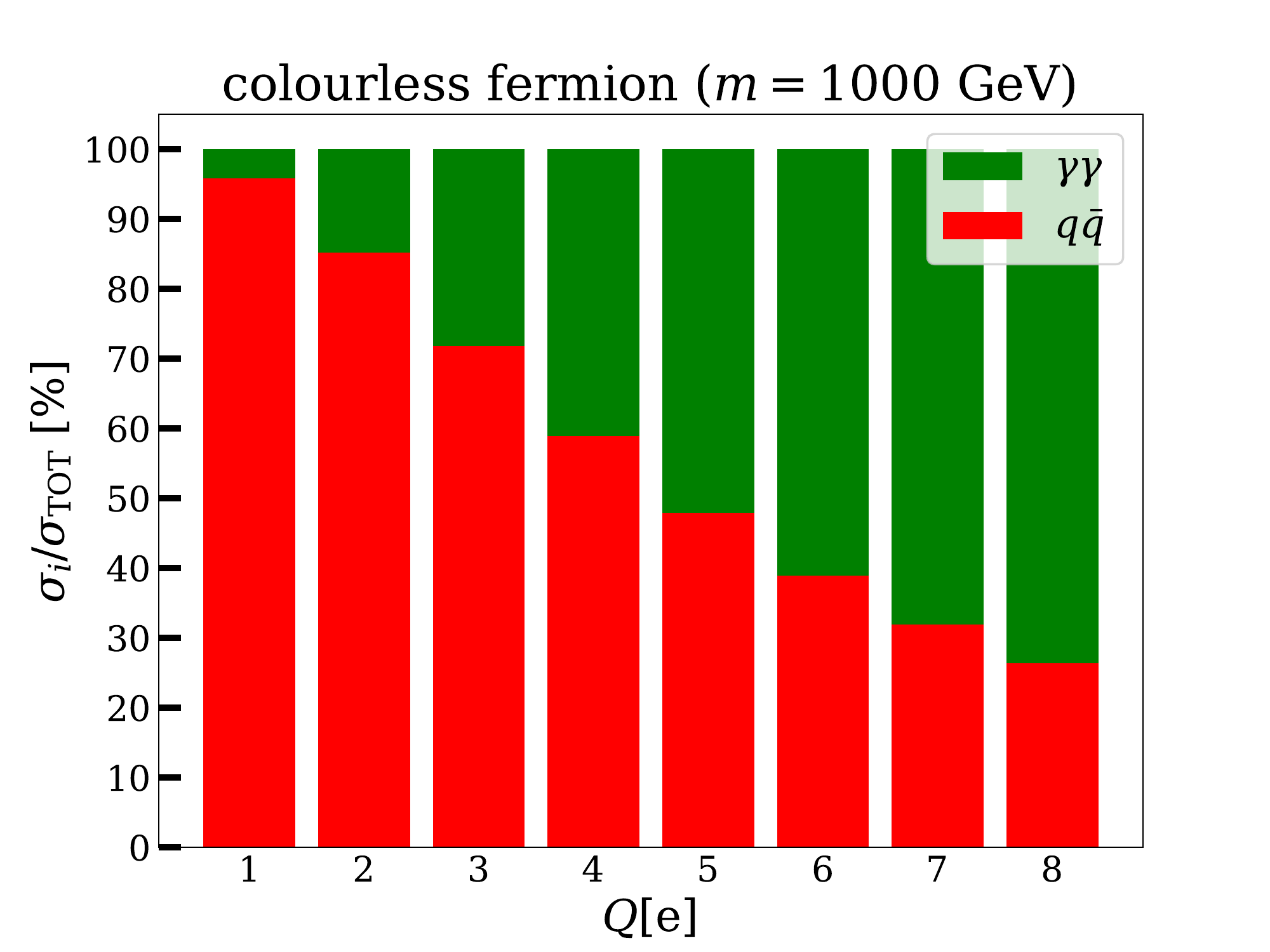}
\caption{\small\label{fig:frac} Contributions of the $q \bar q$ (red) and $\gamma \gamma$ (green) initial states to the production rate for the colourless scalars (left) and fermions (right) with $m=1$ TeV.} 
\end{figure}

Figure~\ref{fig:frac} shows the relative contributions 
from the Drell-Yan and photon fusion
channels to the total cross section
for scalars (left) and fermions (right) with $m = 1$ TeV. 
We see that the photon fusion is 
subdominant for $Q = 1e$ due to the photon PDF suppression. 
Moving to larger $Q$, 
the relative importance of photon fusion 
rapidly grows since 
$\hat \sigma_{\gamma \gamma}/\hat \sigma_{q \bar q} \propto Q^2$
as mentioned above,
and becomes dominant for $Q/e \gtrsim 3$ (5) for scalars (fermions). 
This growth is faster for scalars due to the following reason.
The Drell-Yan production 
(with off-shell $s$-channel $\gamma/Z$ exchange)
from the $q \bar q$ initial state
necessarily has $J=1$ ($p$-wave).
In order for scalars to have non-zero cross sections,
the final state configuration must have a non-zero angular momentum, 
which indicates the cross section must be suppressed 
by the positive power of the velocity of the final state particles
at the centre of mass frame.
This $p$-wave suppression 
is absent for the photon fusion channel,
which makes this contribution relatively more important and quickly dominates 
over the Drell-Yan for the scalar case.
The $p$-wave suppression is absent
in the fermion production both for
Drell-Yan and photon fusion channels.

\subsection{Colour-triplet particles and a hadronisation model}

For colour-triplet particles,
the following processes contribute to the open production:
\begin{itemize}
\item
Drell-Yan: $q \bar q$ initial state 
with off-shell $g/\gamma/Z$ exchange in $s$-channel 
\item
gluon fusion: $g g$ initial state 
with QCD interaction
\item
gluon-photon fusion: $g \gamma$ initial state with 
mixed QCD and QED $t$-channel interaction 
\item 
photon fusion: $\gamma \gamma$ initial state with pure QED $t$-channel interaction
(and the 4-point interaction for scalars)
\end{itemize}

\begin{figure}[b!]
\centering
\includegraphics[width=0.48\textwidth]{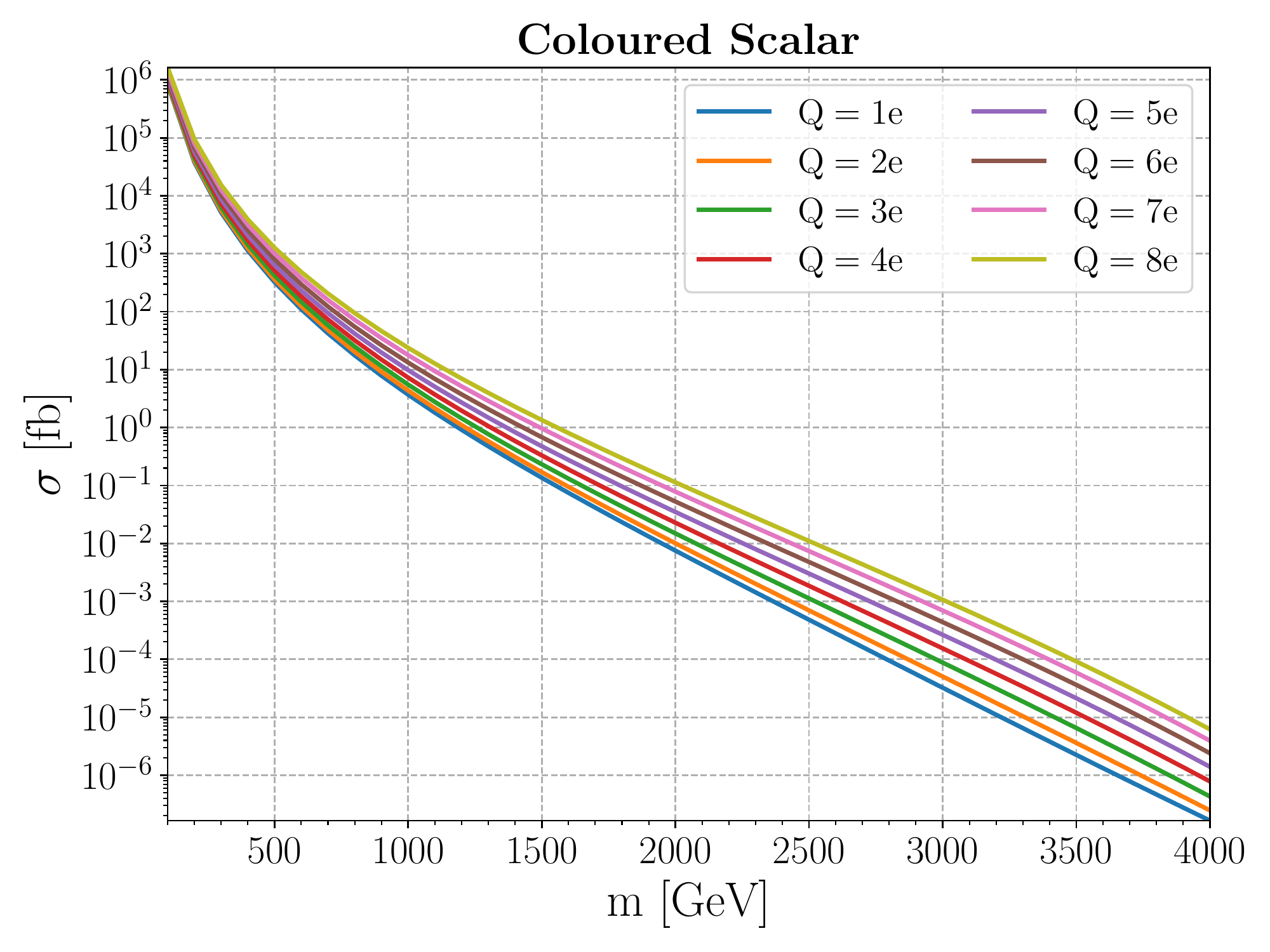}
\includegraphics[width=0.48\textwidth]{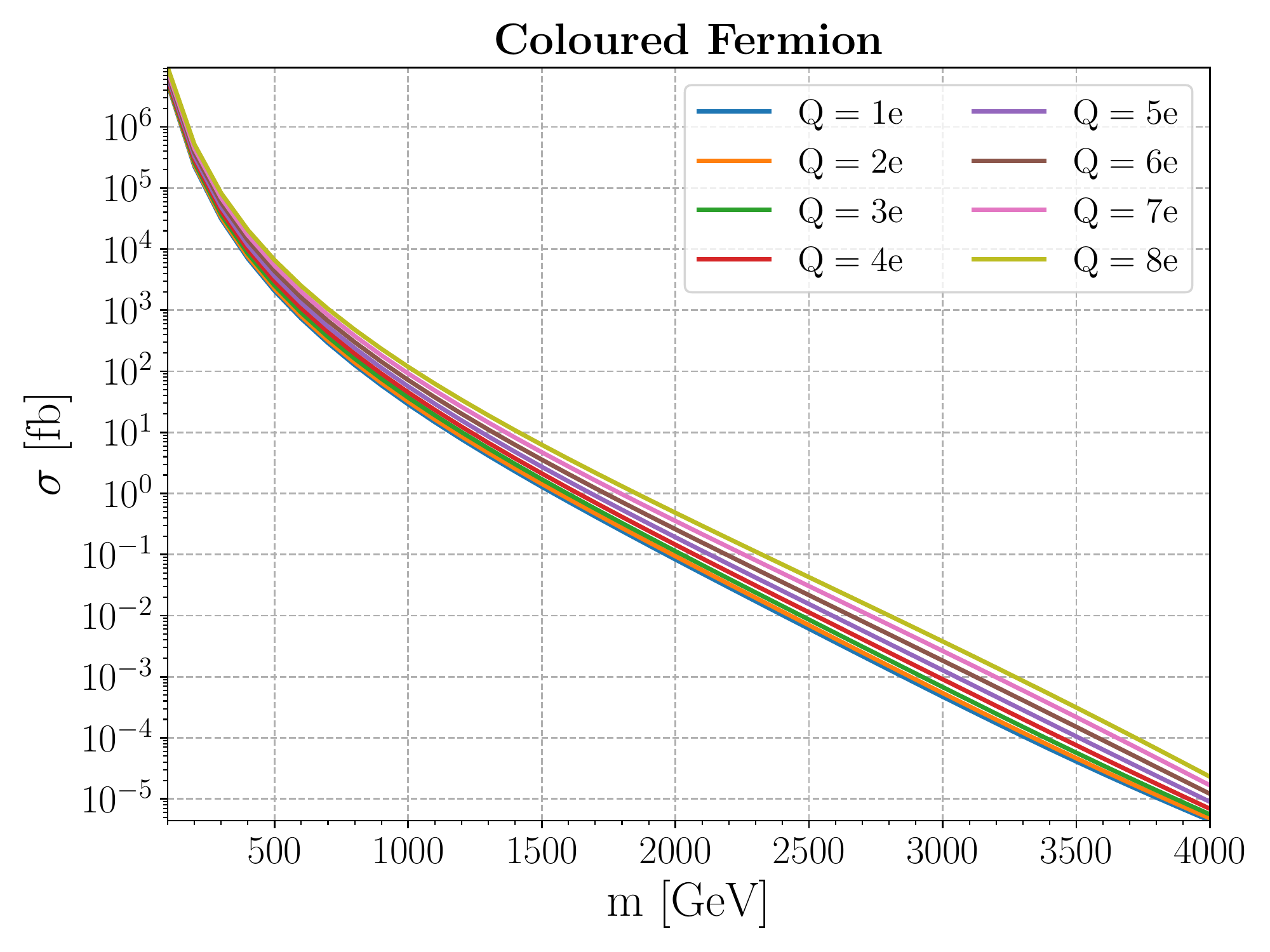}
\caption{\label{fig:xs_c}  
\small Leading order production cross sections, $pp \to \xi^{+Q} \xi^{-Q}$, for
coloured scalars (left) and fermions (right). 
The curves with different colours correspond to different electric charges.
}
\end{figure}
\begin{figure}[t!]
\centering
\includegraphics[width=0.49\textwidth]{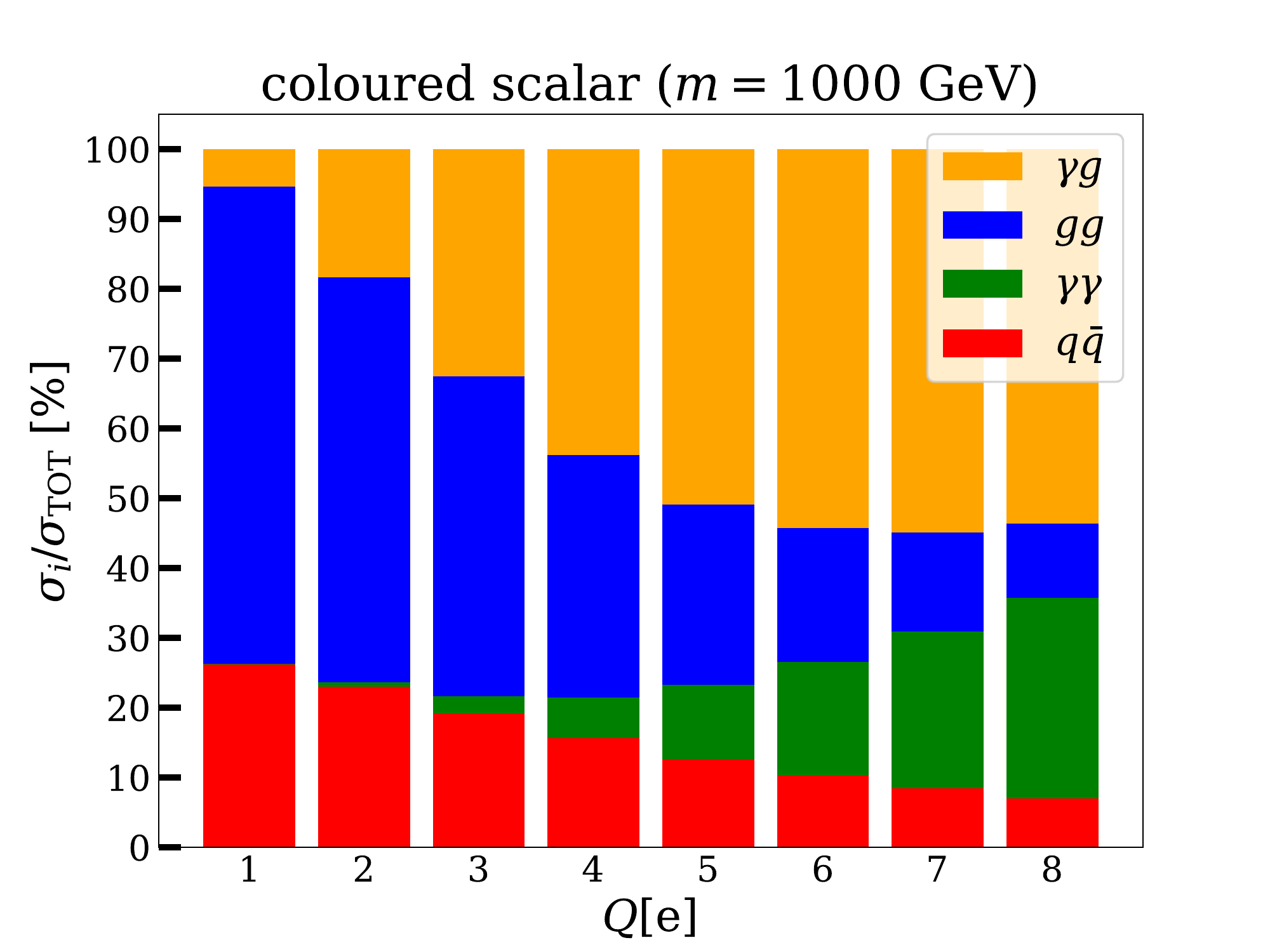}
\includegraphics[width=0.49\textwidth]{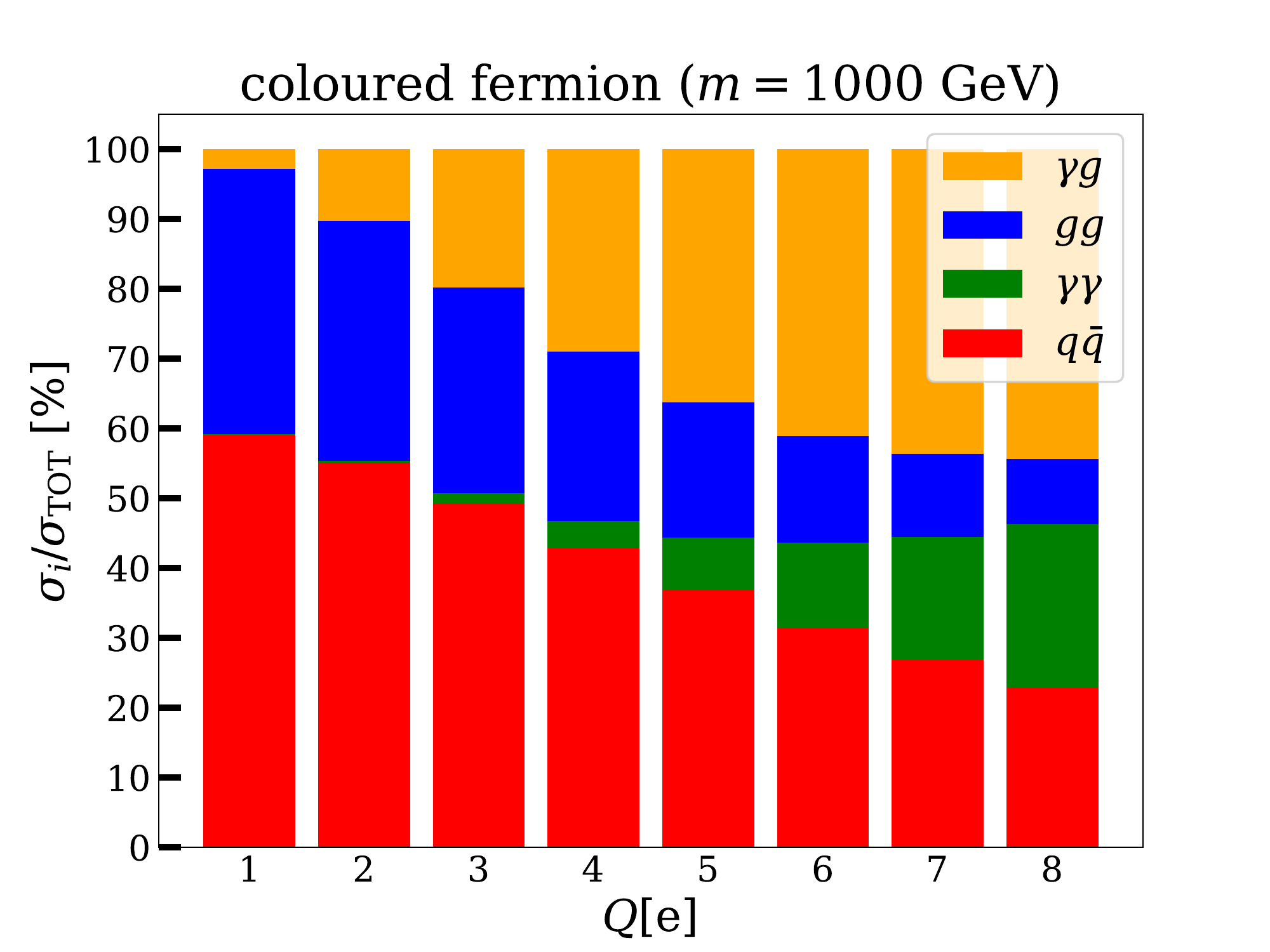}
\caption{\small\label{fig:cfrac} Contributions of various initial states, 
$q \bar q$ (red),
$\gamma \gamma$ (green),
$gg$ (yellow),
$\gamma g$ (blue),
to the production rate of coloured scalars (left)
and fermions (right) with $ m= 1$ TeV.} 
\end{figure}

Figure~\ref{fig:xs_c}
shows the leading order cross section 
for the open production channel
$pp \to \xi^{+Q} \xi^{-Q}$
for colour-triplet scalars (left)
and fermions (right)
as a function of the mass.
Compared with the colour-singlet case,
the $Q$-dependence of the cross section 
is much weaker 
particularly for small values of $m$.
This is because the 
production channels with QCD interaction 
(Drell-Yan with gluon $s$-channel exchange, $gg$ and $g \gamma$ fusions)
take up a large fraction of the total cross section
for all values of $Q/e$ (up to 8),
as can be seen in Fig.~\ref{fig:cfrac},
and they are independent or only weakly dependent on $Q$.
The QCD interaction is relatively more important for
lighter $\xi^{\pm Q}$ due to the enhancement of gluon PDF
at a small $x$ (momentum fraction) region.
As mentioned above, electromagnetic higher order corrections for $Q>1e$ particles are not available to date. 
For consistency, we use the leading-order cross sections also for the QCD productions throughout this paper.

If colour-triplet particles have a sufficiently long lifetime,
they draw quark-antiquark pairs from the vacuum so that  they
dress them and hadronise into colour-singlet states.
The electric charge of the colour-singlet heavy hadrons $\widetilde Q$
gets modified by $\Delta Q$ from the original charge $Q$ of $\xi^{+Q}$
due to the extra quarks; $\widetilde Q = Q + \Delta Q$.
We estimate this effect using a simple hadronisation model,
assuming that a colour-triplet particle goes into one of the mesonic states with
probability $k$ and one of the baryonic states otherwise.
The model assigns an equal probability to every state among the mesonic (baryonic) states.
The colour-singlet states in this hadronisation model are listed 
in Tables \ref{tb:had_scalar} and \ref{tb:had_fermion} for a scalar ($\phi^{+Q}$)
and a fermion ($\psi^{+Q}$) particle, respectively, together with the charge shift $\Delta Q$ and
the probability $p$ assigned to each state.
In estimating the sensitivity of the ATLAS, CMS and MoEDAL experiments, we vary the free parameter $k$ from 0.3 to 0.7 
for a ball-park estimation of the uncertainty originated from the hadronisation effect. 

%
\begin{table}[!h]
\centering
\vspace{3mm}
\begin{tabular}{|c|c|c|}
\multicolumn{3}{c}{\small spin $\frac{1}{2}$ mesons}\\
\hline
state & $\Delta Q/e$ & $p$ \\
\hline
$\phi^{+Q} + \bar u_{L/R}$ & $- \frac{2}{3}$ & $\frac{k}{2}$ \\
$\phi^{+Q} + \bar d_{L/R}$  & $+ \frac{1}{3}$ & $\frac{k}{2}$ \\ 
\hline
\end{tabular}
\hspace{3mm}
\begin{tabular}{|c|c|c|}
\multicolumn{3}{c}{\small spin 0 baryons}\\
\hline
state & $\Delta Q/e$ & $p$ \\
\hline
$\phi^{+Q} + u_L u_R$ & $+ \frac{4}{3}$ & $\frac{1-k}{6}$ \\
$\phi^{+Q} + d_L d_R$  & $- \frac{2}{3}$ & $\frac{1-k}{6}$ \\ 
$\phi^{+Q} + u_L d_R$  & $+ \frac{1}{3}$ & $\frac{1-k}{6}$ \\ 
$\phi^{+Q} + d_L u_R$  & $+ \frac{1}{3}$ & $\frac{1-k}{6}$ \\ 
\hline
\end{tabular}
\hspace{3mm}
\begin{tabular}{|c|c|c|}
\multicolumn{3}{c}{\small spin 1 baryons}\\
\hline
state & $\Delta Q/e$ & $p$ \\
\hline
$\phi^{+Q} + u_L d_L$ &$+ \frac{1}{3}$ & $\frac{1-k}{6}$ \\
$\phi^{+Q} + u_R d_R$  & $+ \frac{1}{3}$ & $\frac{1-k}{6}$ \\ 
\hline
\end{tabular}
\vspace{3mm}
\caption{\small A hadronisation model for a colour-triplet scalar particle ($\phi^{+Q}$).
The charge shift $\Delta Q$ and the probability $p$ assigned to each state are shown
in the second and third columns, respectively. 
\label{tb:had_scalar}
}
\end{table}

\begin{table}[!h]
\small
\centering
\vspace{-5mm}
\begin{tabular}{|c|c|c|}
\multicolumn{3}{c}{\small spin 0 mesons}\\
\hline
state & $\Delta Q/e$ & $p$ \\
\hline
$\psi^{+Q} + \bar u_{L}$  & $- \frac{2}{3}$ & $\frac{k}{4}$ \\
$\psi^{+Q} + \bar d_{L}$  & $+ \frac{1}{3}$ & $\frac{k}{4}$ \\ 
\hline
\end{tabular}
\hspace{3mm}
\begin{tabular}{|c|c|c|}
\multicolumn{3}{c}{\small spin 1 mesons}\\
\hline
state & $\Delta Q/e$ & $p$ \\
\hline
$\psi^{+Q} + \bar u_{R}$  & $- \frac{2}{3}$ & $\frac{k}{4}$ \\
$\psi^{+Q} + \bar d_{R}$  & $+ \frac{1}{3}$ & $\frac{k}{4}$ \\ 
\hline
\end{tabular}
\hspace{3mm}
\begin{tabular}{|c|c|c|}
\multicolumn{3}{c}{\small spin $\frac{1}{2}$ baryons}\\
\hline
state & $\Delta Q/e$ & $p$ \\
\hline
$\psi^{+Q} + u_L u_R$ & $+ \frac{4}{3}$ & $\frac{1-k}{5}$ \\
$\psi^{+Q} + d_L d_R$  & $- \frac{2}{3}$ & $\frac{1-k}{5}$ \\ 
$\psi^{+Q} + u_L d_R$  & $+ \frac{1}{3}$ & $\frac{1-k}{5}$ \\ 
$\psi^{+Q} + d_L u_R$  & $+ \frac{1}{3}$ & $\frac{1-k}{5}$ \\ 
$\psi^{+Q} + u_L d_L$  & $+ \frac{1}{3}$ & $\frac{1-k}{5}$ \\ 
\hline
\end{tabular}
\vspace{3mm}
\caption{A hadronisation model for a colour-triplet fermionic particle ($\psi^{+Q}$).
The charge shift $\Delta Q$ and the probability $p$ assigned to each state are shown
in the second and third columns, respectively. 
For fermionic particles, $\psi$, we consider
hadronic states up to spin 1.
\label{tb:had_fermion}
}
\end{table}

\section{Detecting open production at ATLAS and CMS}
\label{sec:open_ATLAS}

Experiments with general purpose detectors, ATLAS and CMS,
have been actively looking for long-lived charged particles.
Their concrete analyses differ in the details 
but the general strategy is the same.
The most stringent and general constraint
on the open production mode
comes from so-called ``large $dE/dx$''
searches, which look for 
tracks with anomalously high energy deposition 
in the inner tracker.
ATLAS analysed their full Run 1 data
(8 TeV, 20.3 fb$^{-1}$) 
and interpreted their result
for spin-$\frac{1}{2}$ particles with
$2 \le Q/e \le 6$ \cite{ATLAS:2015hau}.
This search was updated 
with the 13 TeV 36.1 fb$^{-1}$ data
in ref.~\cite{ATLAS:2018imb},
where the result was interpreted for
fermions
with $Q/e$ ranging from 2 to 7.
CMS
performed their analysis on 
a combined dataset
of 7 and 8~TeV $pp$ collisions with
integrated luminosity of
5.0 and 18.8 fb$^{-1}$, respectively \cite{CMS:2013czn}.
The most recent result using 
the 13 TeV, 2.5 fb$^{-1}$ data
has also been published
\cite{CMS:2016kce}.
The results of these analyses are  
interpreted for 
charged fermions with $|Q/e| \le 2$ and
long-lived supersymmetric particles.

For multiply charged particles 
with $Q/e \ge 2$,
the above ATLAS and CMS analyses have performed  only for colourless fermions.
Moreover, their analysis considered only Drell-Yan production channels.
Ref.~\cite{Jager:2018ecz}
has recast  
the CMS analysis \cite{CMS:2013czn}
for coloured scalars and fermions
and improved the original analysis by
including the photon-fusion\footnote{Importance of the photon-fusion channel 
in the multiply charged particle searches 
has been pointed out in Ref.~\cite{Barrie:2017eyd}.}
as well as gluon-gluon and gluon-photon initiated processes. 
In this article 
we follow the work of ref.~\cite{Jager:2018ecz}
and extend it by adding
colourless scalars and estimation of projected sensitivities for Run~3 and HL-LHC.

As outlined in ref.~\cite{Jager:2018ecz},
our recasting is based on the CMS analyses \cite{CMS:2013czn,CMS:2016kce},
in particular
the
``Tracker + Time of Flight (TOF)''
selection.
The signal region is comprised 
of several kinematical cuts,
such as \pt and $\eta$ of charged tracks,
as well as more subtle requirements, 
e.g.\ conditions on the number of pixel hits and the uncertainty of $1/\beta$ measurements.
We observe that the class of former cuts
is sensitive to the spin and production mechanisms and its efficiency may be estimated by Monte Carlo (MC) simulations. 
On the other hand, the latter depends more crucially on the masses and charges of the particles.
In order to obtain the final signal efficiency, 
$\epsilon$, we first estimate the efficiency of 
the online (muon trigger) selection, $\epsilon_{\rm on}$,
by MC simulations.
We then take the cut-flow tables 
for the offline selection
provided in ref.~\cite{thesis}
for different masses and charges.
The offline efficiency is given by
the product of relative efficiencies 
of the cuts with respect to the previous ones 
in the cut-flow table;
$\epsilon_{\rm off}(m, Q) = \prod_{i} \epsilon_i(m,Q)$,
where 
$\epsilon_i \equiv ($\# of events passed all cuts up to $i$)/(\# of events passed all cuts up to $i-1$).\footnote{
The condition $i$ appears earlier than the condition $j$ in the cut-flow table if $i < j$.
The relative efficiency for the first 
offline cut, $\epsilon_1$, is defined with respect to the online selection.}
We estimate the relative efficiencies of kinematical cuts by MC and replace the values in the cut-flow table.
Since the relative efficiencies of non-kinematical cuts are not very sensitive to the spin, production mechanism and the collider energy, 
we use the same values provided in ref.~\cite{thesis}.
In the end the final signal efficiency is given by
$\epsilon = \epsilon_{\rm on} \cdot \epsilon_{\rm off}$.

There are several caveats in estimating the efficiencies.
In the ATLAS and CMS analyses 
the momentum of charged particles 
is estimated
from the curvature 
of the track assuming $|Q| = 1 e$.
Since the curvature radius is proportional to $p_T/Q$ of the particle, 
this leads to underestimation of the momentum
of the charged particle
by a factor $1/|Q/e|$.
For example, the online selection 
requires at least one muon candidate 
with the measured transverse momentum larger than 50 GeV \cite{CMS:2016kce}.  This translates to
the condition $\pt > 50 \cdot |Q/e|$ GeV
with \pt being the true transverse momentum of the charged particle.
Due to this effect, the efficiency 
of the \pt cut tends to be smaller for 
particles with larger $|Q|$.

There is an implicit assumption in the muon trigger requirement;
the track measured in the inner detector (ID) must match the one measured in the muon system (MS) in the same bunch crossing.
If highly charged particles are produced 
and travel through the detector,
they lose their velocity
due to interactions with detector materials,
and may arrive at the MS 
too late after the next bunch crossing.
Following \cite{Jager:2018ecz},
we denote the minimum distance to which a particle must travel so that the momentum is measured at the MS, as $x_{\rm trigger}$, as a function of $\eta$.
We then calculate,
for each produced charged particle 
 in the simulation,
the amount of time needed 
to travel
to the distance $x_{\rm trigger}$
and denote it as $t_{\rm TOF}$.
This calculation is performed 
using the procedure outlined 
in the appendix of \cite{Jager:2018ecz},
which is based on 
the Bethe-Bloch formula \cite{Zyla:2020zbs}
and the information of the detector 
geometry and materials.
Since the bunch crossing at the LHC
occurs every 25 ns, we demand
$t_{\rm TOF} - x_{\rm trigger}/c 
< 25\,{\rm ns}$
to pass the muon trigger,
where the left-hand-side 
represents the time separation between
the detection of the massive charged particle and that of massless particles
in the same bunch crossing.
For charged particles with $|\eta| < 1.6$, the {\it resistive plate chamber (RPC) 
muon trigger} is applied, in which the threshold of time separation is relaxed to be 50 ns.
Notice that the energy loss 
in the Bethe-Bloch formula
is proportional to $Q^2$.
We therefore expect 
that the impact of this condition
is more prominent for 
particles with larger $|Q|$.

In summary, for the online selection (muon trigger) we require at least one candidate with
\begin{equation}
\pt \,>\, 50 \cdot |Q/e| ~{\rm GeV} \,, 
~~~
|\eta|  \,<\,  2.1 \,,
\nonumber
\end{equation}
\begin{equation}
t_{\rm TOF} - \frac{x_{\rm trigger}}{c}
\, < \,
\left\{
\begin{array}{ll}
50 ~{\rm ns} & \cdots~ |\eta| < 1.6  \\
25 ~{\rm ns} & \cdots~ 1.6 \le |\eta| < 2.1
\end{array}
\right..
\end{equation}
We then estimate and replace 
relative efficiencies of the following kinematical cuts in the offline selection:
\begin{eqnarray}
\pt &>& 65 \cdot |Q/e| ~{\rm GeV} \,, \nonumber \\
\frac{1}{\beta_{\rm MS}} & \equiv &  \frac{c \cdot t_{\rm TOF}}{x_{\rm trigger}} 
~ > ~ 1.25\,,
\end{eqnarray}
where $\beta_{\rm MS}$ is the velocity
measured at the MS.
As mentioned above, 
for the relative efficiencies 
of non-kinematical cuts
we use the values 
reported in the cut-flow tables 
in \cite{thesis}.
These tables provide the information 
only up to $m = 1$ TeV.
Since all $\epsilon_i(m,Q)$ of non-kinematical cuts, $i$, reach 
plateaus before $m = 1$ TeV,
we use $\epsilon_i(1\,{\rm TeV},Q)$
for particles heavier than 1 TeV.

\begin{figure}[t!]
    \centering
    \includegraphics[width=0.49\textwidth]{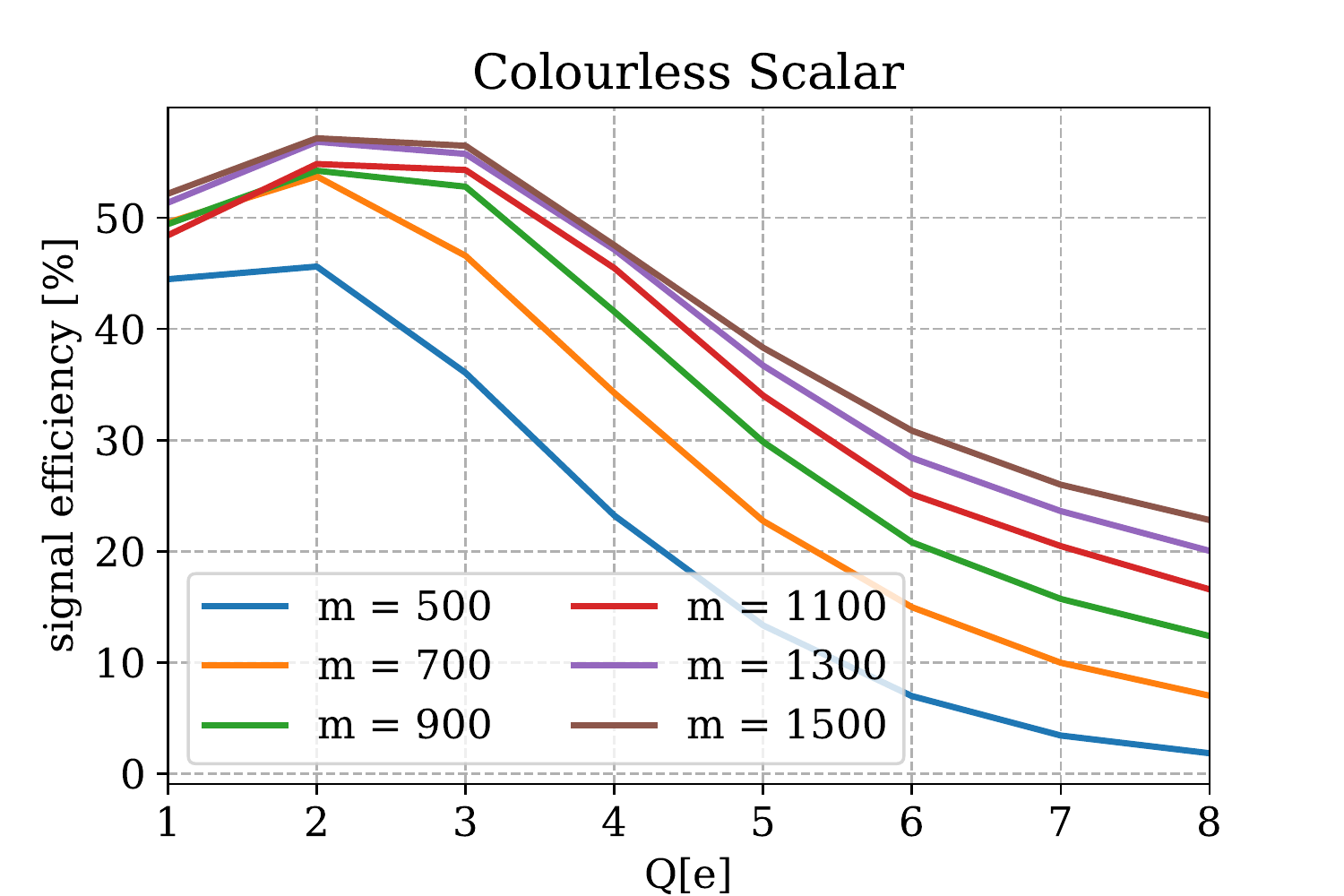}
    \includegraphics[width=0.49\textwidth]{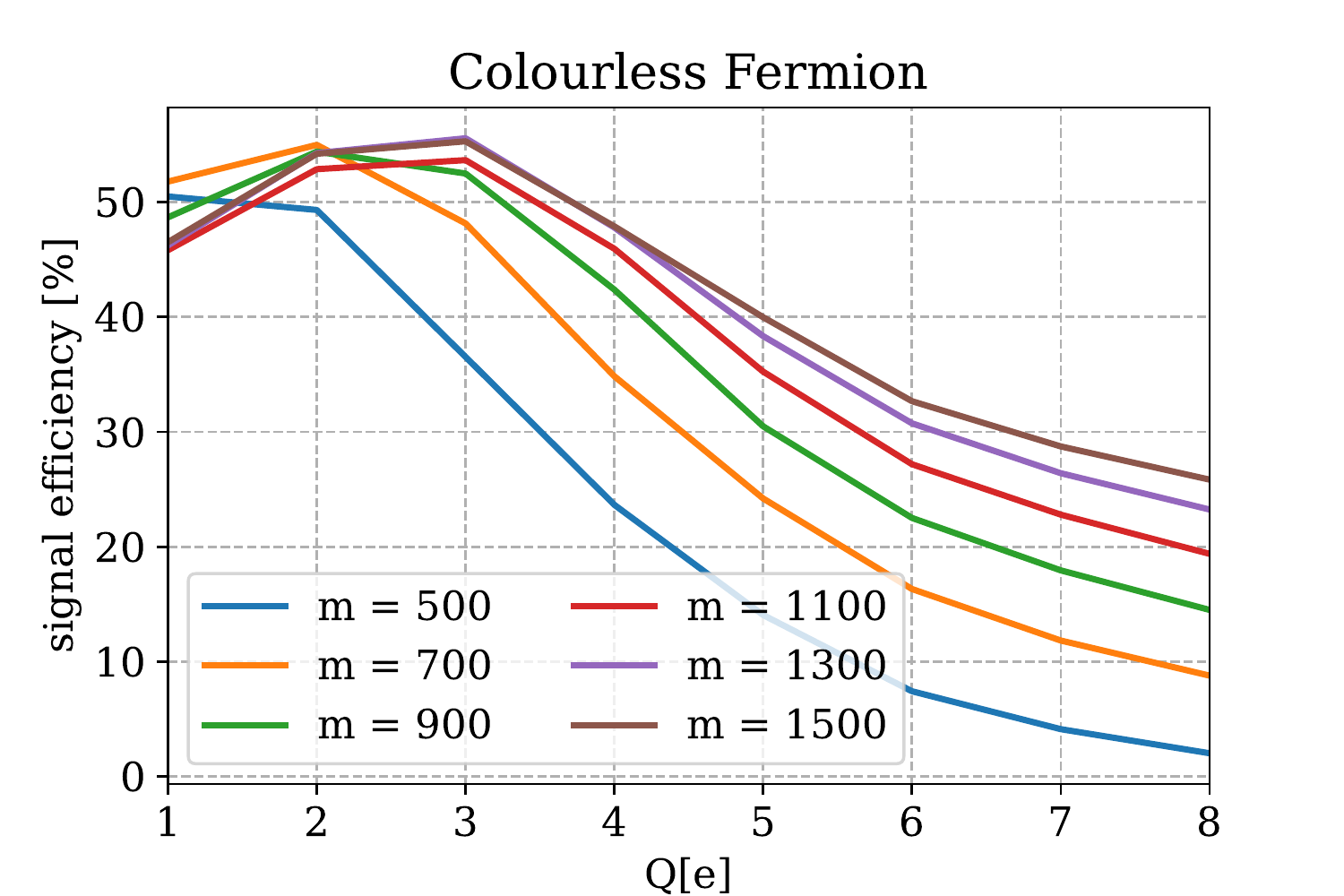}
    \includegraphics[width=0.49\textwidth]{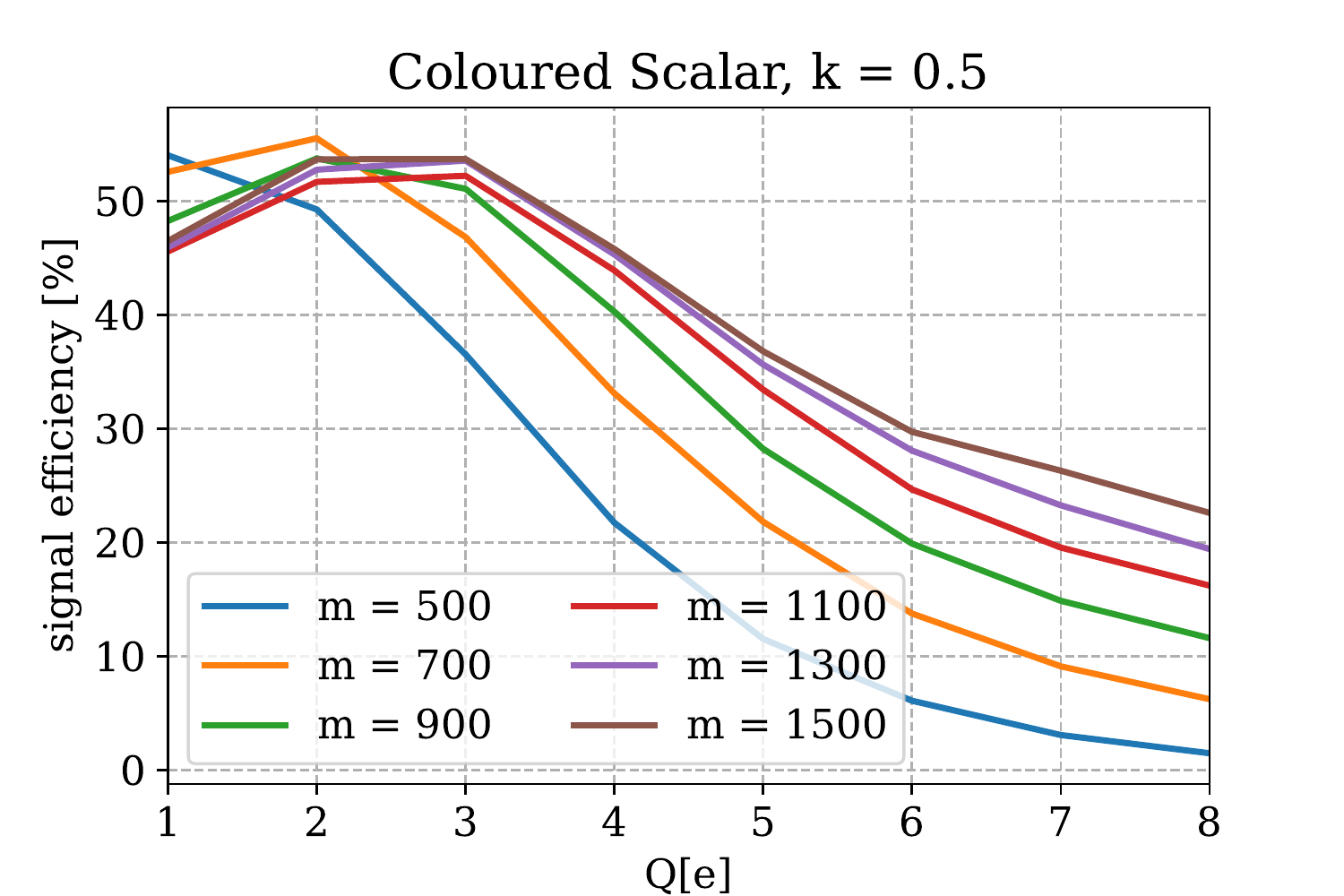}
    \includegraphics[width=0.49\textwidth]{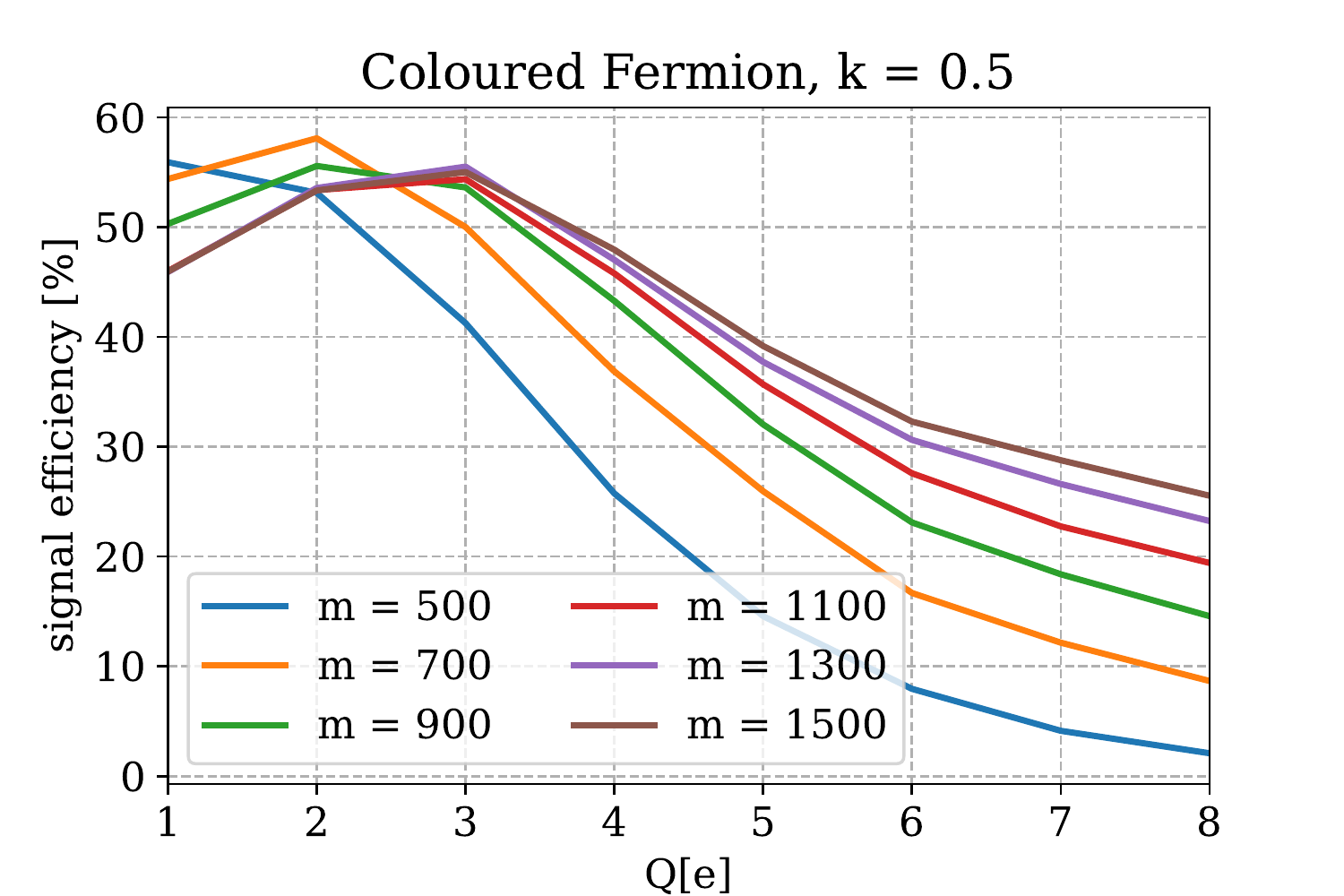}
    \caption{\small Top left: the signal efficiencies for different colourless scalars masses (in GeV), versus charge. Top right: same but for colourless fermions. Bottom left(right): same but for coloured scalars (fermions).}
    \label{fig:eff}
\end{figure}

In Fig.~\ref{fig:eff} we show the signal efficiency, $\epsilon = \epsilon_{\rm on} \cdot \epsilon_{\rm off}$,
as a function of $Q$.
In the figure, the left (right) panels are for scalar (fermionic) particles, and the top (bottom) panels correspond to the colourless (coloured) particles.  
The curves with different colours represent 
different masses of the charged LLPs.
For a fixed mass, we see a general tendency that the efficiency increases as moving from $|Q/e| = 1$ to $2$.
This is because charged tracks with $|Q/e|= 2$ deposit more energies 
in the detector
than $|Q/e|= 1$
and lead to a distinctive signature with large $dE/dx$.
However, the efficiency peaks around $|Q/e| = 2-3$ and decreases for larger $|Q|$.
This can be understood as for very large $|Q|$ ($|Q/e| \gtrsim 4$)
the $p_T$ cut ($p_T > 65 |Q/e|$) 
requires the particle to have a significantly large velocity,
which in turn makes it difficult to pass the $1/\beta_{\rm MS} > 1.25$ cut.
Since the velocity is lower for larger masses at a given momentum,
this also explains why the efficiency is higher for larger masses for the large $Q$ region.
On the other hand, we see  that at $|Q/e| = 1$ the efficiency is higher for $m = 500$ GeV
than heavier masses 
apart from the colourless scalar.
We checked that this is because the efficiency of the charged track isolation criterion is generally smaller for heavier particles.  
As we will see later,  
the production velocity of colourless scalars is generally higher compared 
to the other types of particles.   
The efficiency is therefore still smaller for $m=500$ GeV
than heavier masses 
due to the requirement 
$1/\beta_{\rm MS} > 1.25$.

\begin{figure}[t!]
    \centering
    \includegraphics[width=0.49\textwidth]{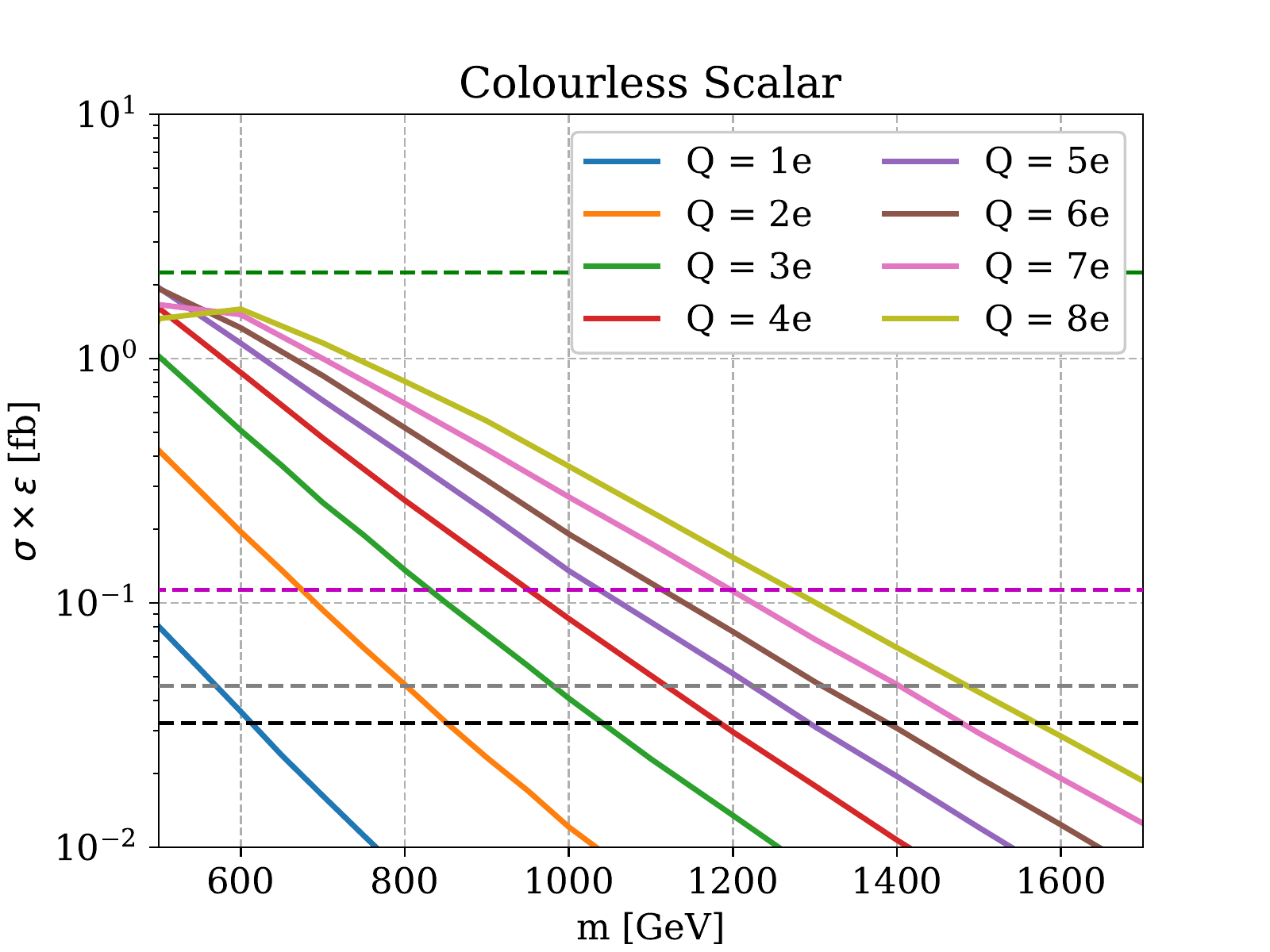}
    \includegraphics[width=0.49\textwidth]{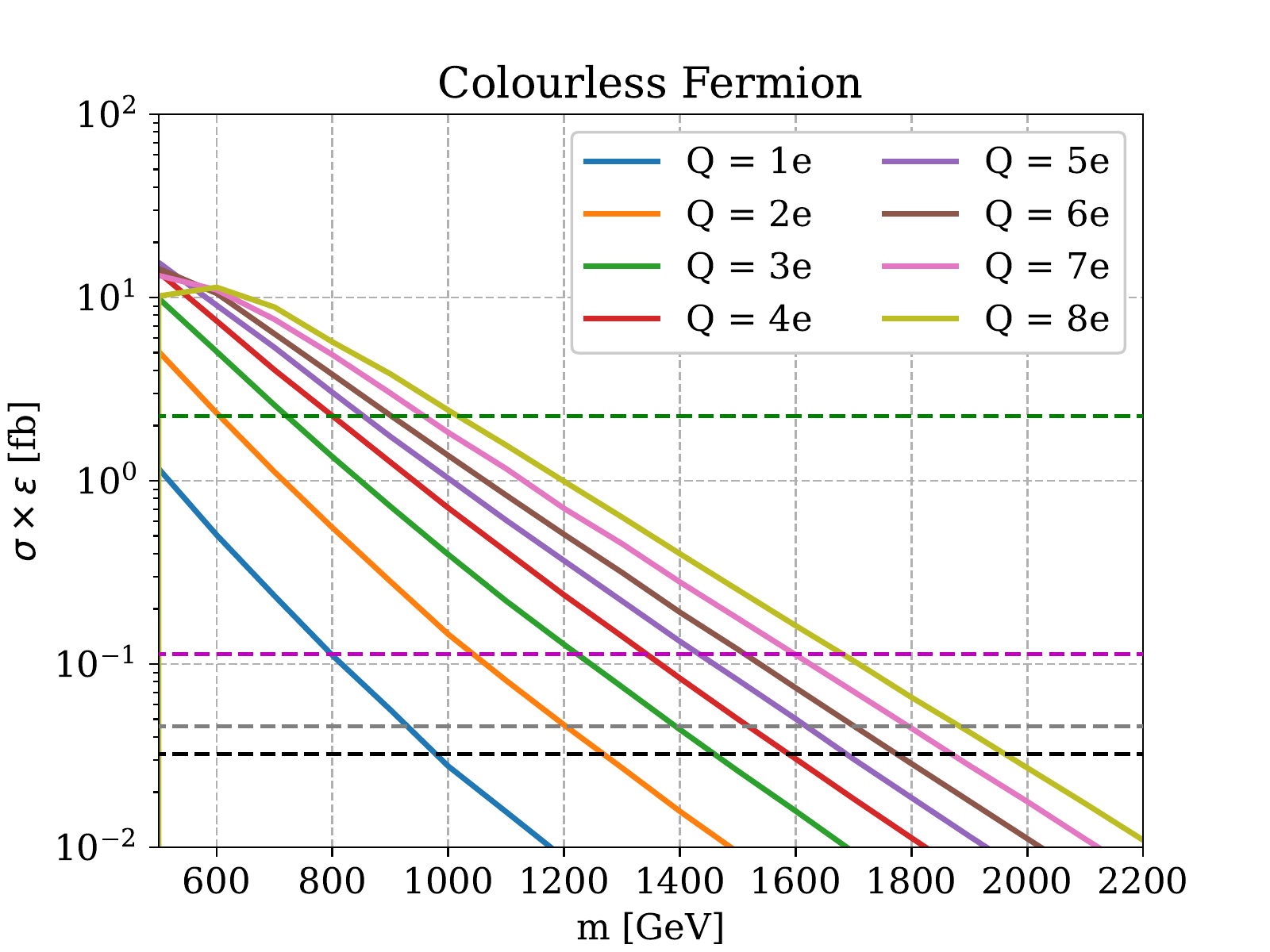}
    \includegraphics[width=0.49\textwidth]{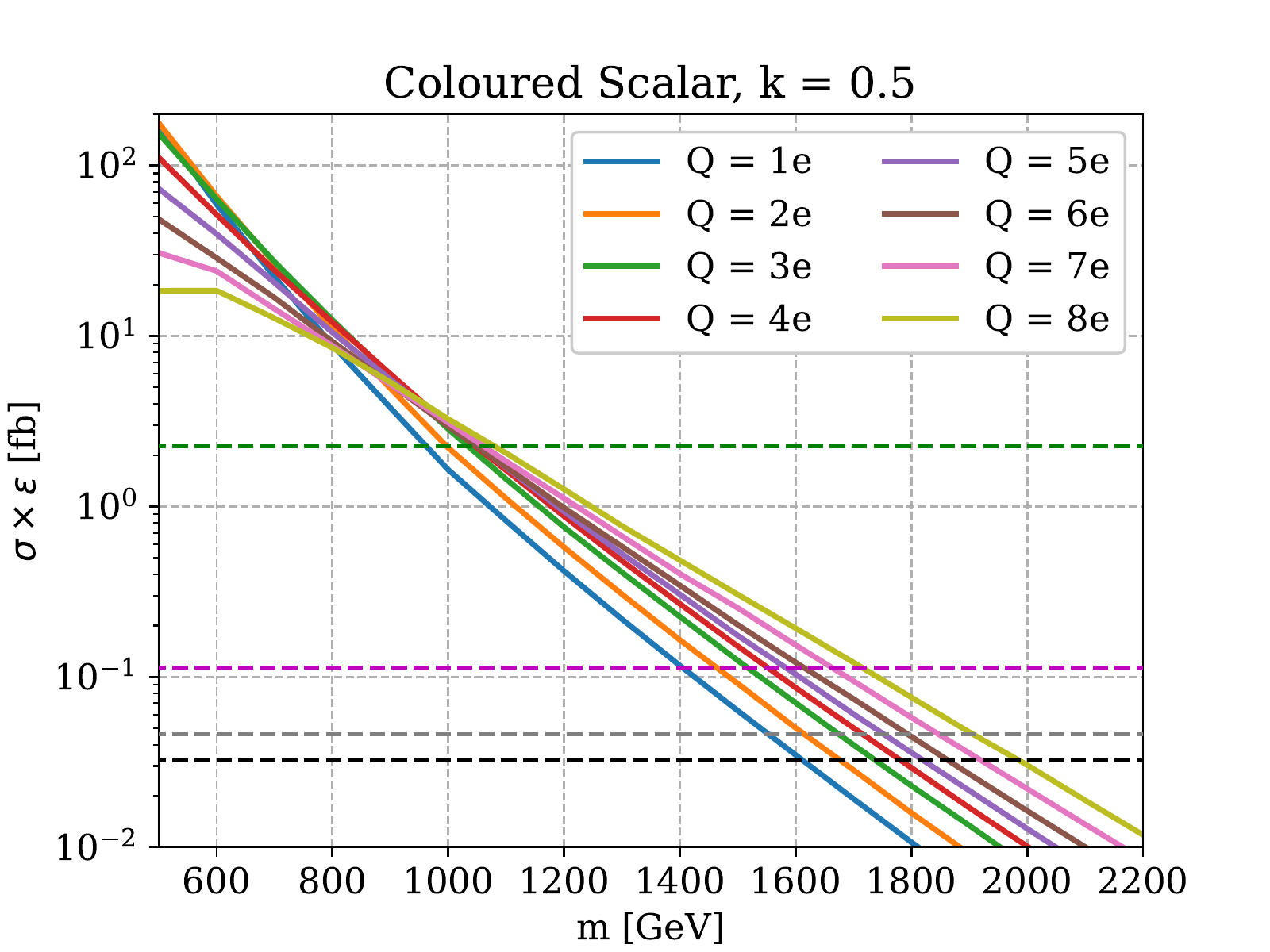}
    \includegraphics[width=0.49\textwidth]{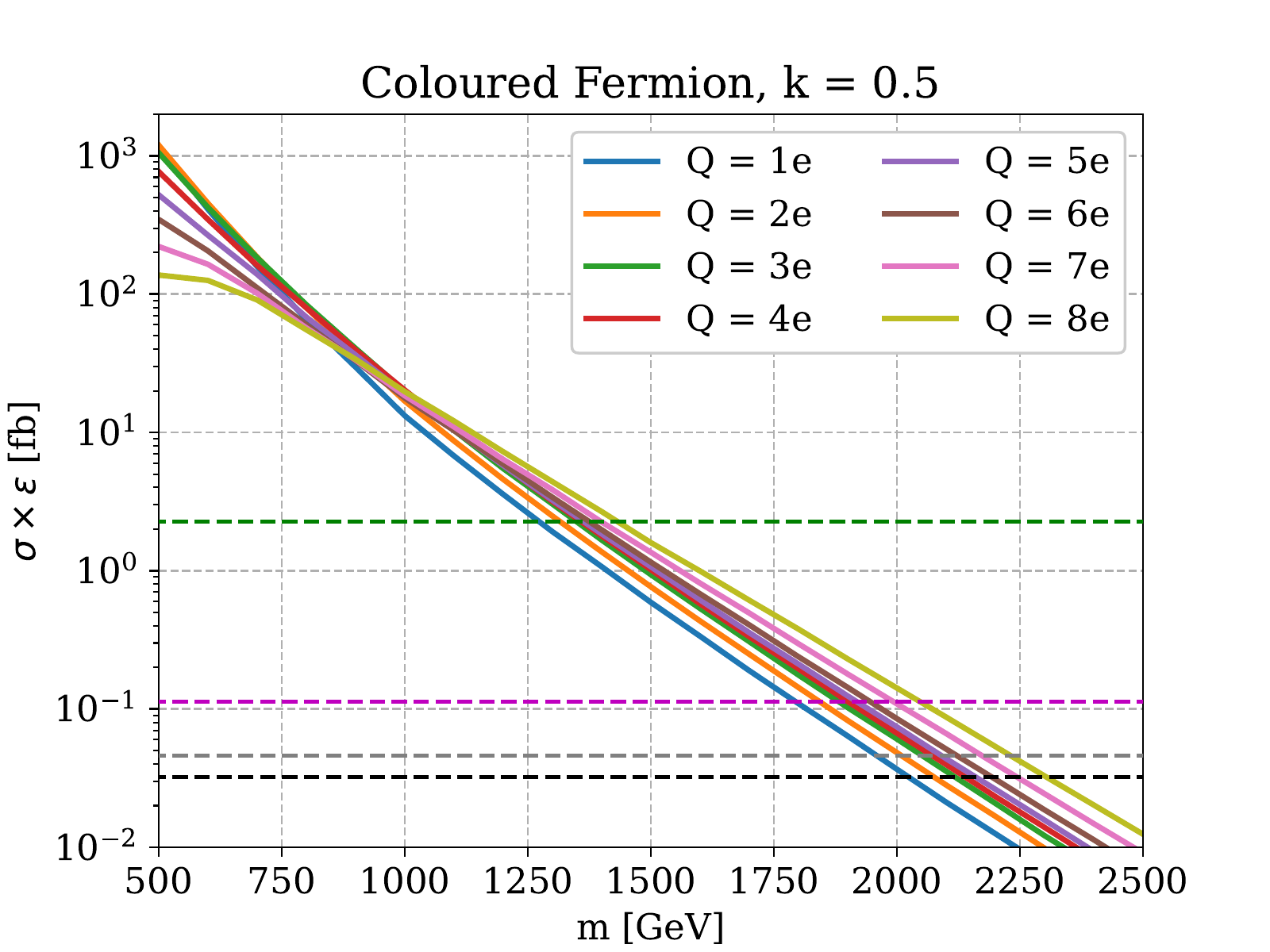}
    \caption{\small Top left: open-production channel signatures for colourless scalars. Solid: effective cross sections $\sigma \cdot \epsilon$ for CMS searches at $\sqrt{s} = 13$ TeV. Dashed: expected upper bounds for ${\cal L} = 2.5$ (green), 138 (magenta), 300 (grey) and 3000 fb$^{-1}$ (black). Top right: same but for colourless fermions. Bottom left: same but for coloured scalars. Bottom right:  coloured fermions.}
    \label{fig:xs_eff}
\end{figure}

In the upper panels of Fig.~\ref{fig:xs_eff}
we show the effective cross sections,
$\sigma_{\rm eff} \equiv \sigma \cdot \epsilon$,
of various colourless particles
(left; scalars, right; fermions).
As can be seen, 
the effective cross section
increases significantly 
for larger $Q$.
This is because the 
photon-fusion process 
is very important for colourless particles,
and its production rate grows with $Q^4$.

The lower panels of Fig.~\ref{fig:xs_eff}
show the effective cross sections 
of colour-triplet particles
(left; scalars, right; fermions).
We see that the $\sigma_{\rm eff}$ 
is less sensitive to $Q$.
This is because the 
production is dominated by
the QCD Drell-Yan, gluon-gluon and gluon-photon 
initiated processes, of which the contribution 
is either independent of $Q$ or grows with $Q^2$ for the gluon-photon fusion.
We note that for $m \lesssim 800$ GeV
the effective cross section  {\it drops} 
as a function of $Q$.
This behaviour is due to the fact that
the signal efficiency decreases for larger $Q$ ($Q \gtrsim 3e$), and this effect is stronger for lower masses, as shown in Fig.~\ref{fig:eff}.

The plots in Fig.~\ref{fig:xs_eff} 
also show the (projected) model-independent upper limits, $\sigma_{\rm eff}^{\rm UL}$,
on the effective cross section.
The dashed green line corresponds to the limit from the latest 13 TeV ${\cal L} = 2.5$ fb$^{-1}$ analysis \cite{CMS:2016kce}, recasted to include all four types of the studied particles and production modes involving initial state photons. 
The gray and black dashed lines correspond to limits scaled to 
${\cal L} = 300 \mathrm{fb}^{-1}$ (Run-3)
and 3000 fb$^{-1}$ (HL-LHC), respectively.
The mass region with $\sigma_{\rm eff} \ge \sigma_{\rm eff}^{\rm UL}$
is 
sensitive (excluded)
in the future (current) analyses.

\FloatBarrier

\section{Detecting open production at MoEDAL}
\label{sec:open_MoEDAL}

In this section we outline the details of the MoEDAL detector and describe the procedure to estimate the expected
number of the
signal events of the open production mode 
observed at each stage of the LHC run by MoEDAL.

\subsection{MoEDAL detector}
\label{sec:moedal}

The \textit{Monopole and Exotics Detector at the LHC (MoEDAL)}~\cite{Pinfold:2009oia,Mitsou:2021vhf} is a unique and mostly passive experiment located at 
the
Intersection Point~8 (IP8),
in the vicinity of the VErtex LOcator
(VELO), in the LHCb detector cavern of the LHC. MoEDAL comprises three main sub-detector systems: \textit{nuclear track detectors, magnetic monopole trappers}, and \textit{TimePix},
each following a different principle of operation.

The main sub-detector is a large array of nuclear track detectors \textit{(NTDs)} surrounding IP8.
Each panel comprises of three layers of \textit{CR39\textsuperscript{®}}, three layers of \textit{Makrofol\textsuperscript{®}}, and three layers of \textit{Lexan\textsuperscript{®}} polymers, with 1.5 mm, 0.50 mm and 0.25 mm thickness respectively. All layers are placed inside 125 μm thick aluminium bags.   
Currently, the \textit{Lexan\textsuperscript{®}} foil is used only for protection, while six other polymer layers are used for detection of highly-ionising particles, such as magnetic monopoles~\cite{MoEDAL:2021mpi}. 
When such particles traverse an NTD panel, they deposit energy in plastic material leading to invisible damage along particles' trajectories.
After one year of exposure, NTD panels are transported to INFN Bologna, where they are dismantled and placed in an etching solution. 
The solvent dissolves polymer material anisotropically, leading to formation of cone-shaped etch-pits revealing a trajectory of a highly-ionising particle.
After etching, NTD panels are scanned with an optical microscope in order to find etch-pits aligned in all sheets of polymer material. 
Signature of a signal is a string of etch-pits present in all 6 analysed layers of polymer from a given NTD panel, where all etch-pits have a shape of a cone, and are present on both sides of plastic layers. Careful analysis of signal pattern allows to assess the charge of the highly ionising particle, and can be used to discover multiply charged long-lived BSM particles.  

One drawback of MoEDAL's LLP searches 
with respect to those at large experiments, i.e.\ ATLAS and CMS,
is that the data available at MoEDAL is roughly 10 times smaller than that collected at the large experiments.
This is because MoEDAL shares the same interaction point 
with the LHCb experiment, which opts for lower 
instantaneous luminosity for their analyses. 

Another property of the MoEDAL experiment is its sensitivity to slowly moving particles. While large experiments are sensitive only to relativistic particles moving typically with $\beta > 0.6$, MoEDAL can detect particles with $\beta < 0.15 \cdot |Q/e|$.
Therefore, if a particle has charge $Q=-3e$, then it can be detected by MoEDAL only if it moves slower than 45\% of speed of light.
Also, the NTD panels are placed at the distance 
of $\sim 2 \,$m away from the interaction point.
This means that the MoEDAL is sensitive only to 
the charged long-lived particles with the lifetime typically $\tau \gtrsim 2 [{\rm m}] / c$ with $c$ being 
the speed of light.

This leads to two interesting consequences: (a) MoEDAL provides a complementary detection phase-space when compared to large LHC experiments; (b) the acceptance and sensitivity of MoEDAL to BSM particles grows with the magnitude of electric charge. It is expected that due to large mass and significant energy loss caused by ionisation, new multiply charged BSM particles will have lower velocities than their SM counterparts.
Therefore, it is possible that despite lower luminosity, MoEDAL can exhibit a comparable, or in some cases even better, sensitivity with respect to ATLAS and CMS, when searching for particles with high electric charges.
Hence, we include MoEDAL in our analysis and compare its predicted sensitivity with that of large LHC experiments. 

Another important property that differentiates MoEDAL from other LHC experiments is the fact that it is basically background free. 
Firstly, MoEDAL can directly detect only highly ionising particles.
Secondly, heavy charged particles in the Standard Model are unstable and decay before reaching MoEDAL's NTD panels placed approximately two meters away from IP8. 
Thirdly, in the Standard Model meta-stable particles that can reach the NTD panels have much larger velocities 
and cannot satisfy the detection criteria, $\beta < 0.15 \cdot |Q/e|$.
Moreover, background radiation from the cavern surrounding the detector is actively monitored by TimePix devices. 
Finally, possible spallation or neutron recoil products will not produce a detectable signature, which corresponds to a particle going through whole NTD panel without much stopping.
It is therefore evident, that MoEDAL can be treated as a background-free experiment.

\subsection{Estimation of expected signal events at MoEDAL}
\label{sec:moedal_signal}

In this study we follow the approach developed in
\cite{Acharya:2020uwc, Felea:2020cvf, Hirsch:2021wge}, 
where it has been proven that MoEDAL has a large   
potential for discovering long-lived BSM particles.
The number of expected events observable at MoEDAL
with the data corresponding to the integrated luminosity, ${\cal L}$, is given by
\begin{linenomath*}
\begin{equation}
N_{\text{sig}} = \sigma \cdot {\cal L} \cdot \epsilon,
\end{equation}
\end{linenomath*}
where $\sigma$ is the production cross section for a pair of new particles.
The efficiency, $\epsilon$, in the above formula 
represents the average number of particles per event
that hit one of MoEDAL's NTD panels with  
$\beta < 0.15 \cdot |Q/e|$.
This can be estimated with the MC simulation as
\begin{linenomath*}
\begin{equation}
\epsilon = \left \langle 
\sum_{i=1,2} 
\Theta \left( \beta^{\mathrm {thr}}_{Q} - \beta_i \right)
\cdot
 P({\vec \beta_i}, \tau)
\right \rangle_{\rm MC},
\label{eq:effMC}
\end{equation}
\end{linenomath*}
where $\vec{\beta_i}$ is the velocity 
of the $i$-th BSM particle produced in an event,
$\Theta(x)$ is the Heavyside step function ($\Theta(x) = 1$ for $x > 0$ and 0 otherwise)
and $\left \langle \cdots \right \rangle_{\rm MC}$ represents the Monte Carlo average. 
$\beta^{\mathrm {thr}}_{Q} \equiv 0.15 \cdot |Q/e|$
is the threshold value below which the particle can be registered by the MoEDAL detector. 
Finally, $P({\vec \beta_i})$ is the probability that the particle reaches an NTD panel, and given by
\begin{linenomath*}
\begin{equation}
 P({\vec \beta_i}, \tau)
=
\delta ( \vec \beta_i ) \exp{\left( -\frac{l(\vec \beta_i)}{\gamma \beta_i c \tau} \right)},
\end{equation}
\end{linenomath*}
where $\delta (\vec \beta_i )$ is 1 if there is an NTD panel on the trajectory of the particle $i$ and 0 otherwise, $l ( \vec \beta_i )$ is the distance between the interaction point and the NTD panel along the particle's trajectory,  $\gamma \equiv 1/\sqrt{1-\beta_i^2}$\, is the Lorentzian factor and $\tau$ is the lifetime of the particle.
In our numerical analysis, we calculate the average in Eq.~\eqref{eq:effMC}
over the Monte Carlo events generated with {\tt MadGraph\,5} and evaluate the functions $\Theta$ and $P$
using the fast simulation code  
developed for \cite{Acharya:2020uwc, Felea:2020cvf, Hirsch:2021wge},
which contains the information of the location of NTD panels.

A couple of comments are in order
regarding the efficiency calculation.
First, precisely speaking, the detection efficiency 
depends also on the incident angle between
the particle and the NTD panel.
However, 
during Run-3 data taking period, NTD panels are placed directly facing the interaction point.
In this situation incident angles cannot be large 
and this effect is negligible.
Secondly, the condition $\beta_i < 0.15 \cdot |Q/e|$
is trivially satisfied for $|Q/e| \ge \frac{1}{0.15} = \frac{20}{3}$ since $\beta \le 1$ by definition.  
For $|Q/e| \ge \frac{20}{3}$ MoEDAL can detect 
charged particles at any velocities as long as
they hit and penetrate an NTD panel.
Lastly, we assume that the charged particles traverse the LHCb VELO detector without much change of four-momentum, which is a good approximation for the range of charges
we consider in this analysis.
For extremely larger charges $|Q/e| \gg 8$, ionisation loss becomes significant and more precise treatment is necessary to accurately estimate the efficiency. 

\begin{figure}[t!]
    \centering
    \includegraphics[width=0.47\textwidth]{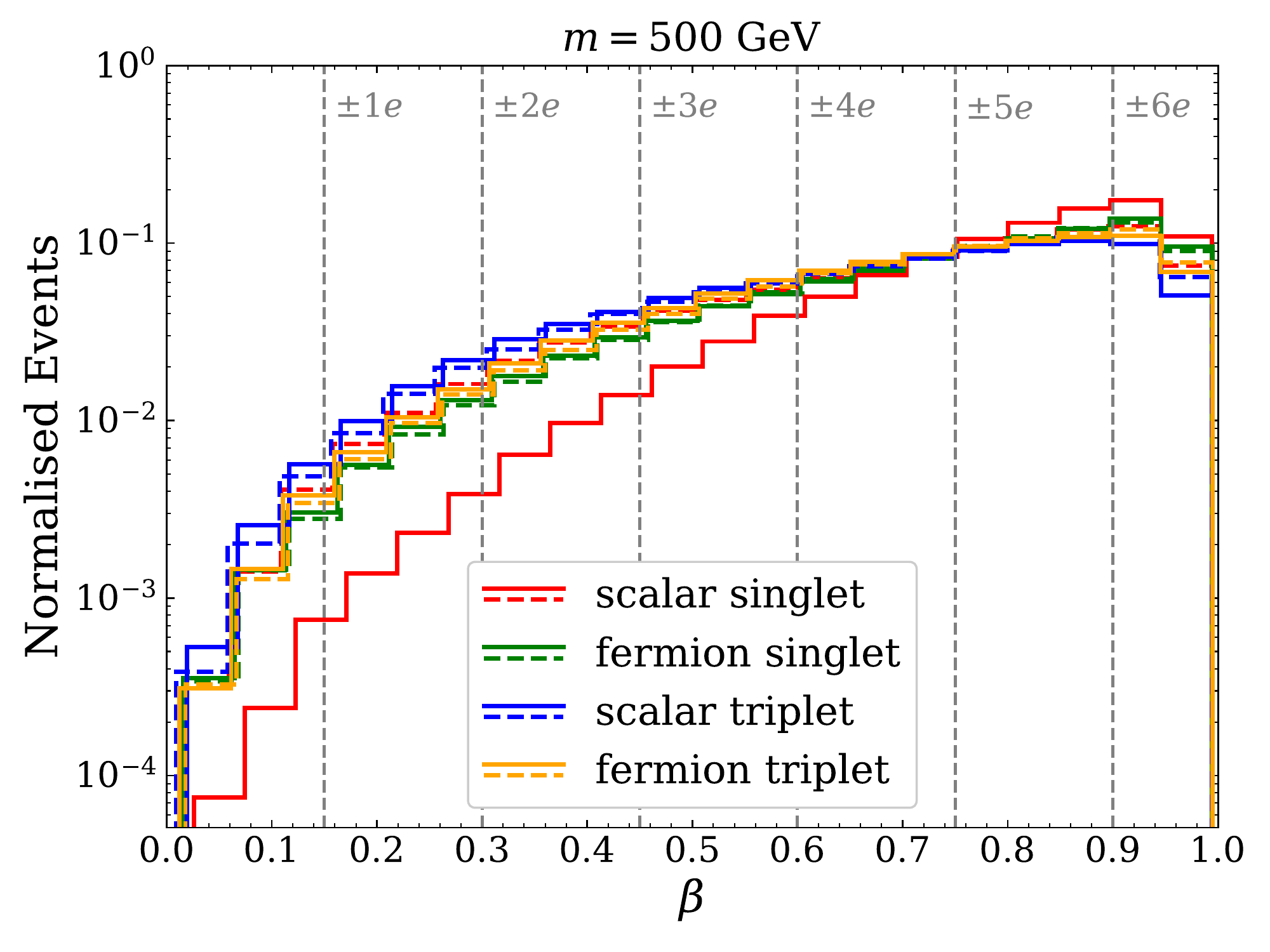}
    \hspace{3mm}
    \includegraphics[width=0.47\textwidth]{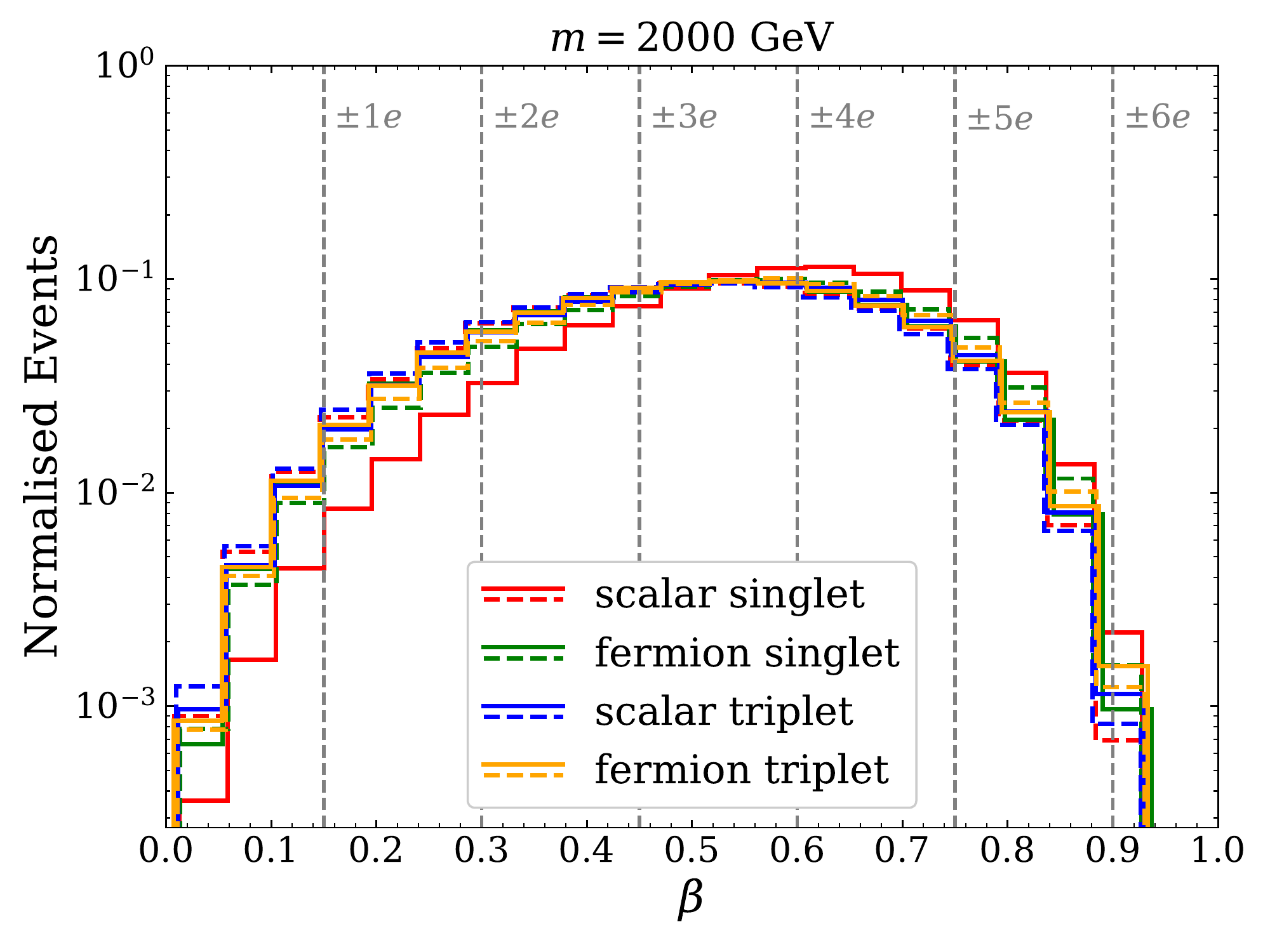}
    \caption{\small
    Velocity distributions of multi-charged LLPs with mass 500 GeV (left) and 2 TeV (right). 
    The solid and dashed histograms 
    correspond to $Q = 1e$ and $8e$, respectively.
    MoEDAL's detection thresholds, $\beta_Q^{\rm thr} = 0.15 \cdot |Q/e|$, are depicted with the gray vertical dashed lines.}
    \label{fig:beta}
\end{figure}

To see how the condition $\beta < 0.15 \cdot |Q/e|$ impacts the signal acceptance of different LLPs, we show $\beta$ distributions of the produced LLPs in Fig.~\ref{fig:beta}.
In the left (right) plot 
we take $m = 500$ (2000) GeV
and the solid (dashed) histograms correspond 
to $Q = 1e$ ($8e$).
Four types of particles are shown:
colourless scalar (red), 
colourless fermion (green)
colour-triplet scalar (blue)
colour-triplet fermion (yellow).
Comparing the left and right plots,
we see the production velocities are  generally larger for smaller $m$.
We also see 
that 
the production velocities of the colourless scalar with $Q=1$ are, on average, significantly larger than those for the other types.
This is because for this type of particles
the dominant production channel is the $q \bar q$ initiated Drell-Yan, whose production rate vanishes in the $\beta \to 0$ limit due to angular conservation.  
For $Q=8$, as we have seen in Fig.~\ref{fig:frac},
the dominant channel is replaced by the photon-photon fusion, in which such a property is absent.
Apart from the $Q=1e$ colourless scalar,
the velocity distributions are similar among all types of particles.
The vertical dashed lines 
in Fig.~\ref{fig:beta}
represent 
MoEDAL's detection threshold, $\beta^{\rm thr}_Q = 0.15 \cdot |Q/e|$.
We see that only a small fraction of LLPs satisfy $\beta < \beta_Q^{\rm thr}$
for $Q = 1e$,
while majority of produced LLPs can be registered by MoEDAL
for $Q = (5-6)e$, depending on the masses and types of the particles.

As explained in section \ref{sec:moedal}, MoEDAL is a background free experiment, hence we consider the thresholds corresponding to $N_{\rm sig} = 1$, 2 and 3
when presenting the sensitivity. 
In Appendix \ref{ap:a} we show contours of those thresholds in the ($m$ vs $c\tau$) planes for each type of particle with ${\cal L}=30$ and 300 fb$^{-1}$ for 
Run 3 and HL-LHC, respectively.

\section{Closed production and diphoton signature}
\label{sec:closed}

\begin{figure}[t!]
\centering
\includegraphics[width=0.36\textwidth]{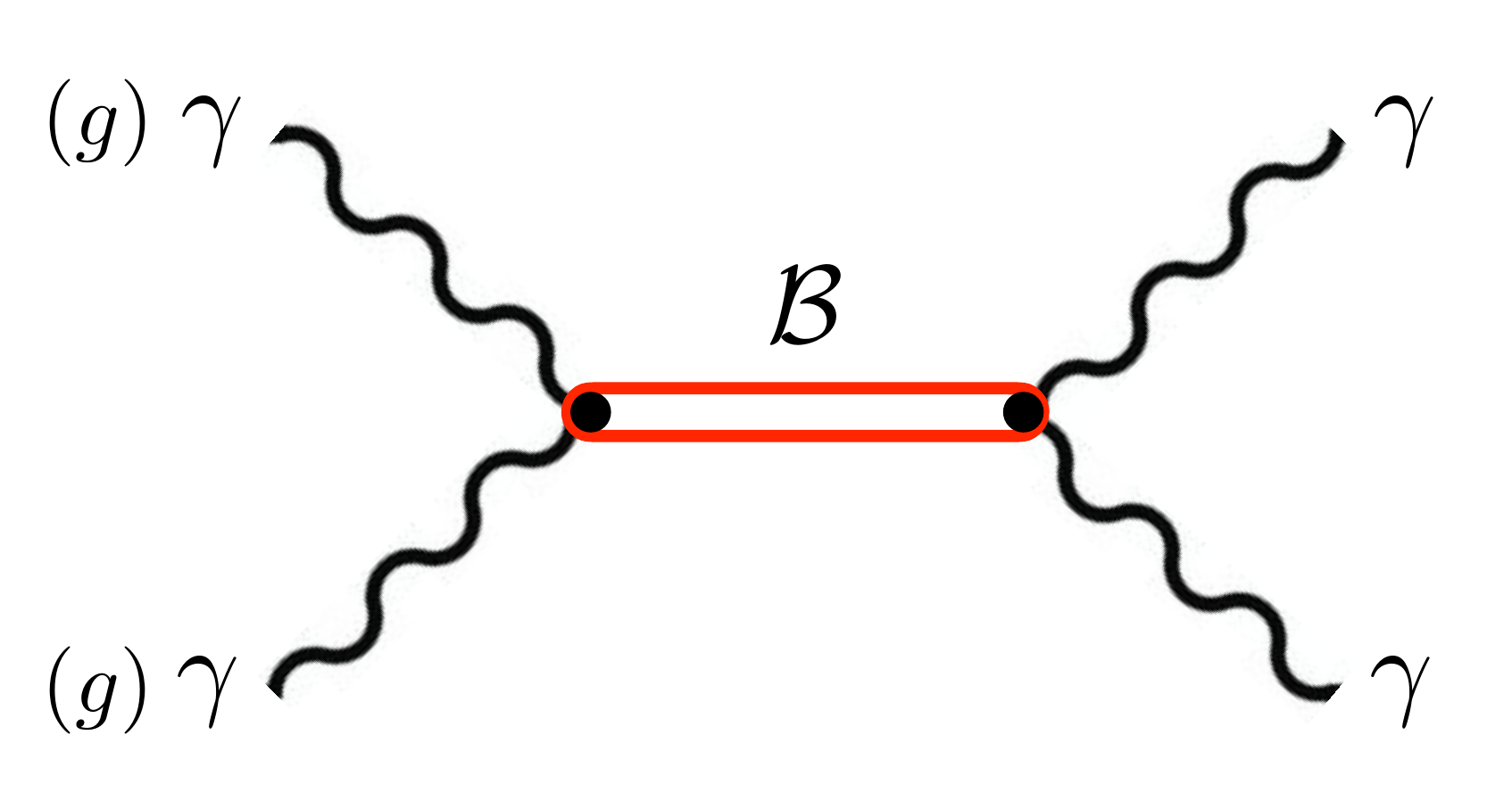}
\caption{\small\label{fig:bound_diagram}
Diagram for the bound state production
and decay into the diphoton channel,
$pp \to {\cal B} \to \gamma \gamma$.}
\end{figure} 
\subsection{Bound state and diphoton decay mode}

For sufficiently large electric charges, 
the Coulomb attraction force between the $\xi^{+Q}$-$\xi^{-Q}$ pair 
becomes important.
At the vicinity of the threshold, the $\xi^{+Q}$-$\xi^{-Q}$ pair
can form a bound state ${\cal B}$ with the mass $M_{\cal B} \simeq 2 m$, as depicted in Fig.~\ref{fig:bound_diagram}.
We call this production channel the {\it closed production mode}.

Depending on the initial partons, the bound state ${\cal B}$ can have different spins.
For example, if the initial state is $q \bar q$ (or $g g$ for coloured $\xi^{\pm Q}$),
the $s$-channel production via the off-shell $\gamma/Z$ (or $g$) exchange
allows ${\cal B}$ to have spin 1.
On the other hand, the $\gamma \gamma$ (or $g g$ for coloured $\xi^{\pm Q}$)  
initial state
allows ${\cal B}$ to have spin 0 or 2,
through the diagrams with the $t$-channel or the 4-point interactions.

The spin of the bound state also dictates the decay.
The spin 0 and 2 bound states can decay into the $\gamma \gamma$, $Z \gamma$ and $Z Z$
final states, as well as the $g g$ final state if $\xi^{\pm Q}$ is colour-triplet. 
On the other hand, the spin-1 bound state can decay into $W^+ W^-$ or any SM fermion-antifermion pair
via the $s$-channel $\gamma/Z$ ($g$) exchange.
Among these final states, the diphoton channel, $p p \to {\cal B} \to \gamma \gamma$, 
has by far the strongest experimental sensitivity and we therefore concentrate on this channel in this study
to constrain the open production mode \cite{Jager:2018ecz}.

If the elementary particle $\xi^{\pm Q}$ is colour-triplet,
the $\xi^{+Q}$-$\xi^{-Q}$ bound state may be colour-singlet or octet, since
${\bf 3} \otimes {\bf \bar 3} = {\bf 1} \oplus {\bf 8}$.
However, at least QCD is concerned, the octet configuration does not lead to an attractive force,
which renders the production rate subdominant compared to the production of colour-singlet states.
In any case, the octet state cannot decay into the diphoton final state
and we therefore do not consider it in this study.

The $\xi^{+Q}$-$\xi^{-Q}$ system near the threshold can be treated in a good approximation
by the non-relativistic quantum mechanics.
The calculation of wave functions of energy eigenstates goes along the same line with the calculation for the hydrogen atom. 
Letting $\vec{r}$ ($r = |\vec{r}|$) be the displacement from the position of $\xi^{+Q}$
to that of $\xi^{-Q}$, the static potential between the two particles is given by 
\be
V(\vec{r}) \,=\, - \frac{ C \alpha_s(r^{-1}) + Q^2 \alpha }{r},
\ee
where $\alpha_s(r^{-1})$ is the running strong coupling evaluated at the scale $r^{-1}$, 
$\alpha$ is the structure constant of electromagnetism,
$C = (C_1 + C_2 - C_{\cal B})/2$
and $C_{1/2}$ and $C_{\cal B}$ are the quadratic Casimir of $SU(3)_C$
for the two constituent particles and the bound state, respectively 
($C = 4/3$ (0) for the colour-triplet (singlet) case).
Using the reduced mass, $\mu = (m_{\xi^{+Q}} m_{\xi^{-Q}})/(m_{\xi^{+Q}} + m_{\xi^{-Q}}) = m/2$ (with $m_{\xi^{+Q}} = m_{\xi^{-Q}} \equiv m$),
the Schr{\" o}dinger equation can be brought into the same form as that of the hydrogen atom.
The wave function $\Psi_{nlm}(\vec{r})$ of various energy eigenstates 
can also be found analogously to the calculation of the hydrogen atom.

The decay rate of the bound state is proportional to the probability of finding the two constituent particles
at the same point, $|\Psi_{nlm}(0)|^2$. 
This is non vanishing only for the $s$-wave. 
The wave function ($\Psi_{n} \equiv \Psi_{n00}$) at $\vec{r} = 0$ is given by
\be
\Psi_{n}(0) \,=\, \frac{1}{\sqrt{\pi}} (n r_b)^{-3/2}, 
\quad\quad 
r_b^{-1} = (C \overline \alpha_s + Q^2 \alpha) \mu
\ee
where $r_b$ is the ``Bohr radius'' of this bound state system and
$\overline{\alpha}_s \equiv \alpha_s( r_{rms}^{-1} )$, for ground state $r_{rms}=\sqrt{3}\,r_b$.
With these inputs the probability of finding the two particles at the same point 
can be obtained as
\be
|\Psi_n(0)|^2 \,=\, \frac{(C \overline \alpha_s + Q^2\alpha)^3 m^3 }{ 8 \pi n^3} \,.
\label{eq:psisq}
\ee
We see that the contributions from the radial excitations are suppressed by $1/n^3$.
We therefore include only the ground state ($n=1$) contribution in our analysis, $\Psi(0) = \Psi_1(0)$.

The partial decay rates for ${\cal B} \to \gamma \gamma$ and ${\cal B} \to gg$  
are given by
\be
\Gamma_{ {\cal B} \to \gamma\gamma}
\,=\,
\frac{8 \pi Q^4\alpha^2}{M_{\cal B}^2} n_c n_f |\Psi(0)|^2 \,,
\quad \quad
\Gamma_{{\cal B} \to gg}
\,=\,
\frac{4 \pi C\alpha_s^2}{M_{\cal B}^2}n_f|\Psi(0)|^2 
\ee
where $n_c$ is the number of colours ($n_c = 1$ for singlet and 3 for triplet) and $n_f$ is
the dimension of Lorentz representation ($n_f = 1$ for scalar and 2 for fermion). 
Since the $gg$ final state is not available when $\xi^{\pm Q}$ is colour-singlet,
we set $\Gamma_{{\cal B} \to gg} = 0$ in colour-singlet case.

The partonic cross section for the bound state production is related to the corresponding decay width as
\be
     \hat{\sigma}_{ab \to {\cal B}}(\hat s) \,=\, c_{ab} \cdot 
     \frac{2 \pi (2J_{\cal B}+1) D_{\cal B}}{D_a D_b} \cdot 
     \frac{\Gamma_{{\cal B} \to ab}}{M_{\cal B}} \cdot 2 \pi \delta(\hat{s} - M_{\cal B}^2),
     \label{eq:sig_abB}
\ee
where $\hat s$ is the partonic centre of mass energy,
$c_{ab}$ is a symmetric factor ($c_{ab} = 1$ for $a \neq b$ and 2 for $a = b$),
$D_p$ is the dimension of the colour representation of particle $p$
and $J_{\cal B}$ is the spin of the bound state.

The hadronic production cross section $\sigma_{pp \to {\cal B}}$ for the spin 0 bound state can be obtained 
by convoluting the partonic cross section $\hat \sigma_{ab \to {\cal B}}(s \tau)$ with the luminosity function 
$d{\cal L}_{ab}/d \tau$ as 
\be
\sigma_{pp \to {\cal B}} \,=\, \sum_{ab}^{\gamma \gamma, gg} \int_0^1 
d \tau \, \frac{d {\cal L}_{ab} (\tau) }{d \tau} \, \hat \sigma_{ab \to {\cal B}}(s \tau),
\ee
where $s$ is the $pp$ collision energy and $\tau \equiv \hat s/s$.
We carry out this integral with the Mathematica package {\tt ManeParse} \cite{Clark:2016jgm}
using
{\texttt LUXqed}
PDF set 
({\texttt {LUXqed17}$\_$plus$\_$PDF4LHC15$\_$nnlo$\_$100}) 
\cite{Manohar:2016nzj,Manohar:2017eqh}.

Finally, the signal rate (cross section times branching ratio) for 
the $pp \to {\cal B} \to \gamma \gamma$ process can be obtained using the narrow width approximation as
\be
\sigma_{pp \to {\cal B} \to \gamma \gamma} \,=\, \sigma_{pp \to {\cal B}} \cdot {\rm BR}_{{\cal B} \to \gamma \gamma},
\label{eq:diphoton_signal}
\ee
with
\be
{\rm BR}_{{\cal B} \to \gamma \gamma} \,=\, \frac{\Gamma_{{\cal B} \to \gamma \gamma}}
{ \Gamma_{{\cal B} \to \gamma \gamma} + \Gamma_{{\cal B} \to Z \gamma} 
+ \Gamma_{{\cal B} \to ZZ} + \Gamma_{{\cal B} \to gg} }
\ee
and
\be
{\Gamma}_{{\cal B} \to \gamma Z} \,=\, \Gamma_{{\cal B} \to \gamma \gamma} 
\cdot 2\tan^2 \theta_W \left( 1 - \frac{m_Z^2}{M^2_{\cal B}} \right),
\quad \,
{\Gamma}_{{\cal B} \to Z Z} \,=\, \Gamma_{{\cal B} \to \gamma \gamma} 
\cdot \tan^4 \theta_W
\sqrt{ 1 - \frac{4 m_Z^2}{M^2_{\cal B}} } 
 \,.
\ee

\subsection{Constraints from diphoton resonance searches}
\label{sec:diphoton}

The signal rate of the diphoton channel from the bound state decay obtained in the previous subsection 
should be confronted with the experimental results of diphoton resonance searches conducted by ATLAS and CMS.
The latest CMS result \cite{CMS:2018dqv} is based on the 13 TeV collision data corresponding to 
the integrated luminosity of 36 fb$^{-1}$.
On the other hand, the recent ATLAS analysis \cite{ATLAS:2021uiz}
uses 139 fb$^{-1}$ of 13 TeV data.
Since the latter analysis is based on a larger data set, 
we take the ATLAS analysis \cite{ATLAS:2021uiz} 
and derive the current limit on the bound state production and project it to
the Run 3 (300 fb$^{-1}$) and HL-LHC (3000 fb$^{-1}$). 

In Ref.~\cite{ATLAS:2021uiz}, ATLAS studied the diphoton invariant mass distribution
and did not find significant deviation from the Standard Model prediction.
The result was interpreted for the spin 0 and 2 resonances for several values of decay widths
and upper limits are placed on the production cross section times branching ratio to two photons 
as a function of the resonance mass.
In our simple models, we have $\Gamma_{\cal B } / M_{\cal B} \lesssim 2$\,\% in all region of interest in the $Q$ vs $M_{\cal B}$ plane, which justifies to use the ATLAS upper limit derived using the narrow width approximation. 

\begin{figure}[t!]
\centering
\begin{subfigure}[t]{.48\linewidth}
\includegraphics[width=1.\textwidth]{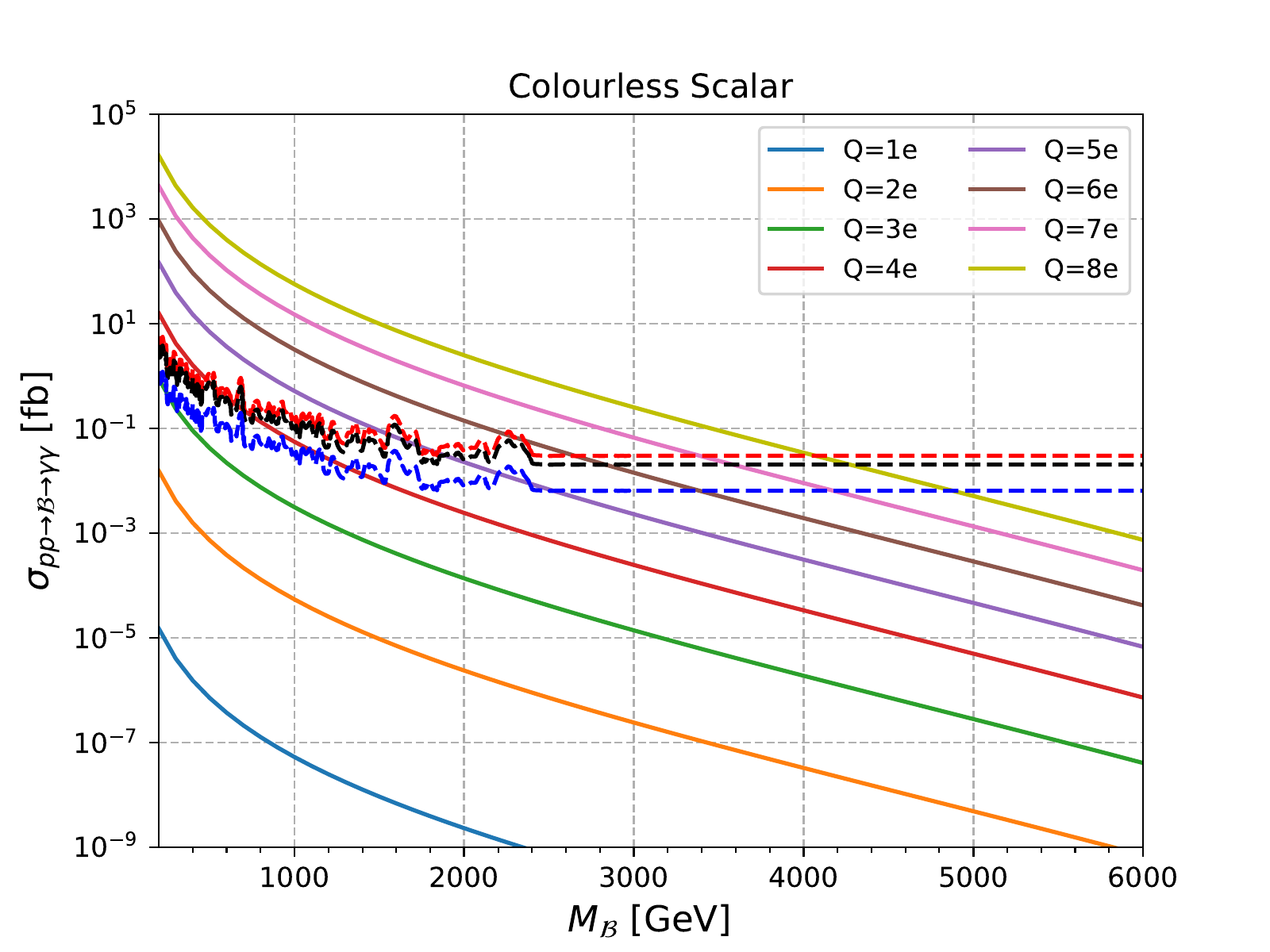}
\end{subfigure}
\begin{subfigure}[t]{.48\linewidth}
\includegraphics[width=1.\textwidth]{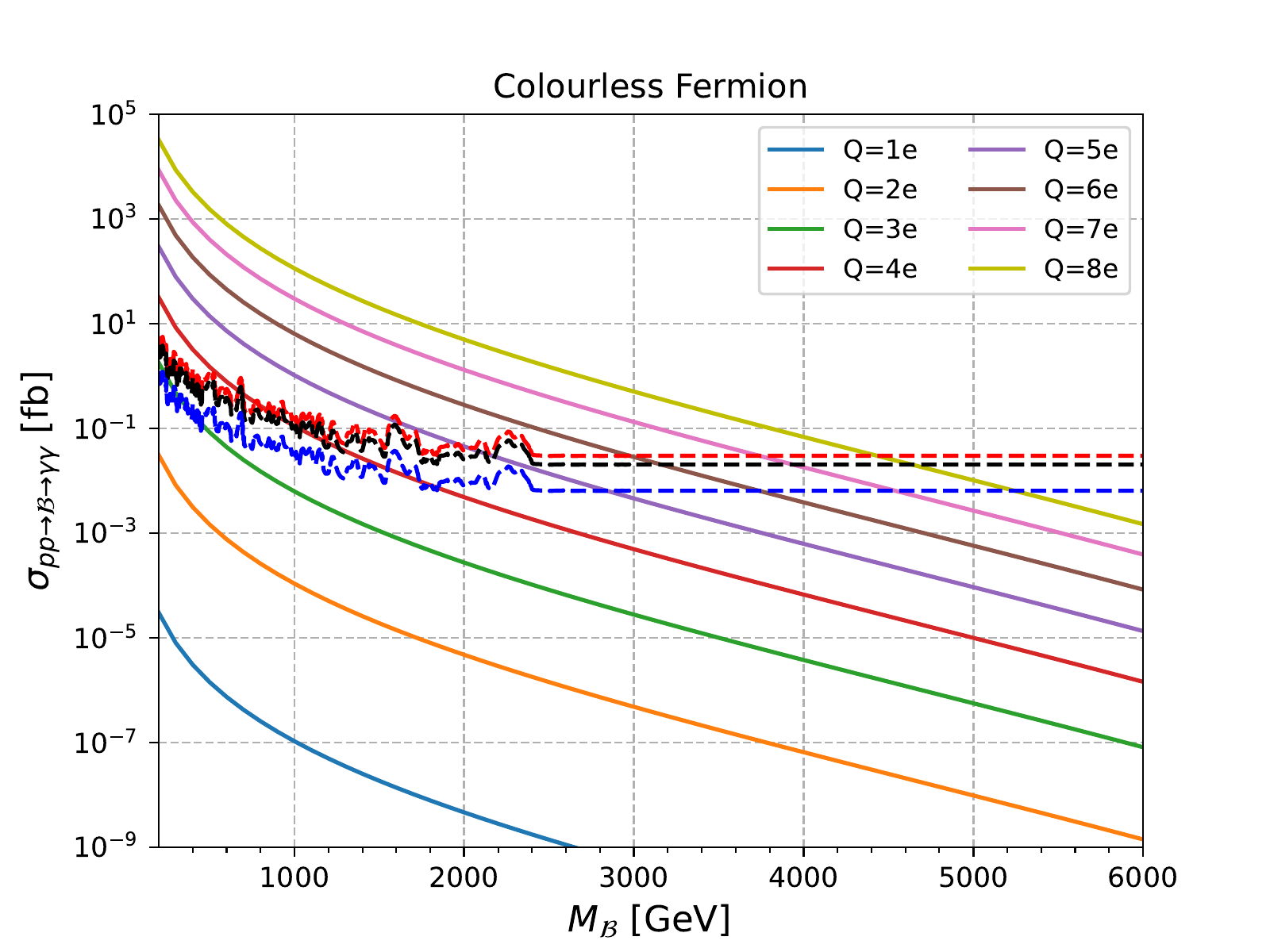}
\end{subfigure}
\begin{subfigure}[t]{.48\linewidth}
\includegraphics[width=1.\textwidth]{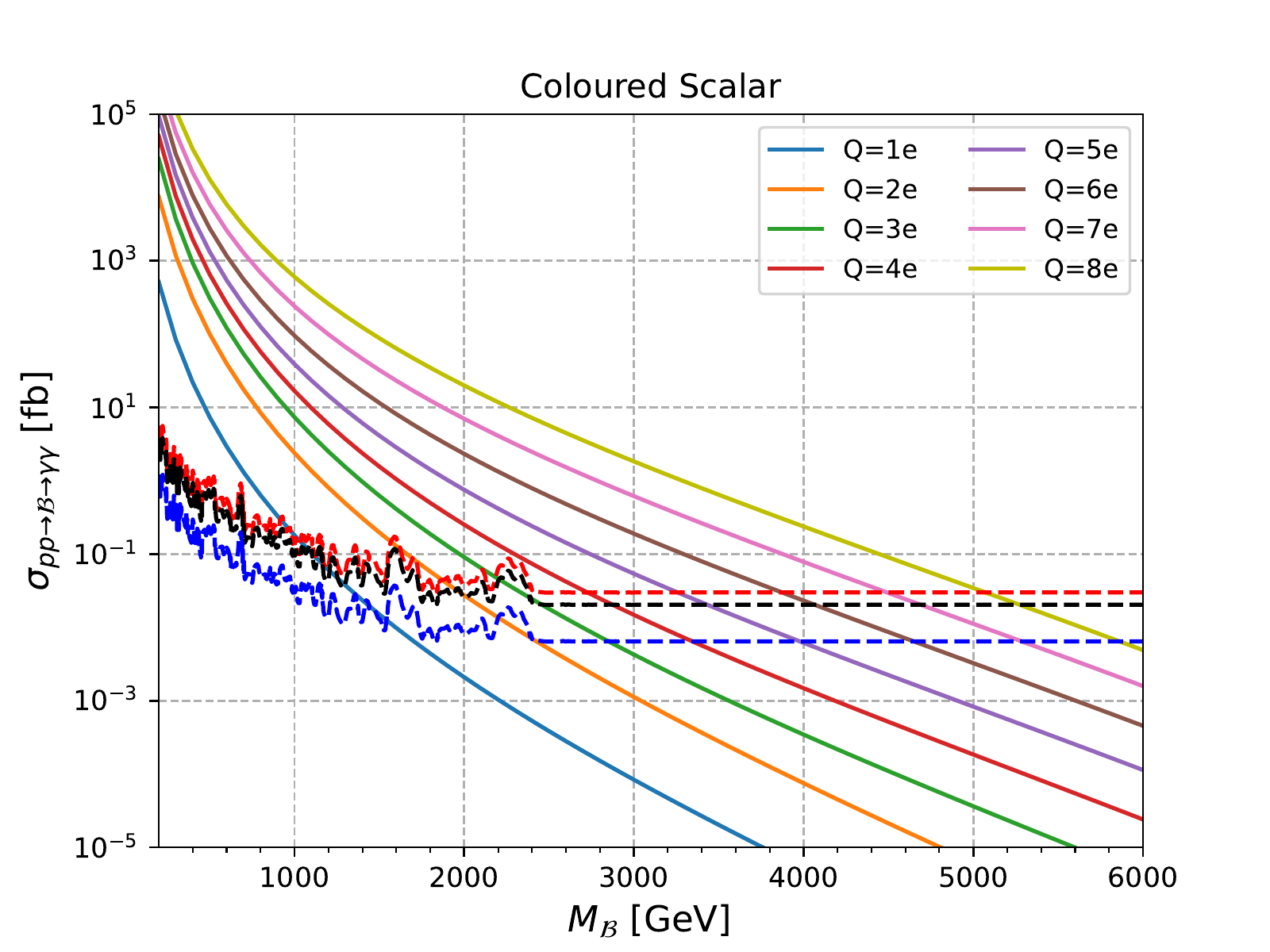}
\end{subfigure}
\begin{subfigure}[t]{.48\linewidth}
\includegraphics[width=1.\textwidth]{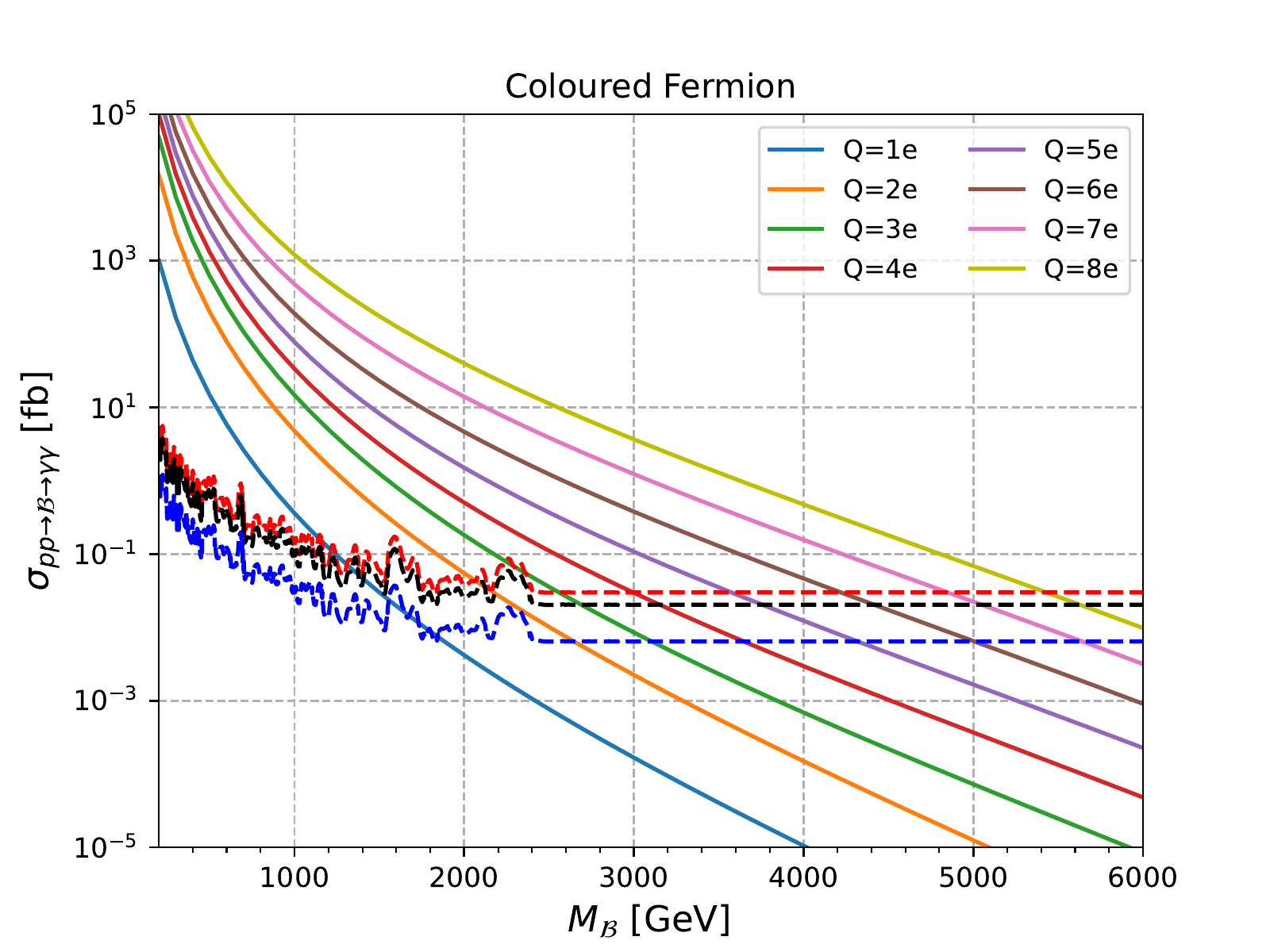}
\end{subfigure}
\caption{\small Diphoton resonant production cross section for colour/colourless fermion and scalar multi charge particles which make bound sate of mass $M_{\cal B} = 2m$. The Dashed lines: expected 95$\%$ CL upper bounds from ATLAS corresponding to an integrated luminosity, $\mathcal{L} = 139$ (red), 300 (black) and 3000 fb$^{-1}$ (blue). 
\label{fig:xs_diphoton}  }
\end{figure}

In Fig.~\ref{fig:xs_diphoton} we show the signal cross section, $\sigma_{pp \to {\cal B} \to \gamma \gamma}$,
for the diphoton channel 
as a function of the bound state mass $M_{\cal B} = 2 m$
for various types of particles.
The upper (lower) panels correspond to the colour-singlet (triplet) $\xi^{\pm Q}$,
and the left (right) panels represent the scalar $\phi^{\pm Q}$ (fermion $\psi^{\pm Q}$) scenarios.
The red dashed line corresponds to the 95\,\% CL upper limit obtained by ATLAS
using 139 fb$^{-1}$ data of 13 TeV $pp$ collision
\cite{ATLAS:2021uiz} using the narrow width approximation.
In this ATLAS analysis the limit is shown only up to $M_{\cal B} = 2.5$ TeV.
In order to constrain the bound states with masses larger than 2.5 TeV,
we adopt the same signal cross section upper limit obtained at $M_{\cal B} = 2.5$ TeV.
This is justified since 
ATLAS observed no event with the diphoton invariant mass larger than 2.5 TeV
and in this region the Standard Model background is below one event.  

To estimate the projected limits for Run 3 (${\cal L}_{\rm Run3} = 300$~fb$^{-1}$)
and HL-LHC (${\cal L}_{\rm HL} = 3000$~fb$^{-1}$), we rescale 
the current upper limit $\sigma^{\rm UL}_0$ at ${\cal L}_0 = 139$ fb$^{-1}$
as $\sigma^{\rm UL}_{\rm Run3} = \sigma^{\rm UL}_0 \sqrt{ {\cal L}_0  / {\cal L}_{\rm Run3} }$
for Run-3 and $\sigma^{\rm UL}_{\rm HL} = \sigma^{\rm UL}_0 \sqrt{ {\cal L}_0 / {\cal L}_{\rm HL} }$
for HL-LHC, 
assuming the signal/background efficiencies and the systematic uncertainty do not drastically change.
Since improving signal-background separation or reducing systematic uncertainties would lead to a better sensitivity, this approach is conservative. 
The projected signal cross section limits (95\% CL) for Run 3 and HL-LHC are shown in the black and blue dashed lines
in Fig.~\ref{fig:xs_diphoton}.

\section{Results}
\label{sec:results}

In this section we collect the results of calculations outlined in the previous sections and compare the expected mass bounds
for the meta-stable multi-charged particles
obtained from the large $dE/dx$ 
and diphoton resonance searches at ATLAS and CMS as well as from the etch-pits analysis at MoEDAL  
assuming the data that will be delivered at Run-3 and HL-LHC. 
For large $dE/dx$ and diphoton resonance searches, Run 3 and HL-LHC correspond to the integrated luminosity of 300 and 3000 fb$^{-1}$, respectively,
and we present the 95\% CL limits,
while for MoEDAL Run 3 and HL-LHC
correspond to $L=30$ and 300 fb$^{-1}$, respectively, 
and we show the expected signal detection $N_{\rm sig} = 1$, 2 and 3, throughout this section.

\begin{figure}[b!]
    \centering
    \includegraphics[width=0.49\textwidth]{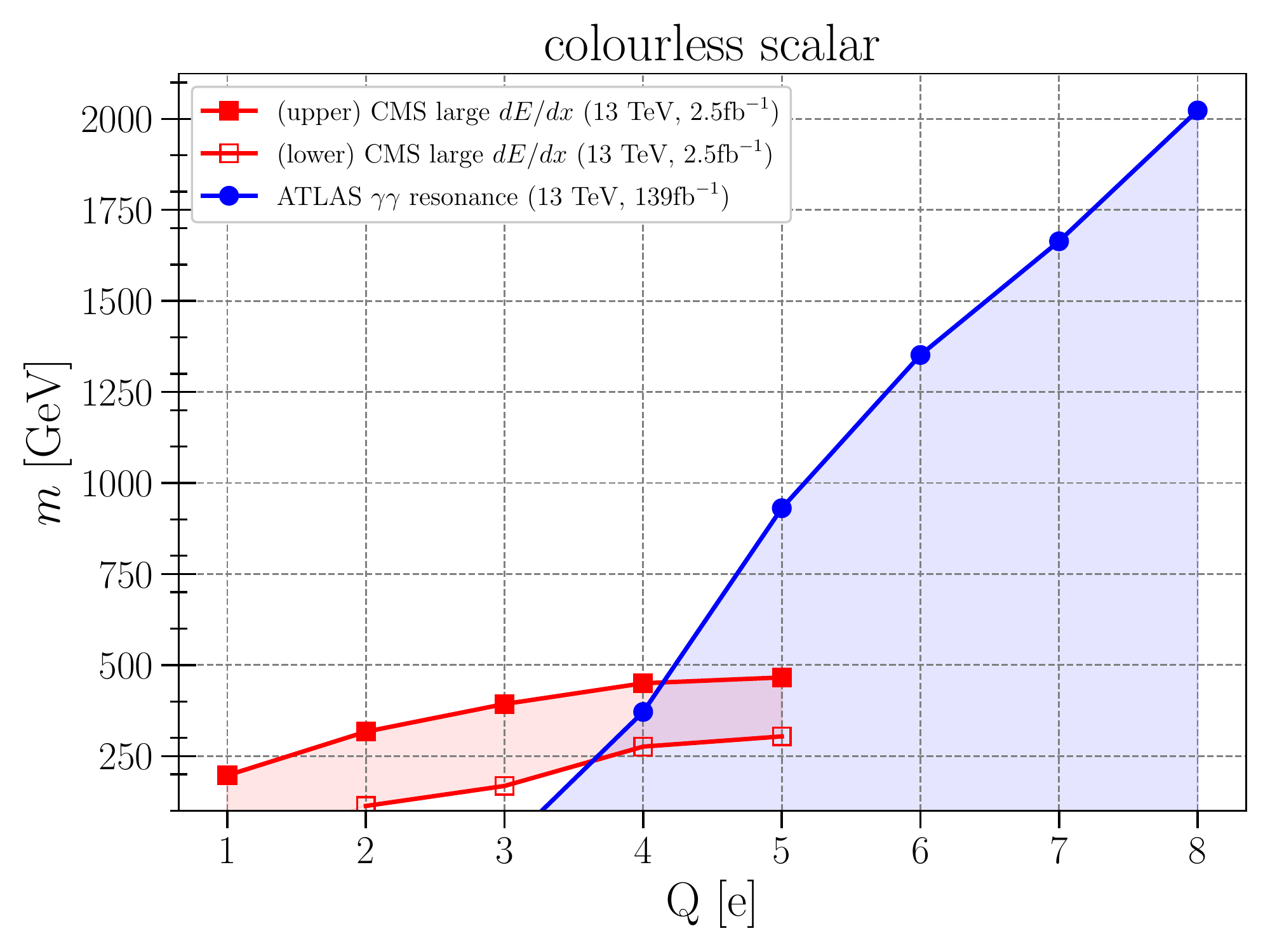}
    \includegraphics[width=0.49\textwidth]{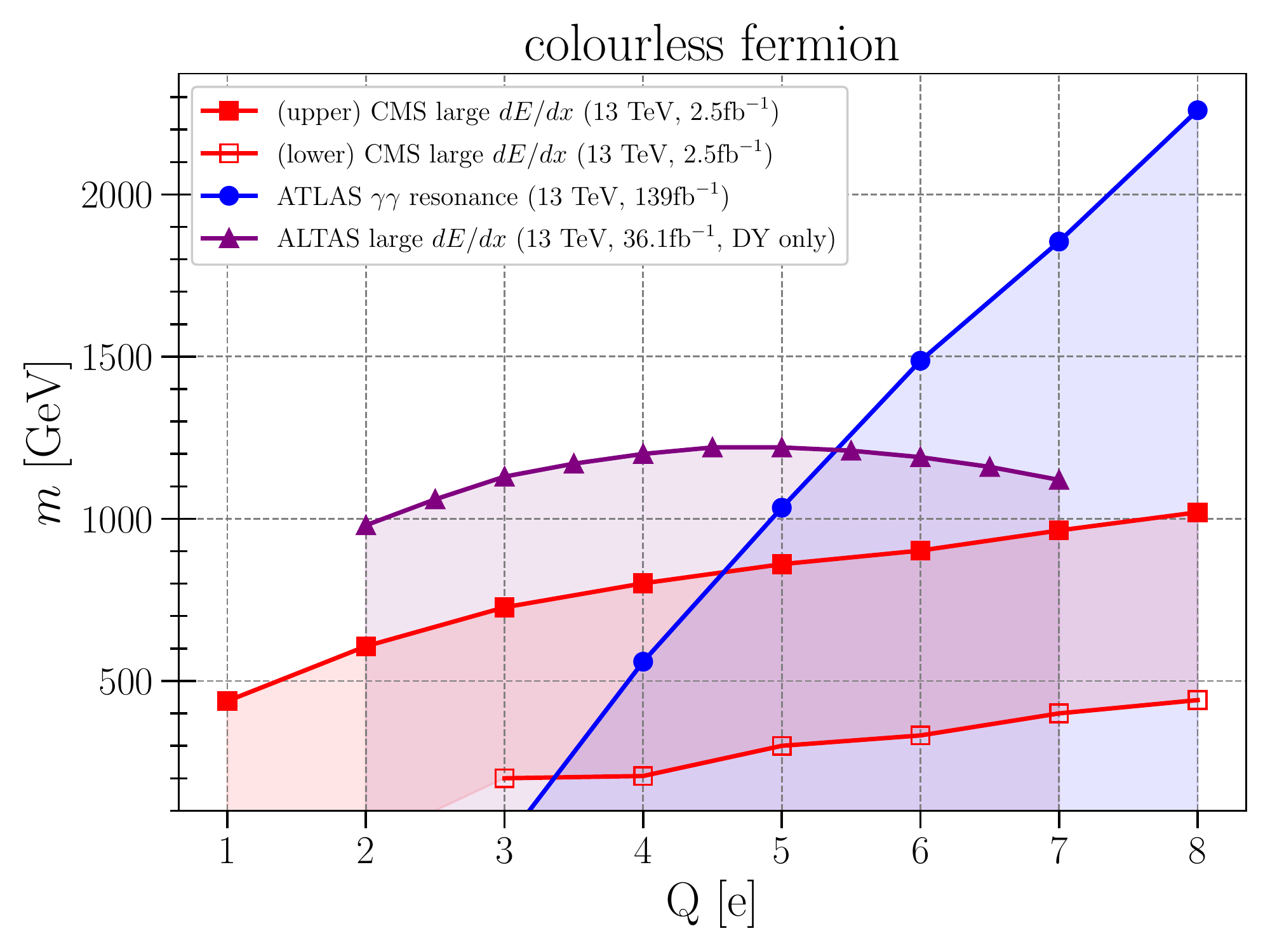}
    \includegraphics[width=0.49\textwidth]{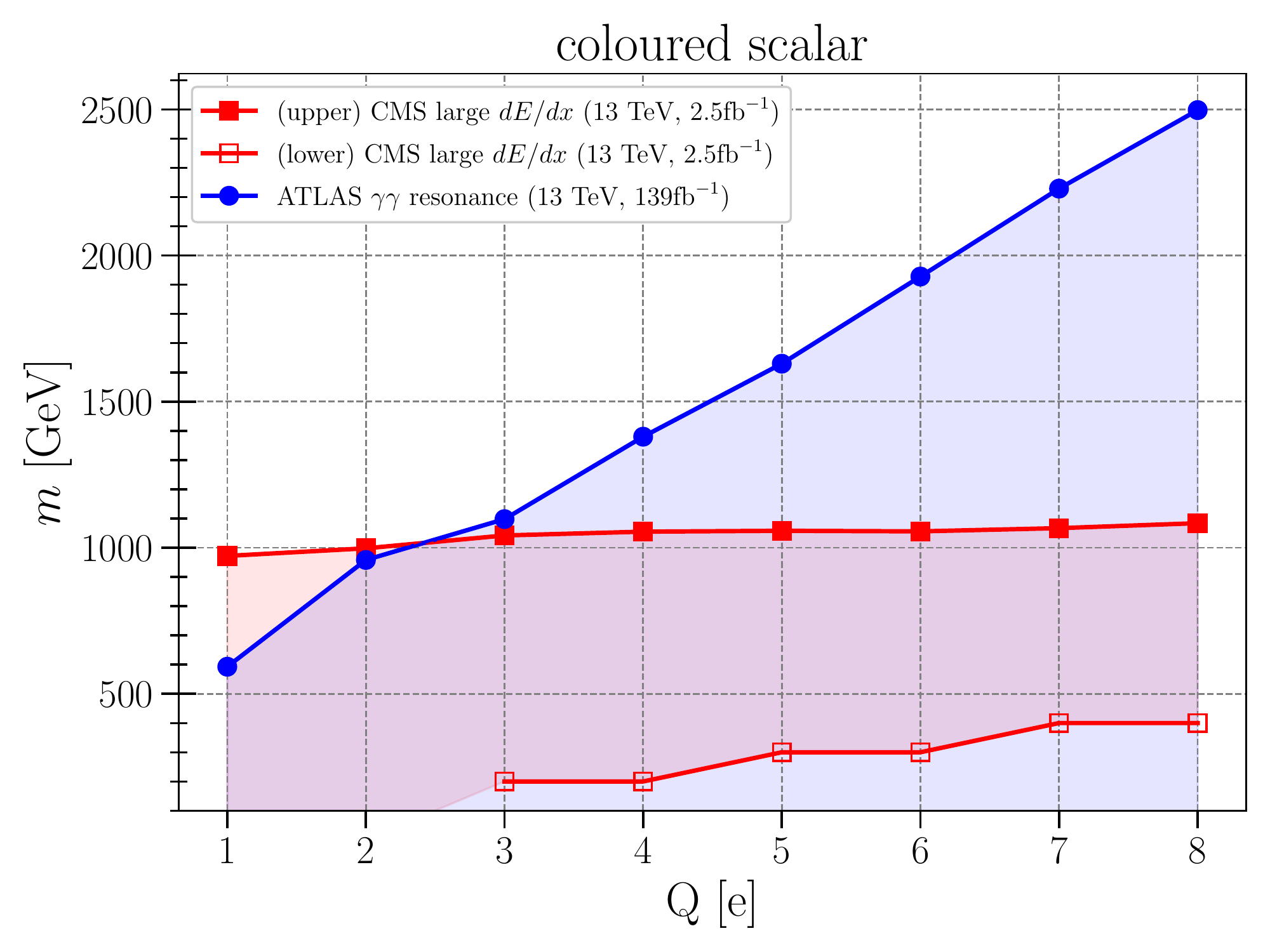}
    \includegraphics[width=0.49\textwidth]{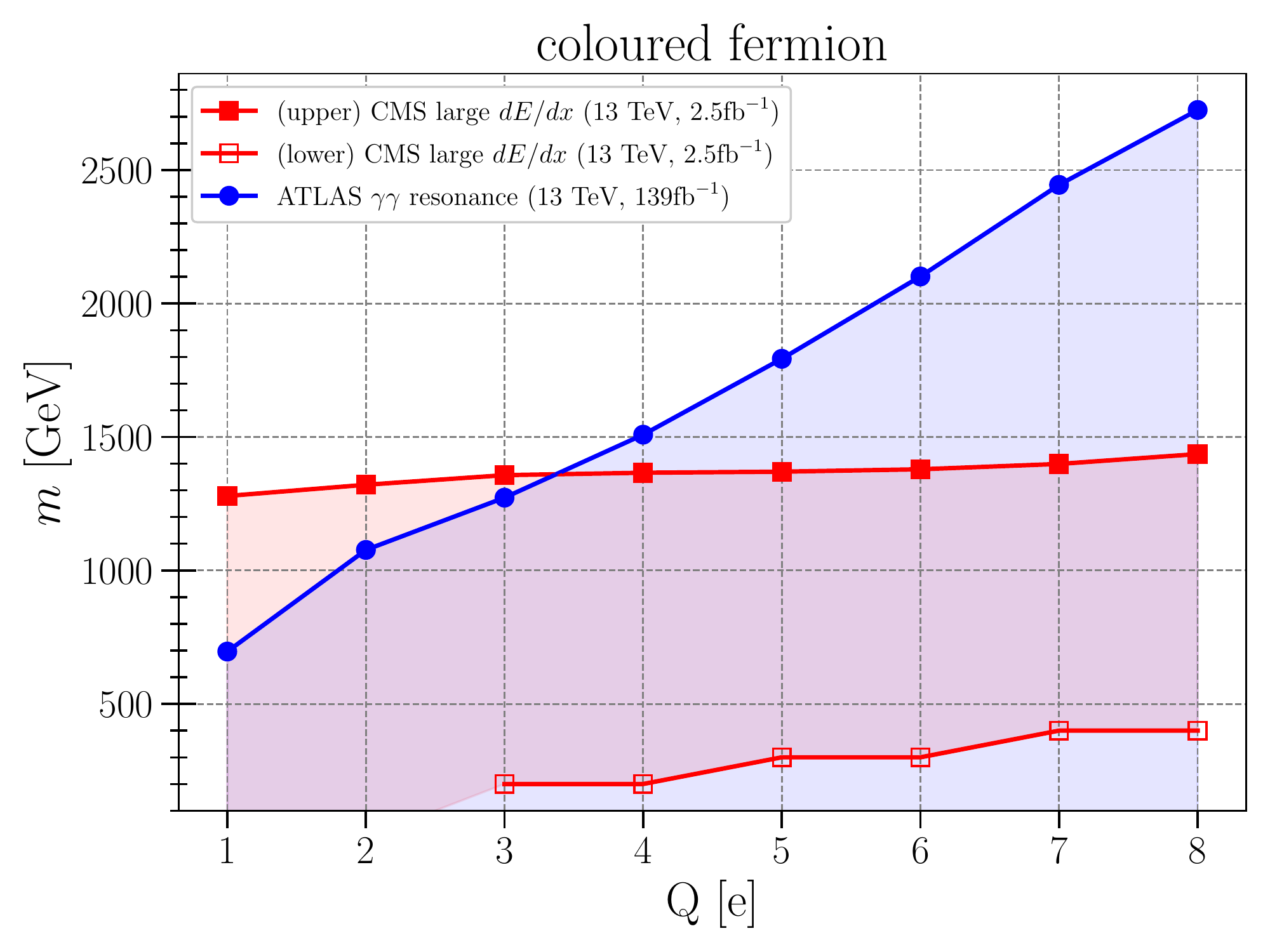}
    \caption{ \small The mass limits on the multi-charged LLPs: colourless scalars (top left), colourless fermions (top right), coloured scalars (bottom left) and coloured fermions (bottom right); recasted to take into consideration photon-induced production processes.
    The region between the two red curves is excluded by 
    the large $dE/dx$ analysis with the 13~TeV 2.5~fb$^{-1}$ data \cite{CMS:2016kce}.
    The purple curve in the top right plot is the mass limit reported in the ATLAS large $dE/dx$ analysis with the 13~TeV 36.1~fb$^{-1}$ data \cite{ATLAS:2018imb}.
    The region below the blue curve is excluded by the
    ATLAS diphoton resonance analysis (13 TeV, 139 fb$^{-1}$) \cite{ATLAS:2021uiz}.
    }
    \label{fig:res_curr}
\end{figure}

We start by showing the current mass limits (at 95\% CL) as functions of the electric charge in Fig.~\ref{fig:res_curr},
where the top (bottom)
panels show colourless (coloured)
particles and 
the left (right) panels represent the scalars (fermions). 
{
The light red area enclosed by the two red curves is the region excluded by the latest CMS result of the large $dE/dx$ analysis with the 13~TeV 2.5~fb$^{-1}$ data \cite{CMS:2016kce}.
The mass limits are re-calculated for various multi-charged particles, including the contribution of initial states involving photons.
The lower boundary of the exclusion comes from the fact that for larger $|Q|$ the $p_T$ cut ($p_T > 65 \cdot |Q/e|$\,GeV)
makes it very difficult to satisfy the slow moving requirement, $1/\beta_{\rm MS} > 1.25$, for lighter particles. 
We see, however, that most of the region below the lower exclusion boundary is already excluded 
by the ATLAS diphoton resonance analysis (blue curve).
We also believe that the remaining mass region below the lower boundary is excluded by 
the previous large $dE/dx$ analyses with the 7 and 8 TeV data \cite{CMS:2013czn}, where the $p_T$ cut is milder.
We do not find any excluded region
for colourless scalars
with $|Q/e| \ge 6$.
This is because the production velocity of scalar particles is generally large and 
almost all events 
are rejected by the
$1/\beta_{\rm MS} > 1.25$
requirement.
} Only for colourless fermions (top right panel) we show by the purple curve the mass limits reported in the ATLAS large $dE/dx$ analysis with the 13~TeV 36.1~fb$^{-1}$ data \cite{ATLAS:2018imb}.
ATLAS has interpreted their result only for colourless fermions with 
$|Q|/e$ from 2 to 7.
We note that those limits are computed based on leading-order cross sections excluding the photonic initial states.
Although the amount of data used in the ATLAS analysis is larger than the latest CMS analysis \cite{CMS:2016kce}, we cannot reinterpret this analysis since the signal efficiency of each cut is not publicly available.

The blue curves correspond to
the mass limits obtained by recasting the ATLAS diphoton resonance analysis (13 TeV, 139 fb$^{-1}$) \cite{ATLAS:2021uiz}
for the bound state decay into diphoton.
We can see that for smaller charges ($Q \lesssim (2-4) e$), the large $dE/dx$ search gives tighter mass limits.
Increasing $|Q|$, the mass bounds from the diphoton resonance rapidly grows  and surpass the large $dE/dx$ limits around $Q \sim (2-4) e$, depending on the type of particles.
In fact, as can be seen in Eqs.~\eqref{eq:psisq}-\eqref{eq:sig_abB},
the production cross section of the bound state grows as $|Q|^{10}$,
which is responsible for the steep rise of the mass limits as moving towards large values of $|Q|$.
On the other hand, the limits from the large $dE/dx$ search only mildly increase.
Although the cross section of open production grows with the charge, the signal acceptance decreases as we have seen in Fig.~\ref{fig:eff},
due to the underestimation 
of the charged track $p_T$s
and the velocity loss 
due to the strong electromagnetic interaction with detector materials
as discussed in section~\ref{sec:open_ATLAS}.

\begin{figure}[t!]
    \centering
    \includegraphics[width=0.49\textwidth]{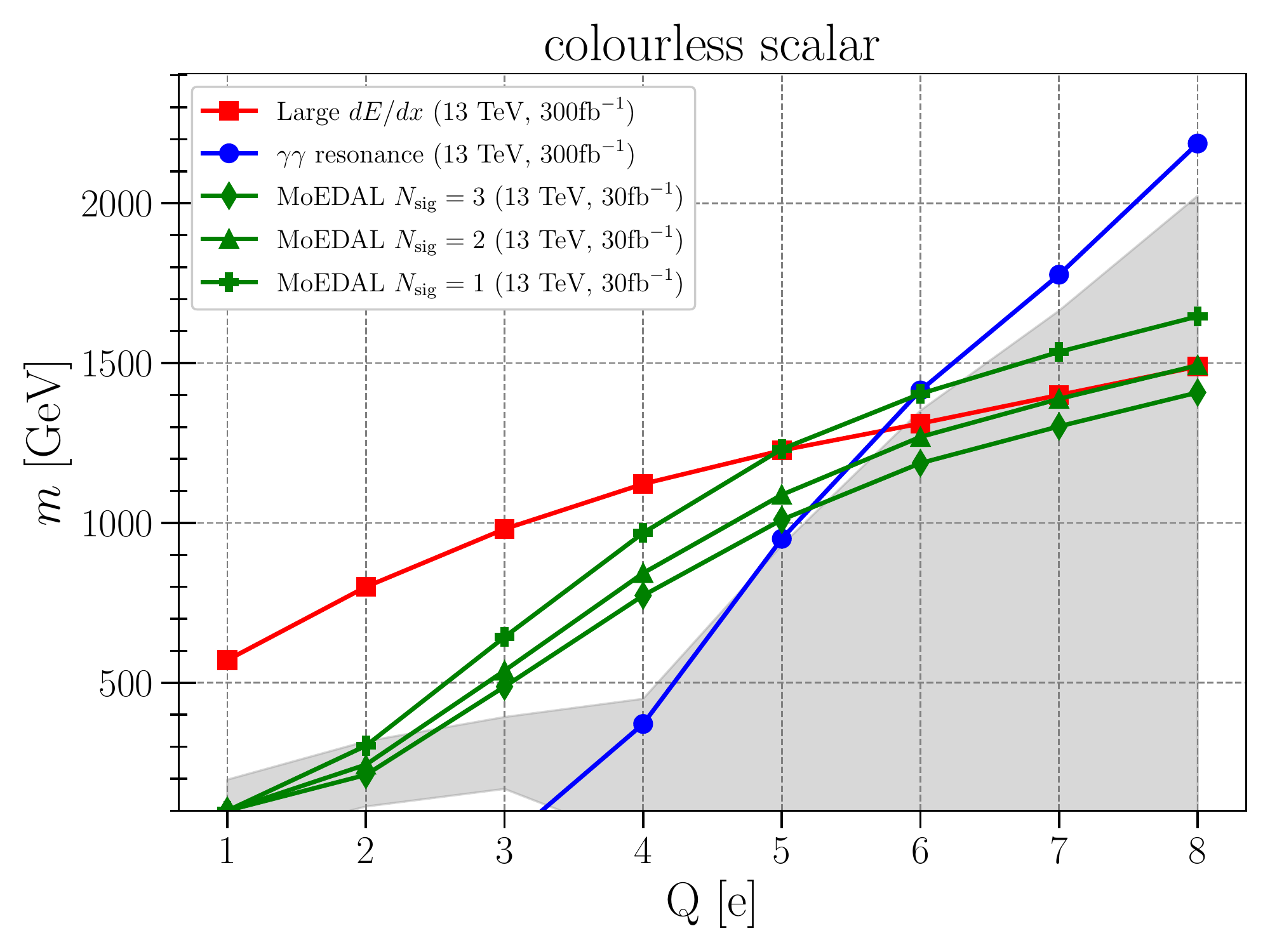}
    \includegraphics[width=0.49\textwidth]{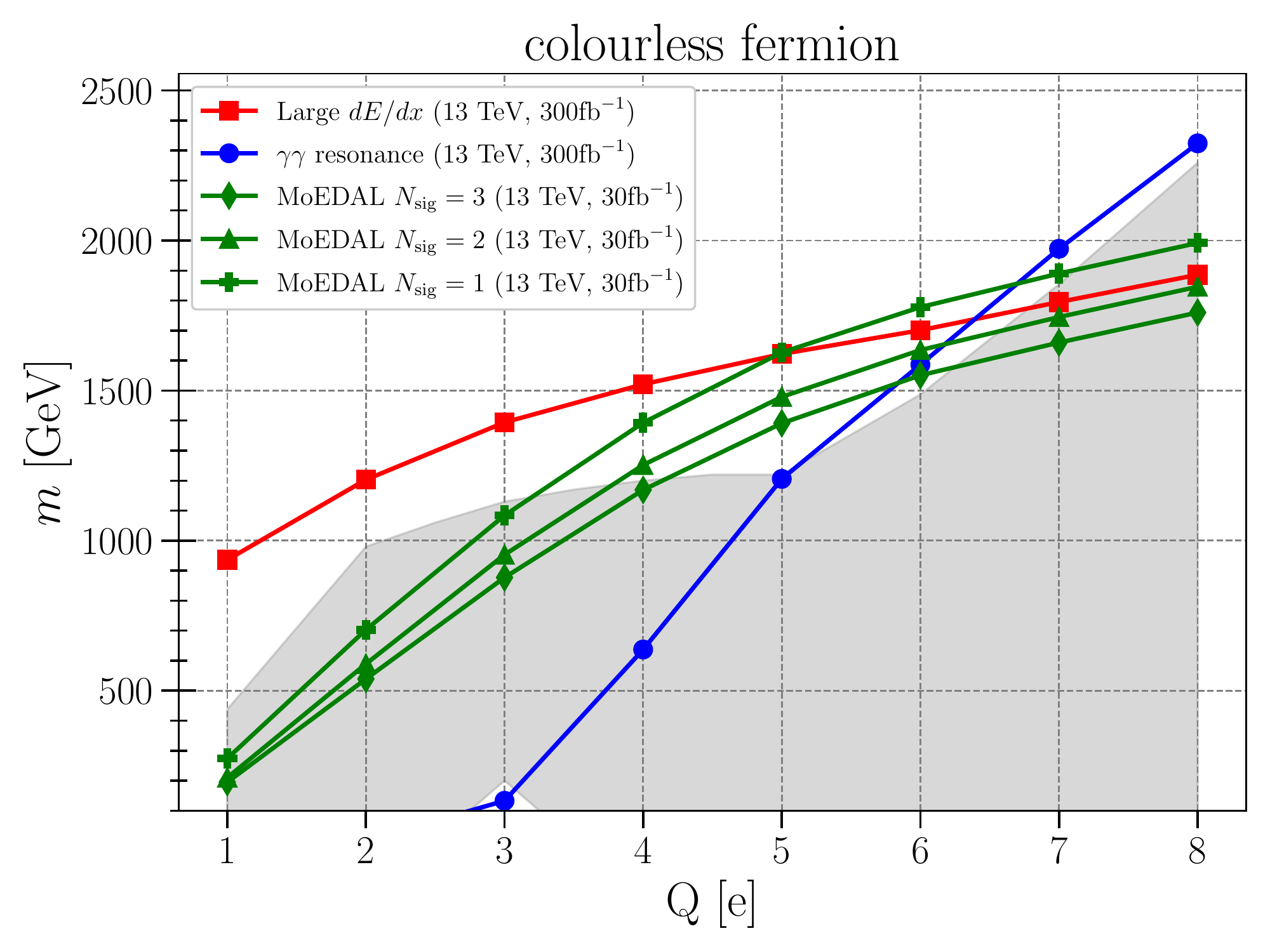}
    \includegraphics[width=0.49\textwidth]{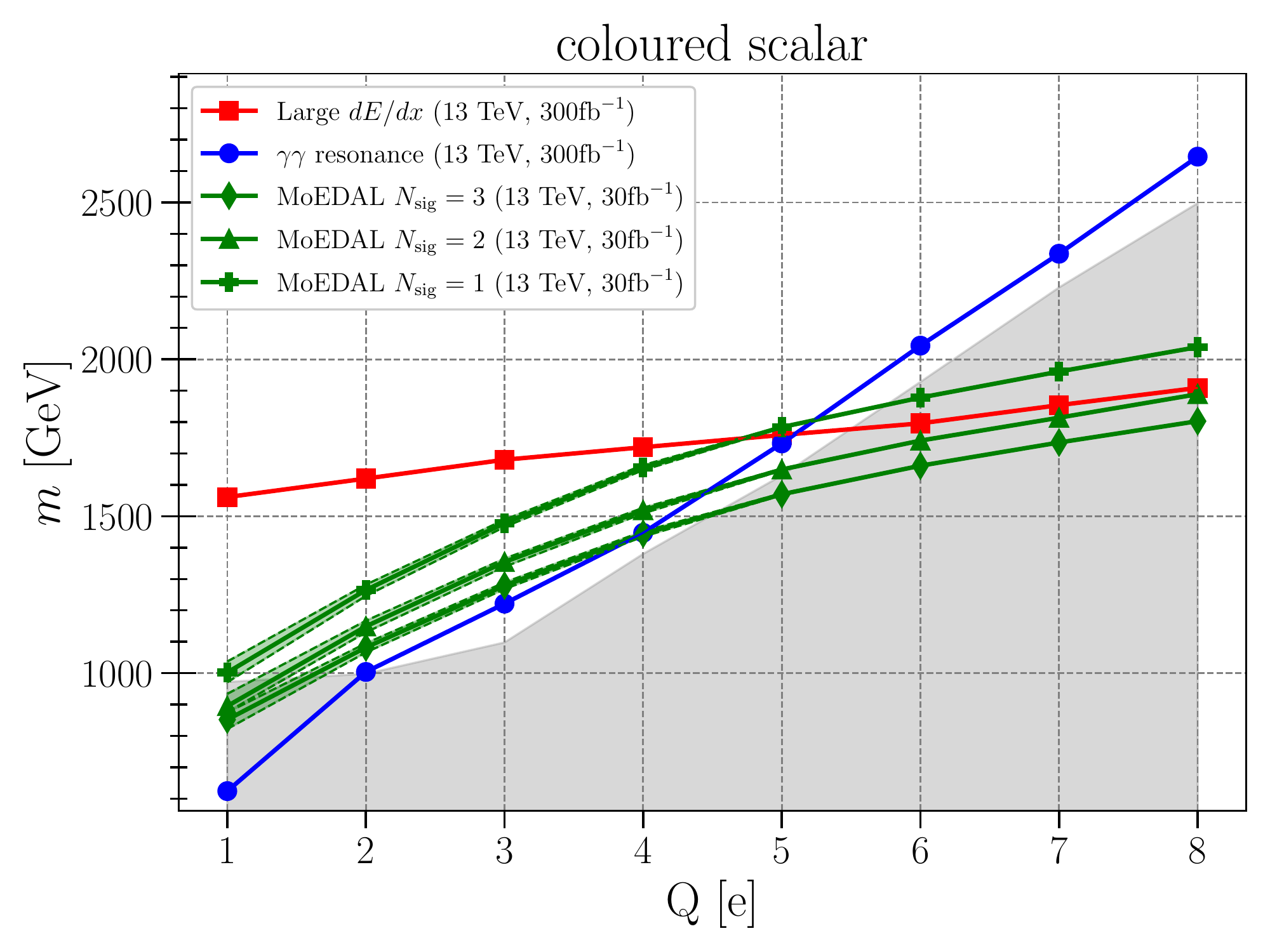}
    \includegraphics[width=0.49\textwidth]{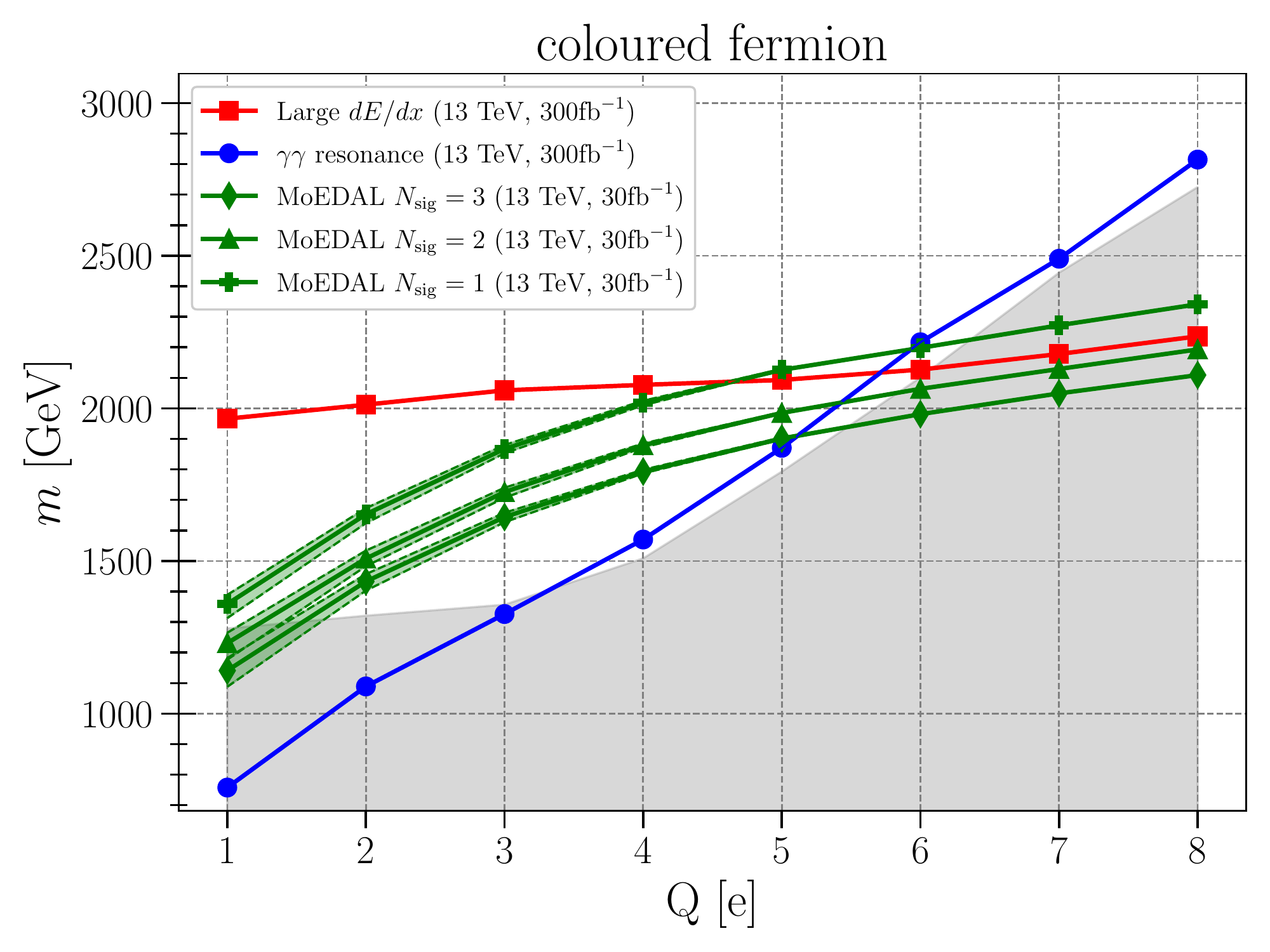}
    \caption{\small Expected lower mass limits on BSM particles: colourless scalars (top left), colourless fermions (top right), coloured scalars (bottom left) and coloured fermions (bottom right); for Run-3 LHC experiments. Integrated luminosity for ATLAS/CMS was $L=300$ $\mathrm{fb}^{-1}$ and $L=30$ $\mathrm{fb}^{-1}$ for MoEDAL.
    Green bounds around MoEDAL limits for coloured particles come from uncertainty of the adopted hadronisation model. For open channel ATLAS/CMS searches uncertainty is too small to be visible.
    }
    \label{fig:res_run3}
\end{figure}

Figure~\ref{fig:res_run3}
shows the projected mass limits 
for Run 3.
The red and blue curves correspond to the 95\% CL limits
from the large $dE/dx$ and diphoton resonance searches, respectively,
while the green curves represent 
the expected signal detection 
at the MoEDAL experiment,
corresponding to $N_{\rm sig} = 1$, 2 and 3.
{We do not show the lower boundary of the 
large $dE/dx$ sensitivity 
since the region below it is already covered (or excluded) by the other analyses, e.g.~the MoEDAL and ATLAS diphoton analyses in the plot as well as the analyses elsewhere with the Run-1 and Run-2 data.}
For coloured scalars and fermions,
we vary the parameter, $k$, in our hadronisation model
from 0.3 to 0.7.
This in principle impacts the sensitivities 
of open production searches, i.e.~the large $dE/dx$ 
and MoEDAL analyses, since the electric charge
of the final state particles are probabilistically shifted.
This effect is visible 
in the MoEDAL's sensitivity 
for smaller $|Q|$ ($|Q| \lesssim 3e$)
as a width of the green curves.
However, for larger charges and for the large $dE/dx$ analysis,
the effect is very small and almost invisible in the plots.
This assures that the uncertainty coming from the hadronisation model is small and the mass limits are robust against this uncertainty.

When extrapolating the mass bounds of large $dE/dx$ analysis
from $L=2.5$ to 300 fb$^{-1}$,
we follow the procedure described in 
\cite{Allanach:2011wi, Sakurai:2011pt}
and obtained the 95\% CL upper limit on the detected signal.
In this procedure we assume the systematic uncertainty stays the same level ($\sim 66 \%$) relative to the total background.  
Our projected limits are conservative in this regard. 
The grey shaded regions show
the currently excluded regions already presented in Fig.~\ref{fig:res_curr}.
We see that the Run-3 diphoton limit improves the current limit only mildly.  This is because the current limit is calculated already with the 139 fb$^{-1}$ data.
This may be contrasted to the large $dE/dx$ limits where the improvement is significant, since the current limit from the large $dE/dx$ search  
in Fig.~\ref{fig:res_curr}
is based on the 2.5 fb$^{-1}$ data except for the colourless fermions 
with $2 \le |Q|/e \le 7$,
for which the limit is based on the 36.1 fb$^{-1}$ data.
As discussed above, 
the large $dE/dx$ analysis constrains smaller charges well,
while the diphoton resonance search is very constraining for large values of $|Q|$.
At Run 3 the sensitivities from these searches become comparable for $|Q| \sim (5-6)e$.
The sensitivities of the MoEDAL experiment typically lie between 
the aforementioned two constraints.
As can be seen, MoEDAL's Run-3 sensitivity with $N_{\rm sig} = 1$ may be as strong as 
those of the large $dE/dx$ and diphoton resonance searches for $|Q| \sim (4-6)e$.

\begin{figure}[t!]
    \centering
    \includegraphics[width=0.49\textwidth]{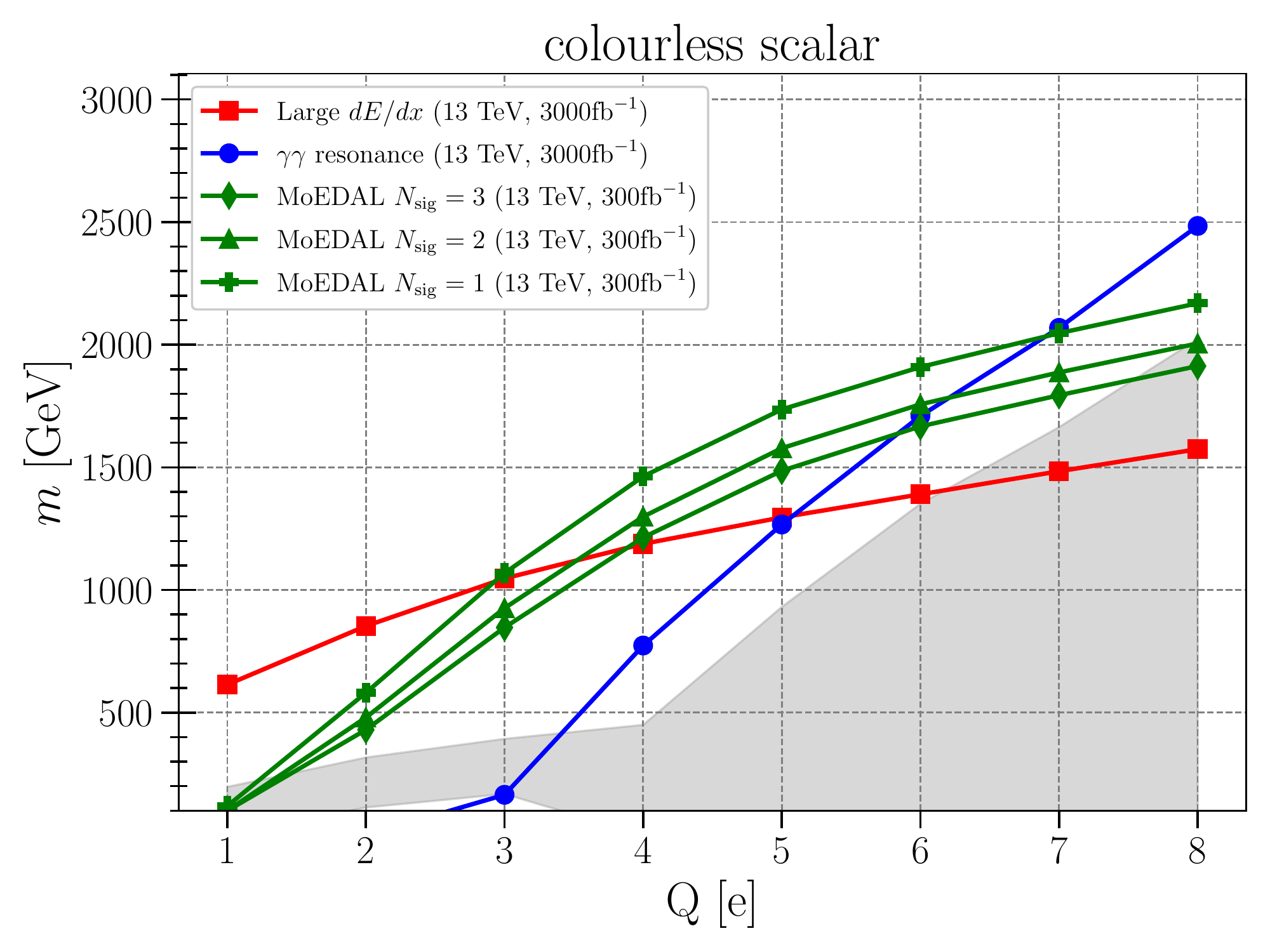}
    \includegraphics[width=0.49\textwidth]{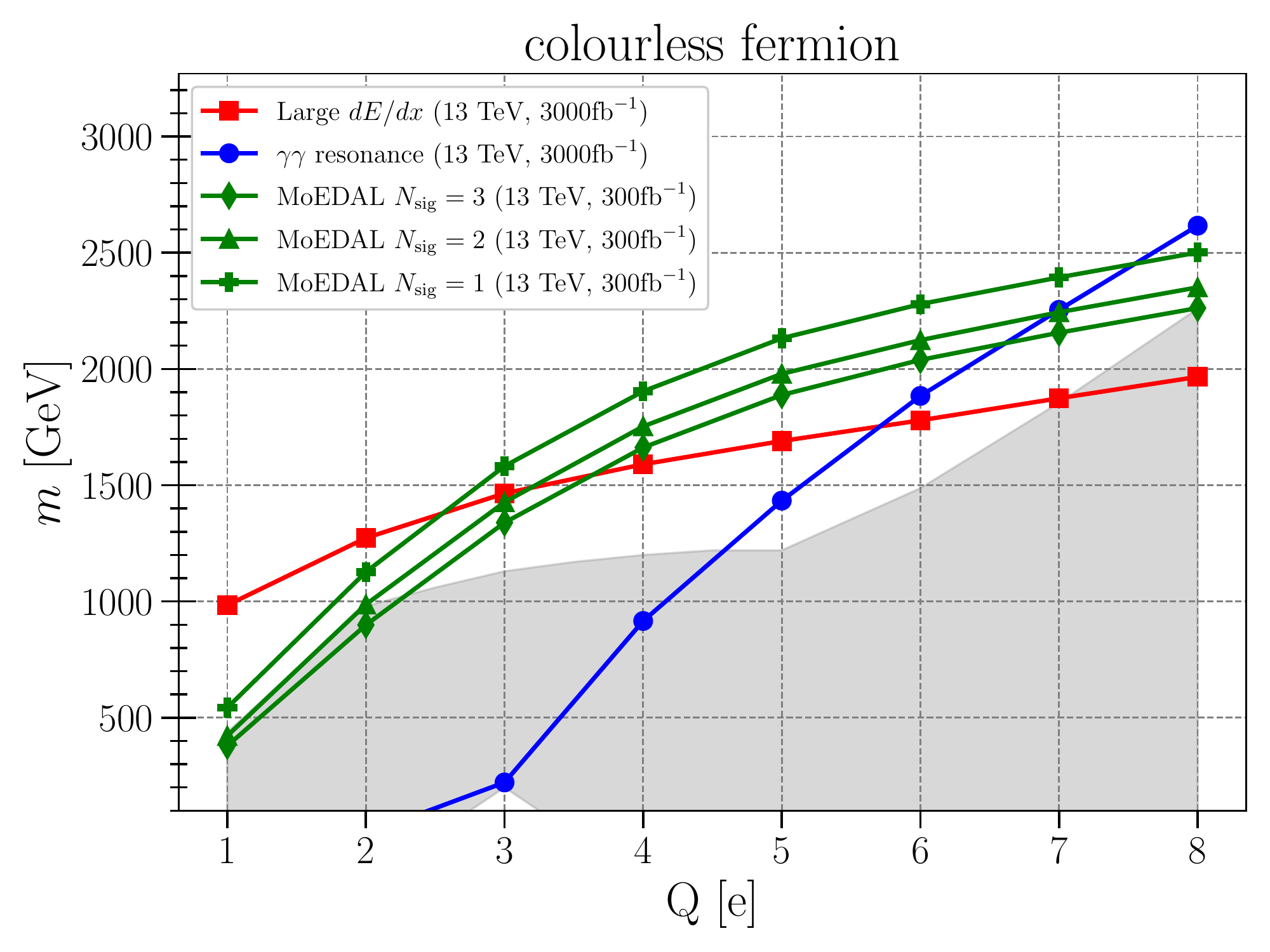}
    \includegraphics[width=0.49\textwidth]{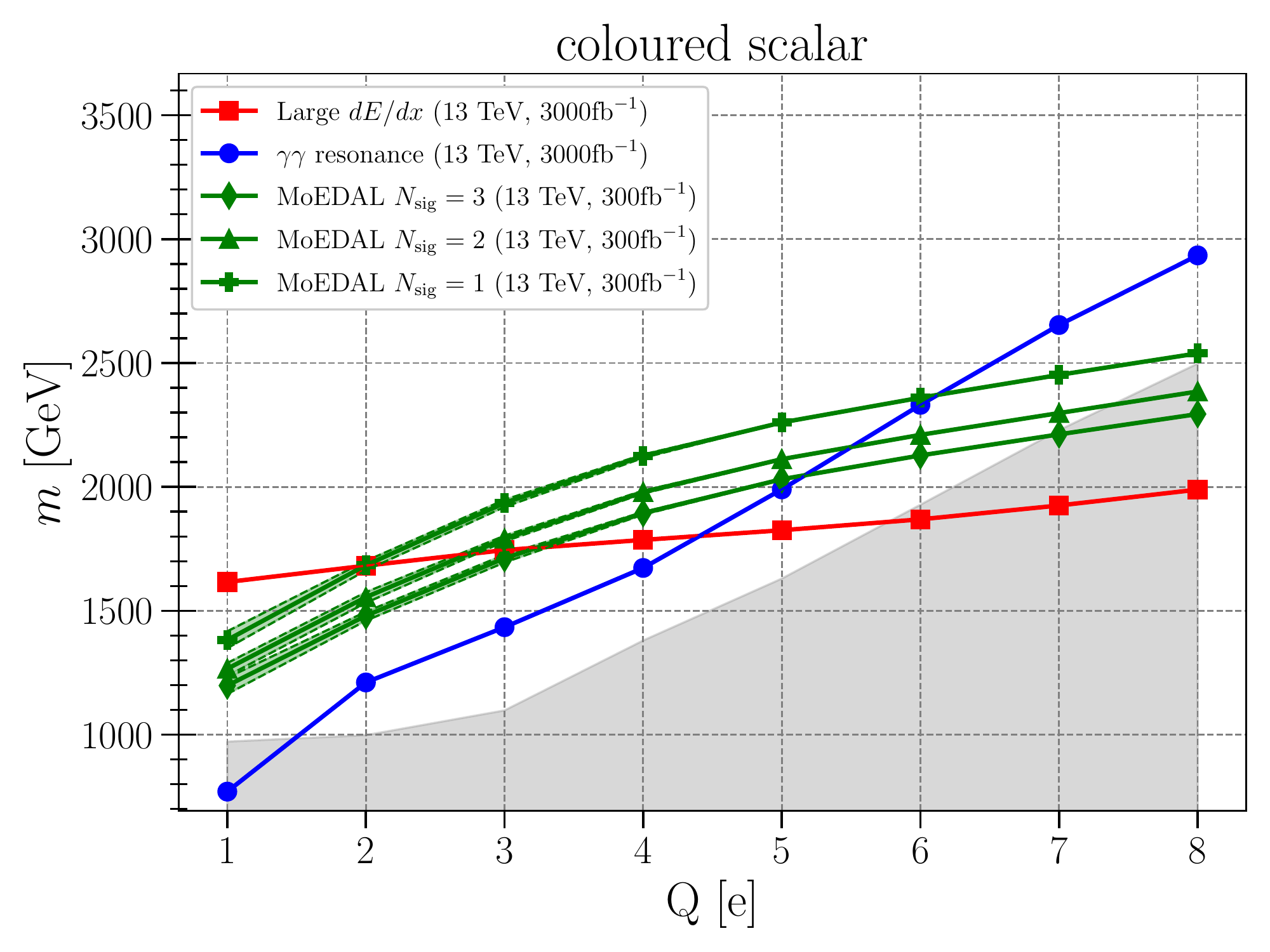}
    \includegraphics[width=0.49\textwidth]{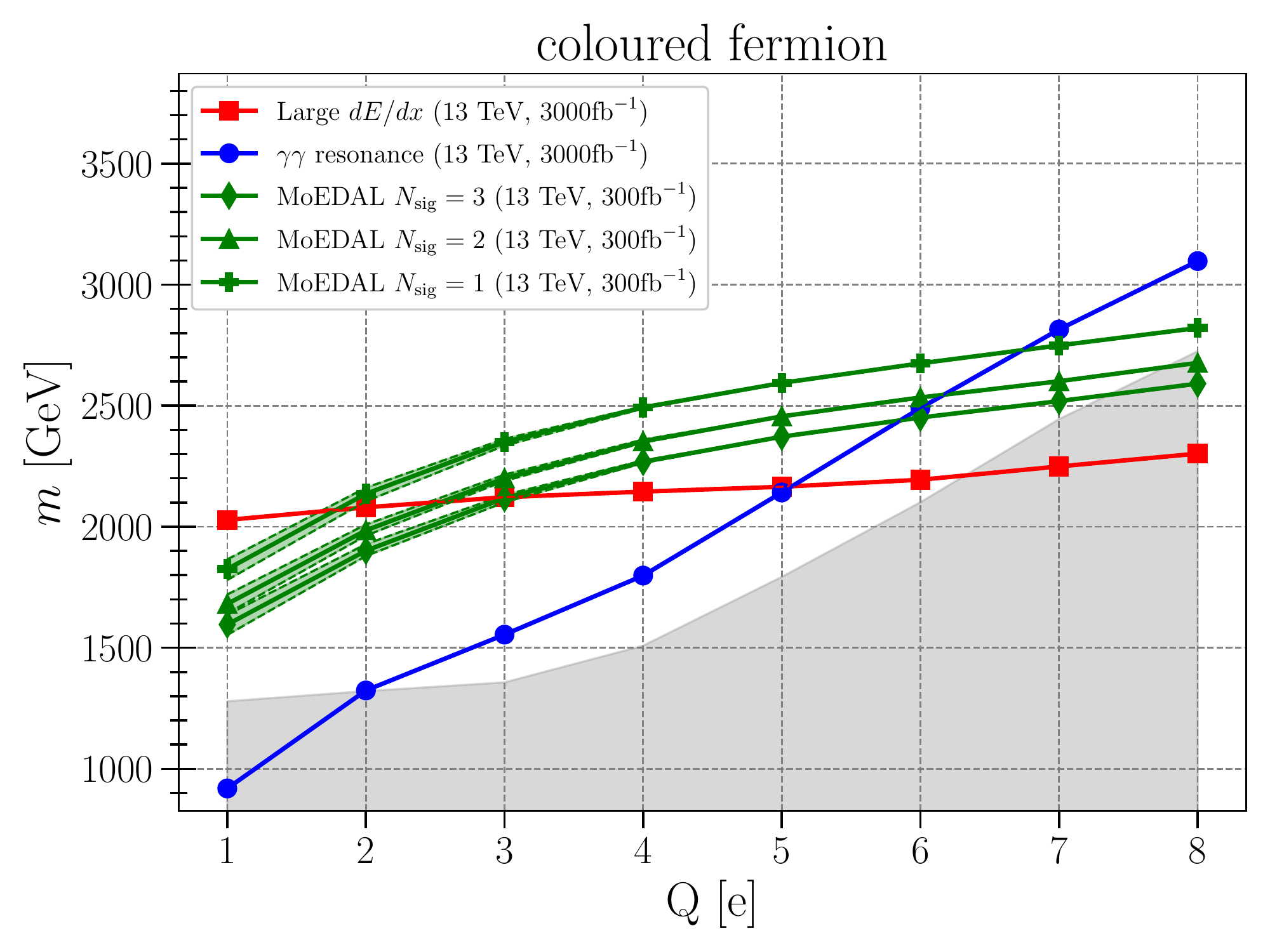}
    \caption{\small Expected lower mass limits on BSM particles: colourless scalars (top left), colourless fermions (top right), coloured scalars (bottom left) and coloured fermions (bottom right); scaled up to HL-LHC luminosity. Integrated luminosity for ATLAS/CMS was $L=3000$ $\mathrm{fb}^{-1}$ and $L=300$ $\mathrm{fb}^{-1}$ for MoEDAL.
    Green bounds around MoEDAL limits for coloured particles come from uncertainty of the adopted hadronisation model. For open channel ATLAS/CMS searches uncertainty is too small to be visible.
    }
    \label{fig:res_hl}
\end{figure}

We finally show the expected sensitivities at the HL-LHC in Fig.~\ref{fig:res_hl}.
We see that the improvement of the large $dE/dx$ limits are very mild.
We employ the same event selection used in \cite{CMS:2016kce},
where CMS estimated 
the background contribution 
to the relevant signal region
to be 0.06 with $L = 2.5$ fb$^{-1}$.
For the HL-LHC, however,
the background contamination to the same signal region is expected to be 72 events, which requires a large 
number of
detected signal events for exclusion. 
We expect that the limit from large $dE/dx$ analyses can be further improved by optimising the event selection, which is beyond the scope of this paper.
Compared to the large $dE/dx$ and 
diphoton resonance searches,
the improvement of the MoEDAL's mass limits 
from Run-3 to HL-LHC is significant. 
This is largely due to the background free nature of the MoEDAL experiment. While for ATLAS and CMS experiments the increase of luminosity results in larger number of expected signal and background events, only the expected signal rate grows for MoEDAL.
We see that for $3 \lesssim |Q|/e \lesssim 7$ MoEDAL may provide superior sensitivities than the other searches by ATLAS and CMS.

\section{Conclusion}
\label{sec:concl}

In this paper we have studied the prospect of
observing long-lived particles with 
high electric charges, ranging $1 \le |Q/e| \le 8$, at the LHC.
In particular, we compared 
Run-3 and HL-LHC sensitives of three different types of searches:
(a) anomalous track searches (i.e.~large $dE/dx$ analyses) at ATLAS and CMS,
(b) charged LLP searches at MoEDAL
and
(c) diphoton resonance searches at ATLAS and CMS.
The former two target the open production, $pp \to \xi^{+Q} \xi^{-Q}$, with detection of charged LLPs within the detectors,
whereas the latter concerns the production of the bound state followed by the decay into two photons, $pp \to {\cal B}(\xi^{+Q} \xi^{-Q}) \to \gamma \gamma$.

For the open production mode,  we demonstrated the importance of the photonic initial states for the cross section, which have been omitted in the current ATLAS and CMS large $dE/dx$ analyses.
Inclusion of photonic initial states is also crucial
to correctly estimate the MoEDAL's detection sensitivity, since it has a large impact on the production velocities of the charged particles especially for scalars.

In order to estimate the current sensitivities of the anomalous track searches and extrapolate them to the future luminosities, we recast the CMS large $dE/dx$ analysis \cite{CMS:2016kce}
for various charged LLPs
including the photonic initial states.
We also studied the effects of underestimation of the charged track $p_T$ as well as 
the velocity loss due to the (strong) electromagnetic interaction with the detector materials. 
Since these two effects grow with $|Q|$ and negatively impact the signal efficiency, we found that the large $dE/dx$ 
is most powerful for smaller $|Q|$.

On the other hand, the sensitivity of 
the diphoton resonance search grows rapidly with $|Q|$, since the event rate of $pp \to {\cal B} \to \gamma \gamma$ scales as $|Q|^{10}$.

We found the MoEDAL's sensitivity is somewhere between those of the anomalous track searches and the diphoton resonance searches.
For Run-3 MoEDAL may be competitive at best around $|Q| \sim (5-6)e$.
Due to the background free nature of MoEDAL,
its sensitivity grows faster relative 
to that of the anomalous track and diphoton resonance searches as the luminosity increases. 
At the HL-LHC, therefore, MoEDAL may give superior sensitivities than ATLAS and CMS for $3 \lesssim |Q| \lesssim 6$.   

Although our conclusion is qualitatively robust, 
the mass bounds are derived based on the leading-order cross sections,
since the electromagnetic higher order corrections for $Q > 1e$ particles have not been calculated.
Also,
our estimate of Run-3 and HL-LHC sensitivities from the large $dE/dx$ and diphoton resonance searches are based on the event selection employed in Run-2 and the conservative assumption on the systematic uncertainty.   
We hope this study encourages more thorough studies on the highly charged LLPs searches at ATLAS and CMS as well as the MoEDAL experiments.

\section*{Acknowledgments}

We thank James Pinfold for useful comments.
The works of M.M.A.\ and P.L.\ are partially supported by the National Science
Centre, Poland, under research grant 2017/26/E/ST2/00135.  
The work of R.M.\ is partially supported by 
the National Science Centre, Poland,
under research grant 2017/26/E/ST2/00135
and 
the Beethoven grant DEC-2016/23/G/ST2/04301.
The work of V.A.M.\ is supported in part by the Generalitat Valenciana via a special grant for MoEDAL, by the Spanish MICIU\,/\,AEI and the European Union\,/\,FEDER via the grant PGC2018-094856-B-I00 and by the CSIC AEPP2021 grant 2021AEP063.
The work of
K.S.\ is partially supported by the National Science Centre, Poland,
under research grant 2017/26/E/ST2/00135 and 
the Norwegian Financial Mechanism for years 2014-2021, grant DEC-2019/34/H/ST2/00707.

\newpage

\appendix

\section{Effect of $Z$-boson induced processes}
\label{ap:a0}

\begin{figure}[t!]
\centering
\begin{feynman}
    \fermion[label=$u$, showArrow=false]{4.00, 4.0}{5.00, 4.00}
    \dashed[label=$\phi$, showArrow=false]{6.00, 4.50}{8.50, 4.0}
    \dashed[label=$\phi$, showArrow=false]{6.00, 6.50}{8.50, 7.0}
    \fermion[label=$u$, showArrow=false]{5.00, 4.00}{7.00, 3.0}
    \electroweak[label=$\gamma/Z$]{5.00, 7.00}{6.00, 6.50}
    \dashed[label=$\phi$, showArrow=false]{6.00, 4.50}{6.00, 6.50}
    \electroweak[label=$\gamma/Z$]{5.00, 4.00}{6.00, 4.50}
    \fermion[showArrow=false, label=$u$]{5.00, 7.0}{7.00, 8.0}
    \fermion[showArrow=false, label=$u$]{4.00, 7.00}{5.00, 7.00}
\end{feynman}
\caption{\label{fig:ZZ}
Example of a $Z/\gamma$ fusion diagram.}
\end{figure}
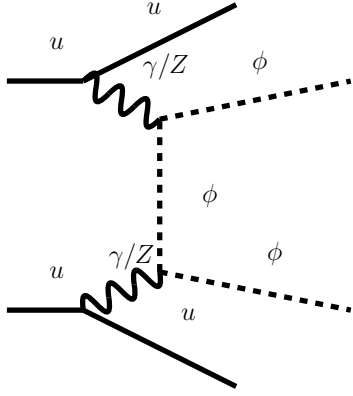

In section \ref{sec:open},
we showed 
that
the
production of multi-charged particles 
from the $\gamma \gamma$ initial state 
has cross sections proportional to $Q^4$
, hence
becomes important for large $Q$.
Since the production rate of the $Z$-boson initiated processes is also proportional to $Q^4$,
we perform a rough estimate to understand 
the significance of the $Z$-boson initiated processes.
Since the $Z$-boson is much heavier than the proton,
the parton distribution functions (PDFs) used for LHC analyses 
do not include the $Z$-boson as a parton.\footnote{Electroweak bosons ($Z$, $W^{\pm}$ and $h$) should be included in the PDFs for future high energy hadron colliders with $\sqrt{s} \sim 100$ TeV.  See e.g.\ \cite{Bauer:2017isx}.}
At LHC energies, the $Z$-boson initiated processes should be better understood as the $Z/\gamma$ fusion processes, 
$qq \to q q \phi^{+Q} \phi^{-Q}$
or
$qq \to q q \psi^{+Q} \psi^{-Q}$.
One of such diagrams is depicted in Fig.~\ref{fig:ZZ}.

\tikzset{every picture/.style={
    scale=0.4, transform shape,
    }}

To evaluate the significance of the $Z$-boson contribution to the $Z/\gamma$ fusion process with respect to the $\gamma \gamma$ fusion, we compare the two cross sections computed with {\tt MadGraph\,5} by the following commands: 
\begin{align*}
    \sigma_{\rm inc}:~ &~ \texttt{generate u u > u u phiQ+ phiQ- QCD=0} \\
    \sigma_{\rm exc}:~ &~ \texttt{generate u u > u u phiQ+ phiQ- QCD=0 / z} 
\end{align*}
The \texttt{QCD=0} option
prohibits QCD vertices from appearing in the diagrams.
With the first command, $\sigma_{\rm inc}$ represents 
the cross section of the $u u \to u u \phi^{+Q} \phi^{-Q}$ process including both $Z$ and $\gamma$ in the intermediate state.
On the other hand, the second command asks for the same process but excludes $Z$-bosons anywhere in the diagrams.
$\sigma_{\rm exc}$ therefore corresponds to the cross section to which only the $\gamma \gamma$ fusion contributes.
From these cross sections, we construct 
\be
\frac{
\sigma_{\rm inc} - \sigma_{\rm exc}
}{
\sigma_{\rm exc} 
}
\nonumber
\ee
and use it to measure the significance of the $Z$-boson contribution with respect to the $\gamma \gamma$ fusion.

Table \ref{tb:Zg_ratio} shows $(\sigma_{\rm inc} - \sigma_{\rm exc})/\sigma_{\rm exc}$ calculated for
$Q=8e$ scalars and fermions at several mass points.
For some cases we find $\sigma_{\rm inc} < \sigma_{\rm exc}$
due to destructive interference. 
As can be seen, in the mass range we are interested,  
$(\sigma_{\rm inc} - \sigma_{\rm exc})/\sigma_{\rm exc} \lesssim 15\,\%$,
which suggests that the $Z$-boson contribution to the $Z/\gamma$ fusion process is subdominant compared to the $\gamma \gamma$ fusion.
This justifies our approximation to neglect the
$Z$-boson initiated productions in the analysis.

\begin{table}

\begin{center}
\begin{tabular}{|c || c | c | c | c|} 
 \hline
 mass [TeV] & 0.5 & 1 & 2 & 3 \\ 
 \hline
 \hline
 scalar & $-0.13$ & $-0.07$ & $0.08$ & $0.15$
 \\ 
 \hline
 fermion & $-0.15$ & $-0.12$ & $-0.04$ & $0.05$ 
\\
\hline
\end{tabular}
\vspace{3mm}
\caption{\label{tb:Zg_ratio}
$(\sigma_{\rm inc} - \sigma_{\rm exc})/\sigma_{\rm exc}$
for scalars and fermions with $Q=8e$.
}
\end{center}
\end{table}

\section{Detection reach for MoEDAL}
\label{ap:a}

\begin{figure}[t!]
\centering
      \includegraphics[width=0.4\textwidth]{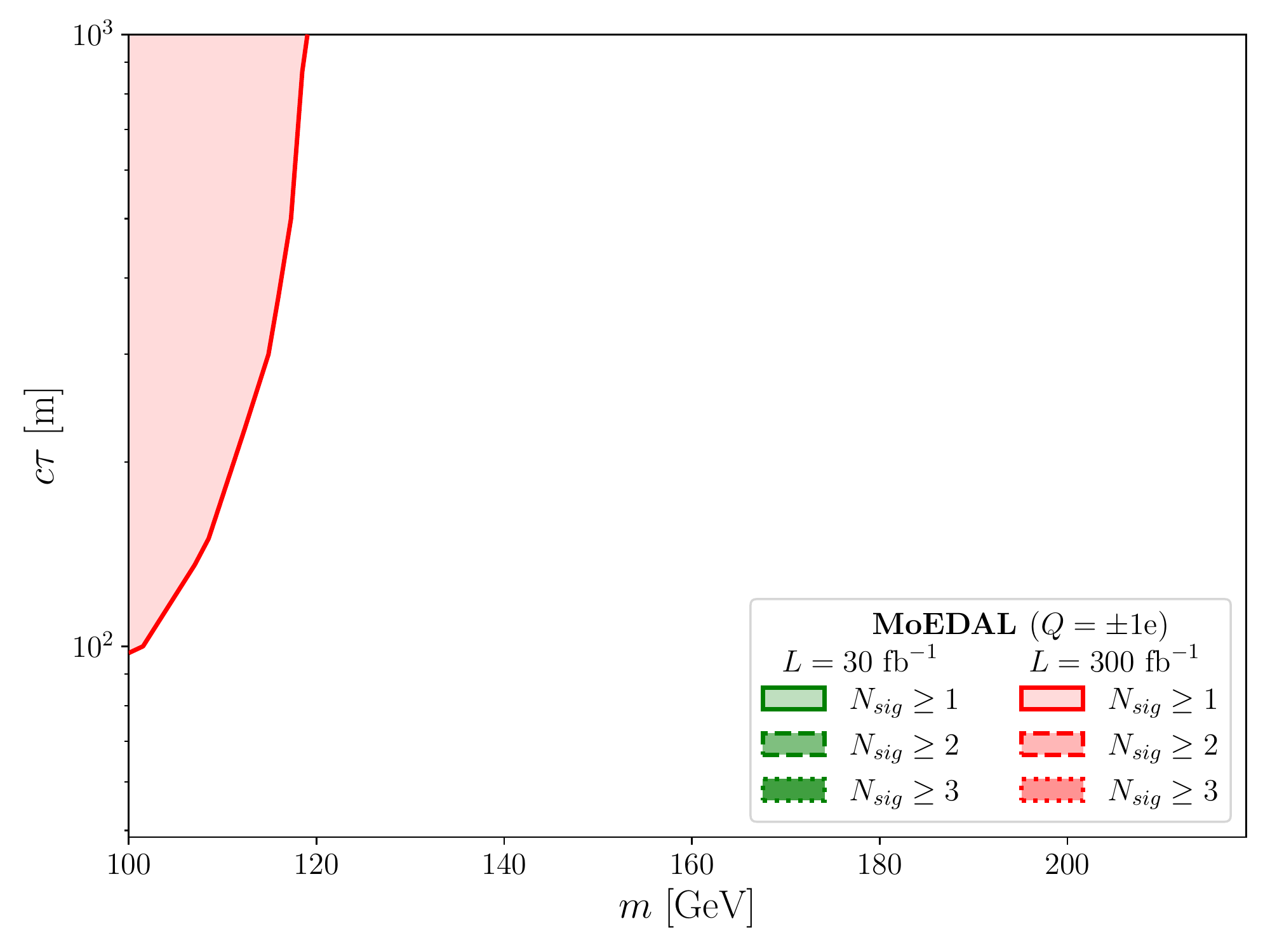} \hspace{5mm}
      \includegraphics[width=0.4\textwidth]{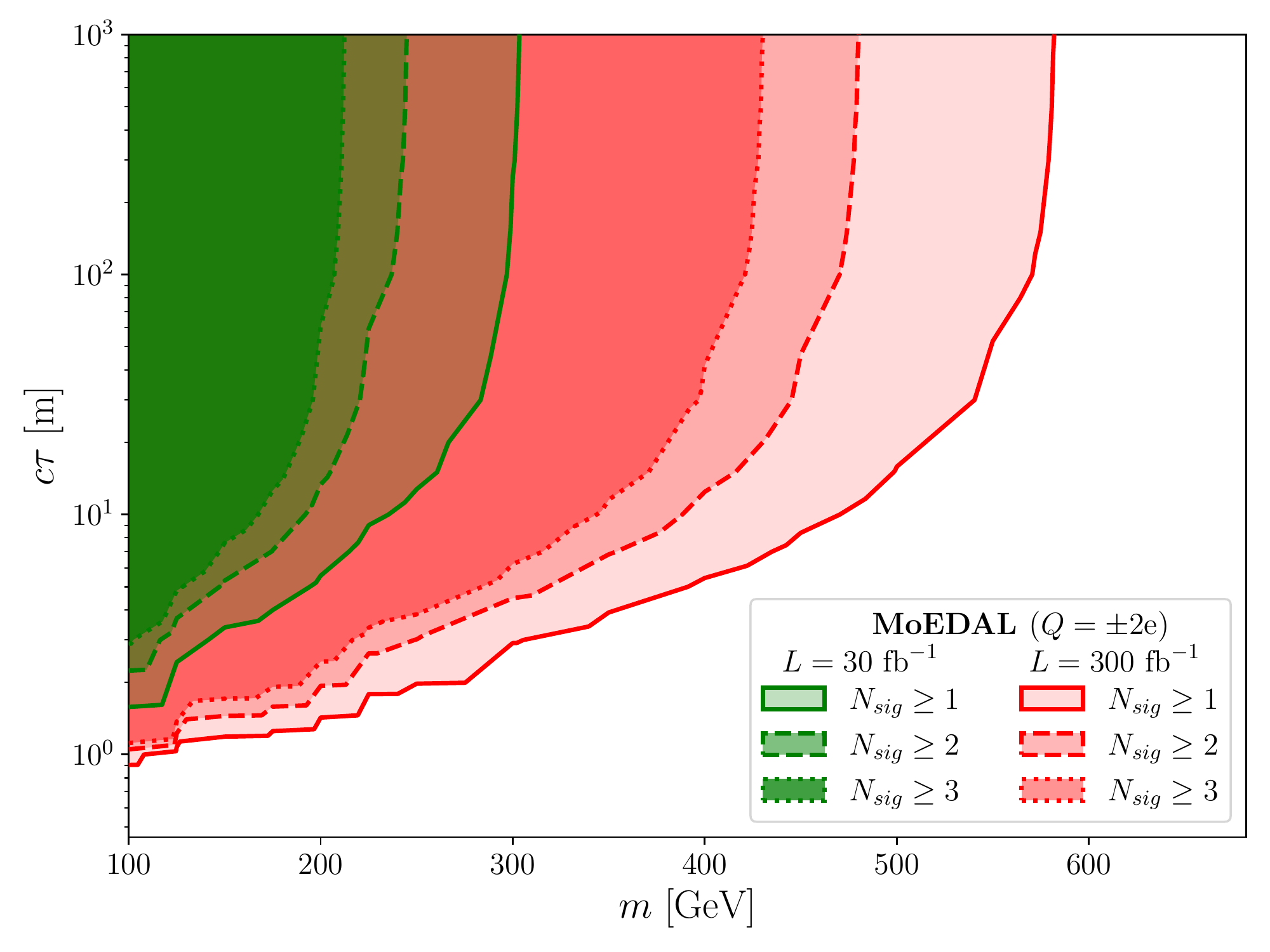}
      \includegraphics[width=0.4\textwidth]{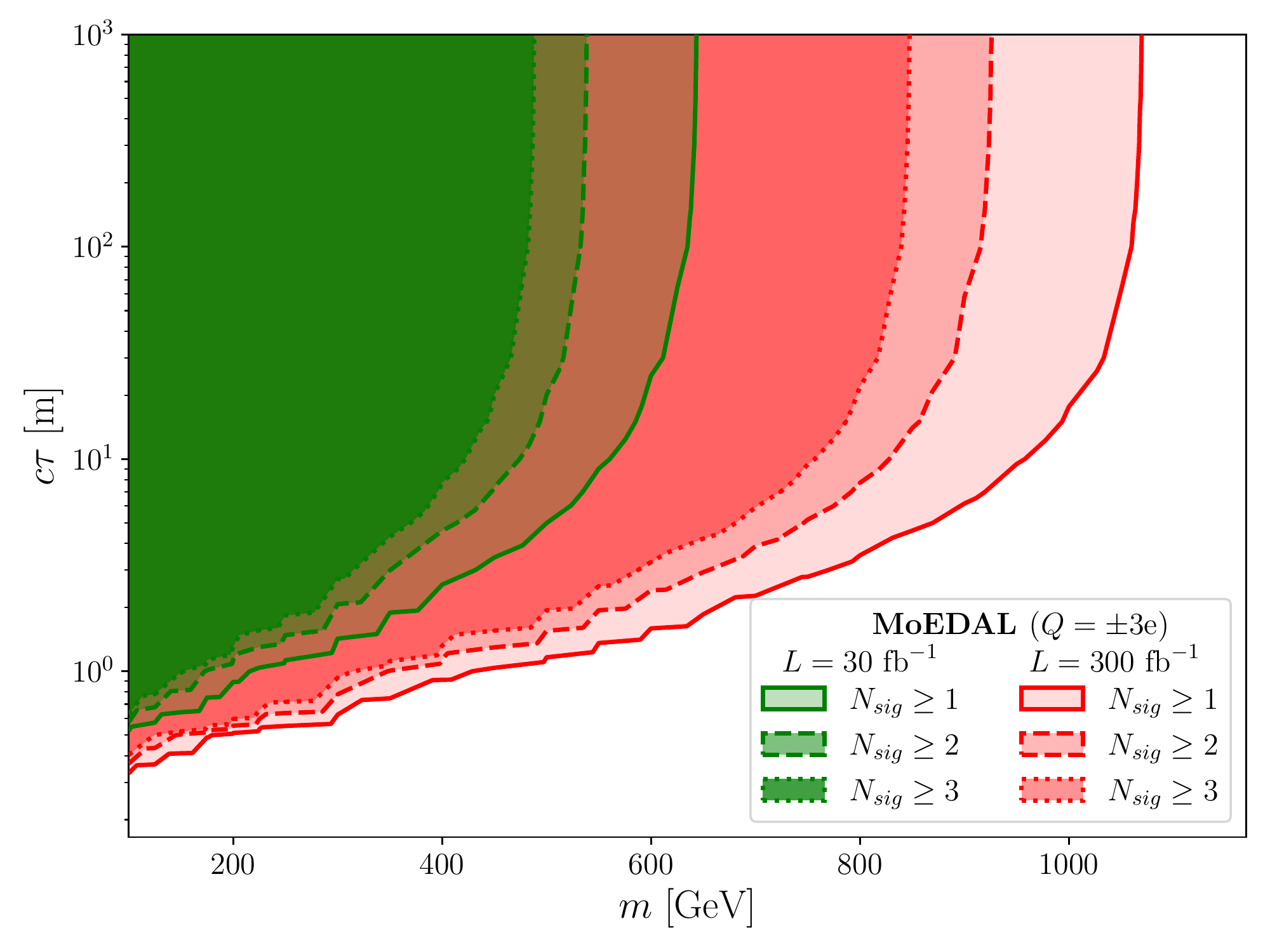}  \hspace{5mm}
      \includegraphics[width=0.4\textwidth]{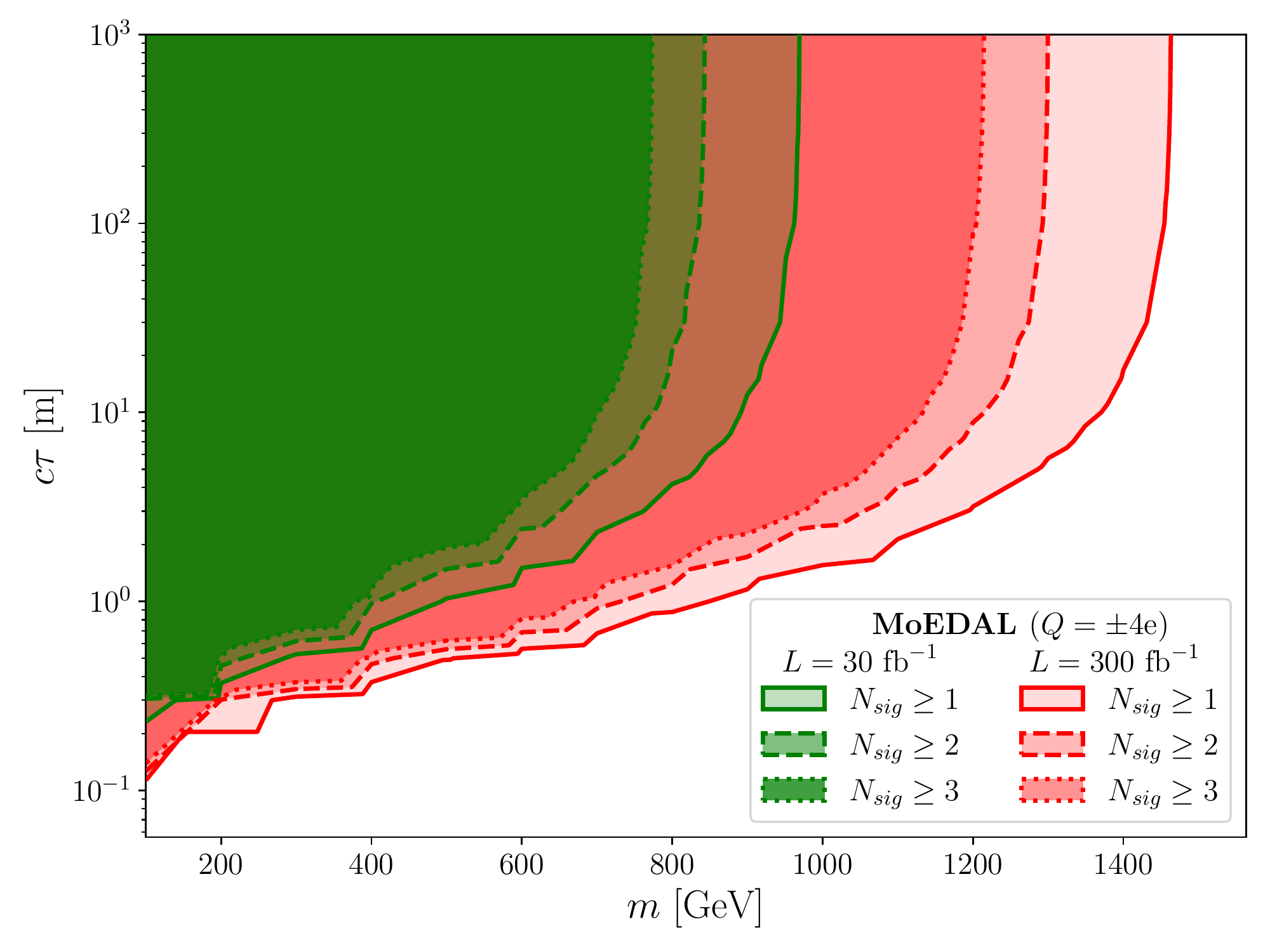}
      \includegraphics[width=0.4\textwidth]{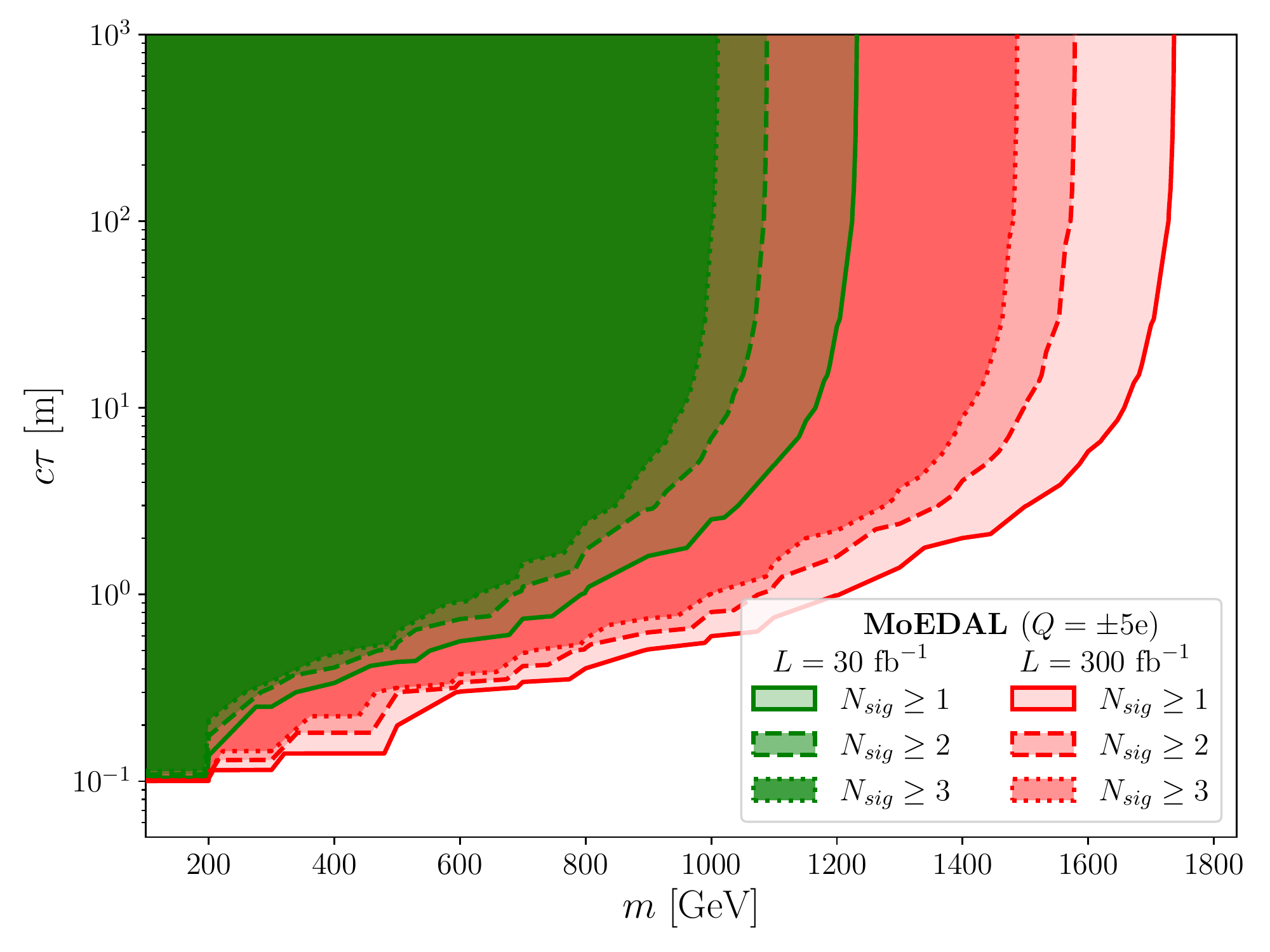}  \hspace{5mm}
      \includegraphics[width=0.4\textwidth]{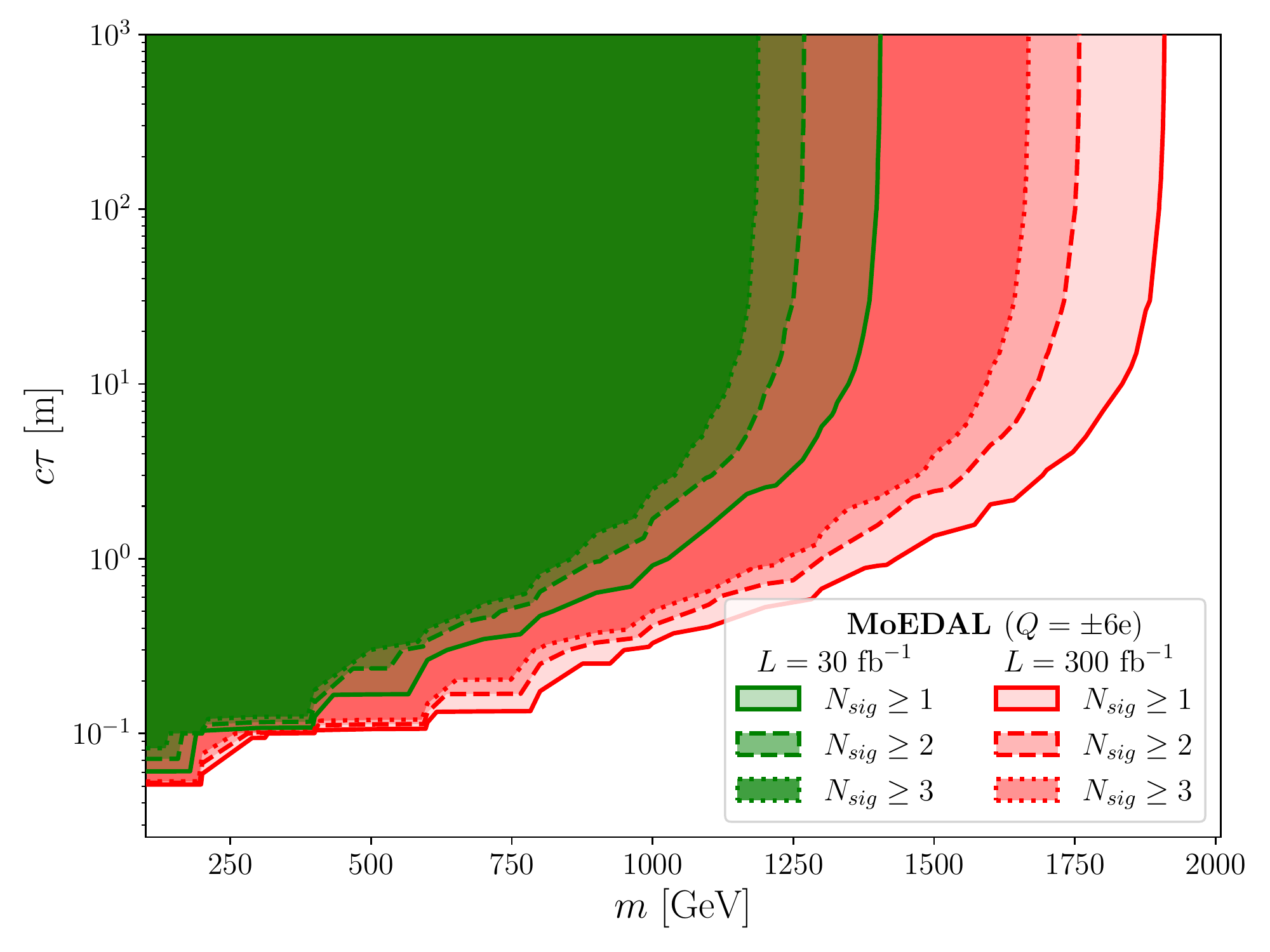}
      \includegraphics[width=0.4\textwidth]{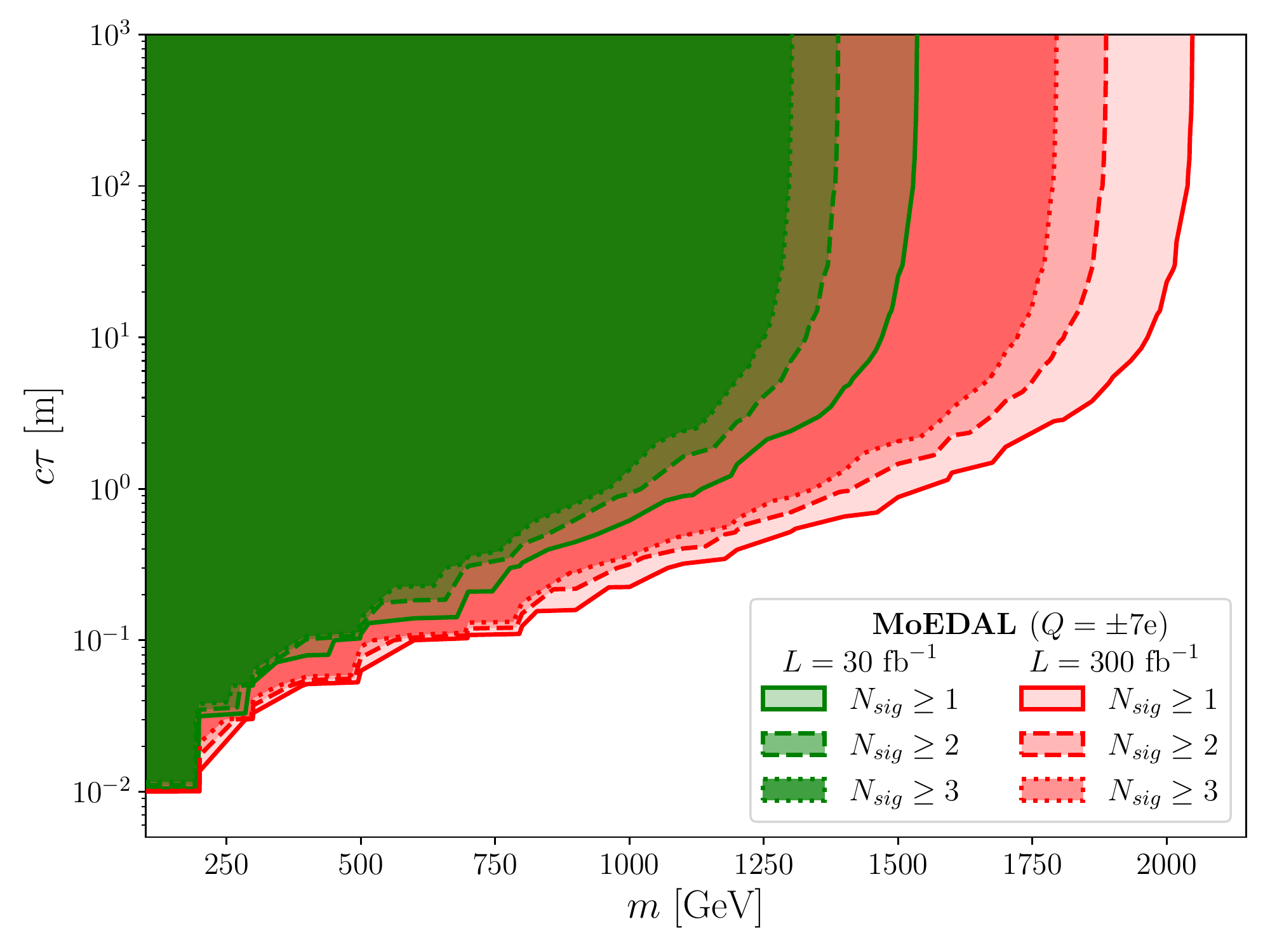}  \hspace{5mm}
      \includegraphics[width=0.4\textwidth]{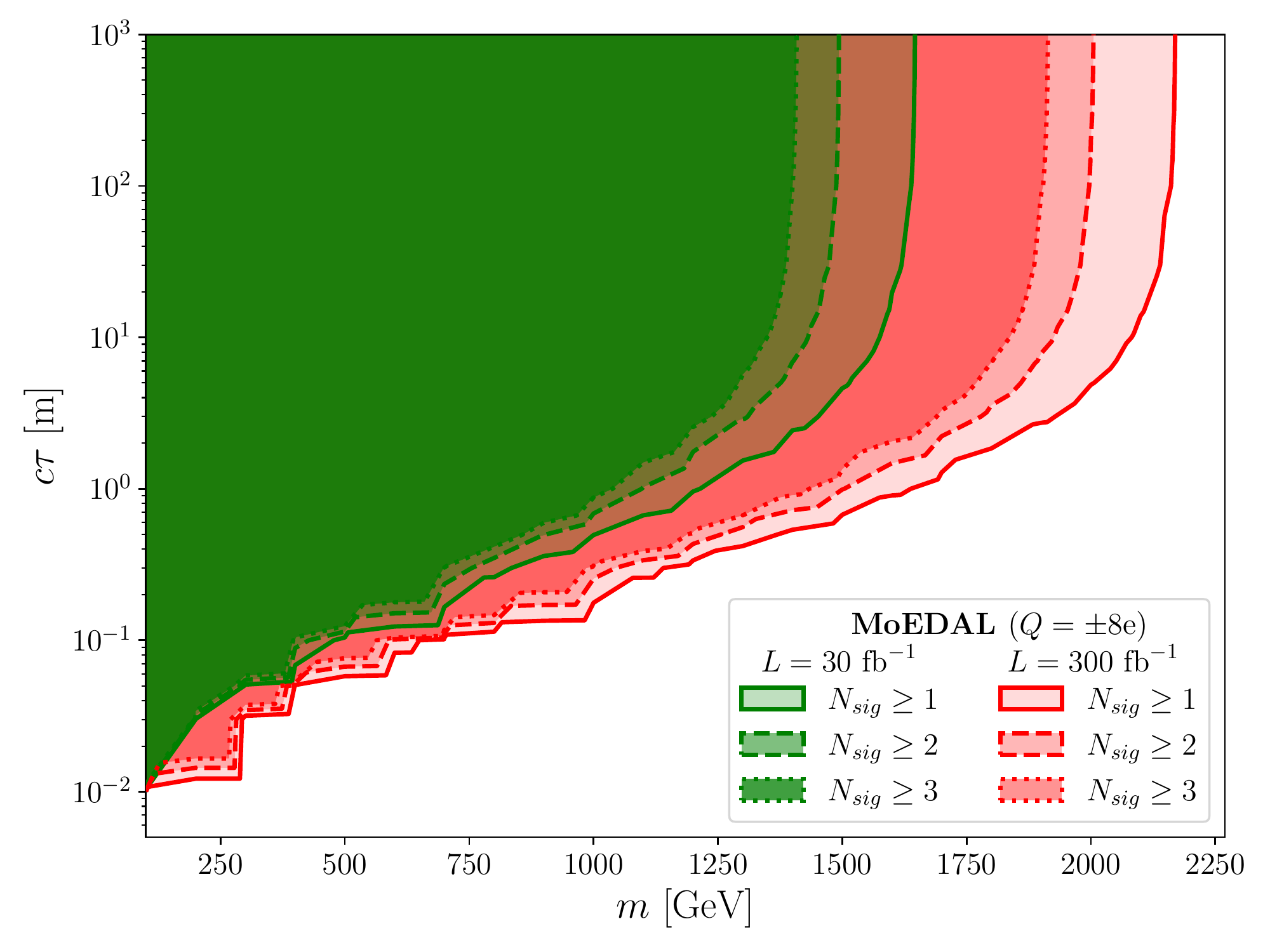}  
\caption{\small Model-independent detection reach at MoEDAL in the ($m$, $c
  \tau$) parameter plane for colourless scalars. The
  solid dashed and dotted contours correspond to $N_{\rm sig} = 1$, 2
  and 3, respectively. Green and red contours correspond to Run-3 
  $(L=30$ $\mathrm{fb}{}^{-1})$ and HL-LHC $(L=300$ $\mathrm{fb}^{-1})$ luminosities respectively.
  }
\label{fig:lim_sHighQ}
\end{figure}

\begin{figure}[t!]
\centering
      \includegraphics[width=0.4\textwidth]{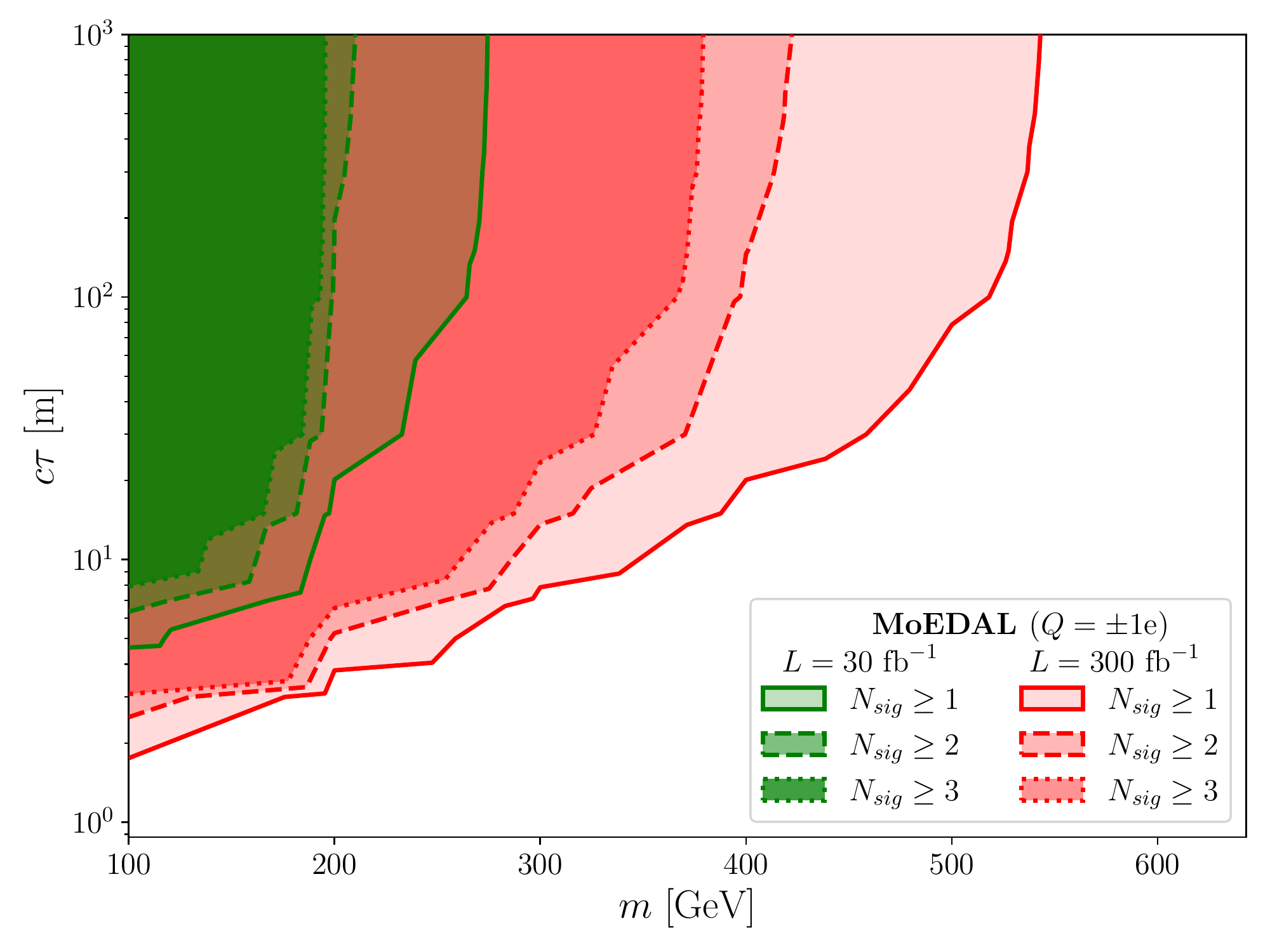} \hspace{5mm}
      \includegraphics[width=0.4\textwidth]{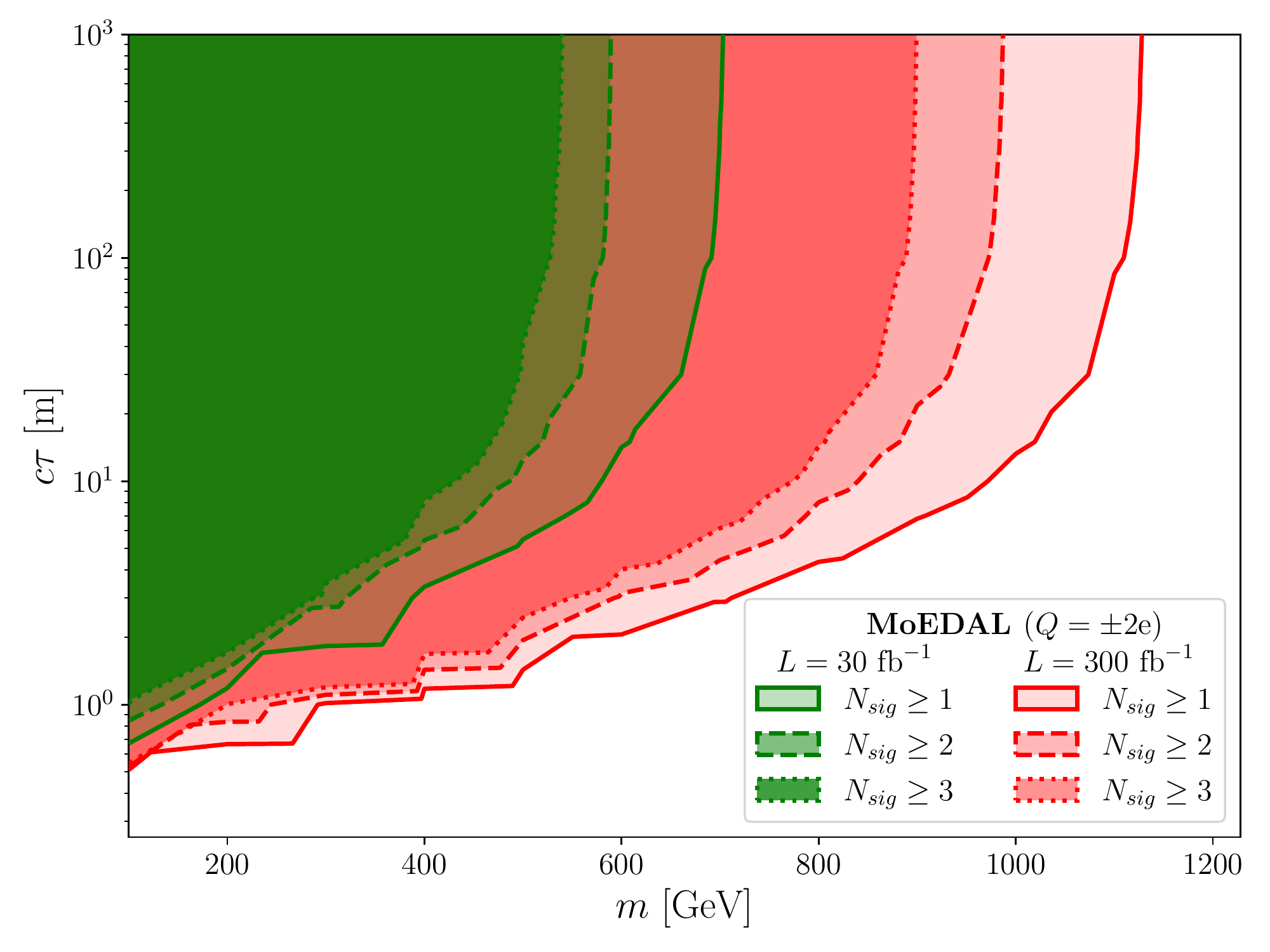}
      \includegraphics[width=0.4\textwidth]{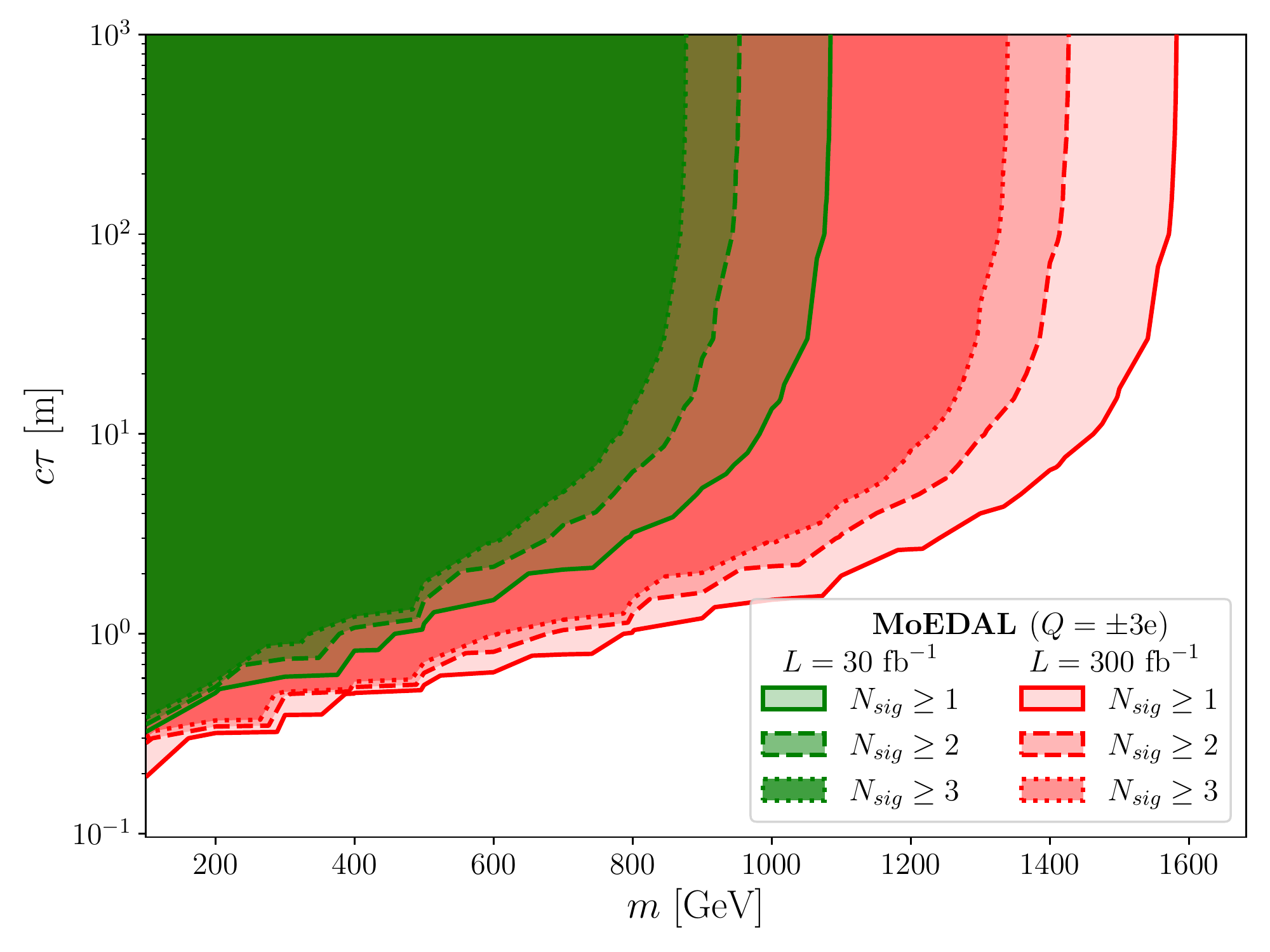}  \hspace{5mm}
      \includegraphics[width=0.4\textwidth]{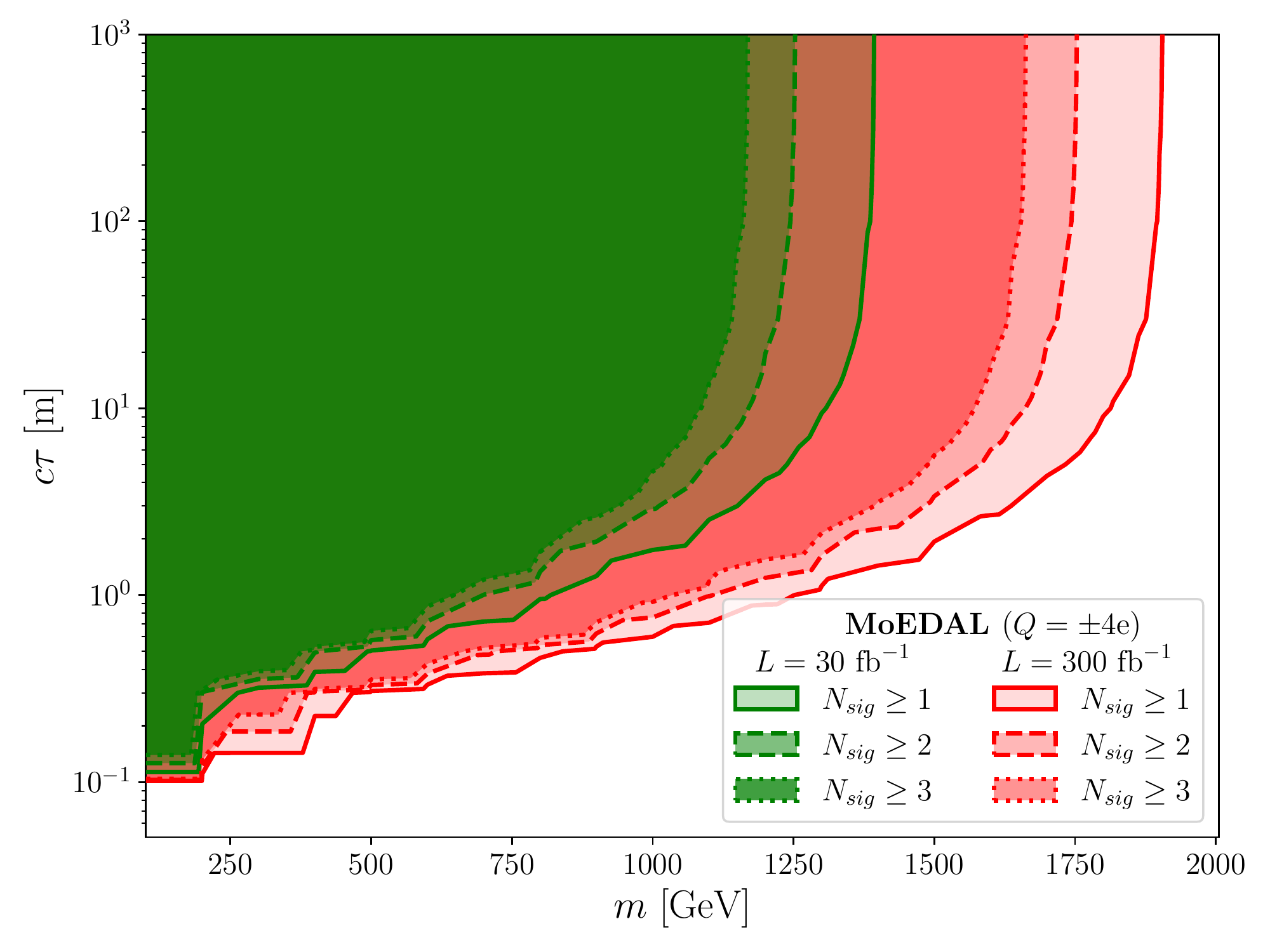}
      \includegraphics[width=0.4\textwidth]{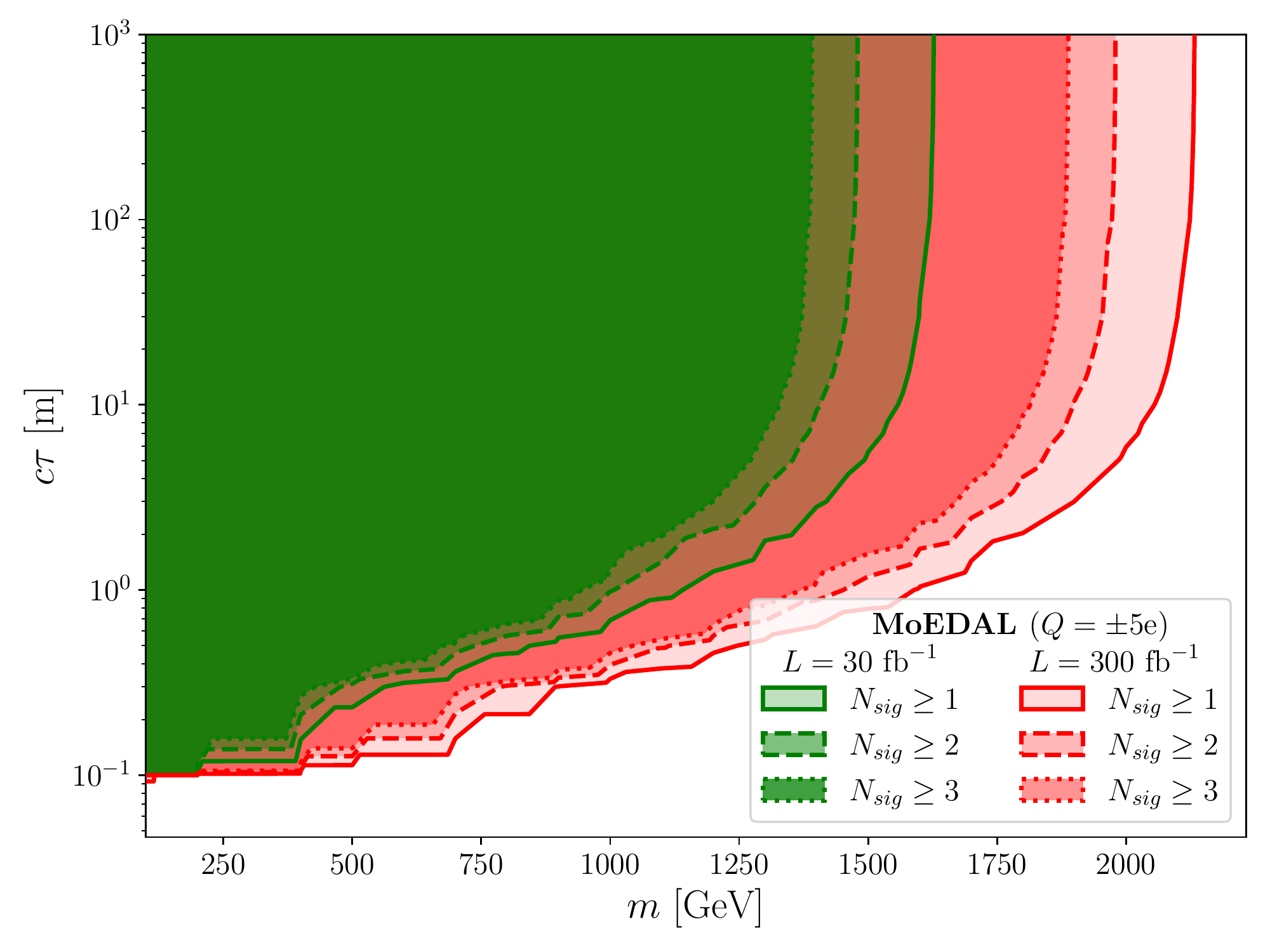}  \hspace{5mm}
      \includegraphics[width=0.4\textwidth]{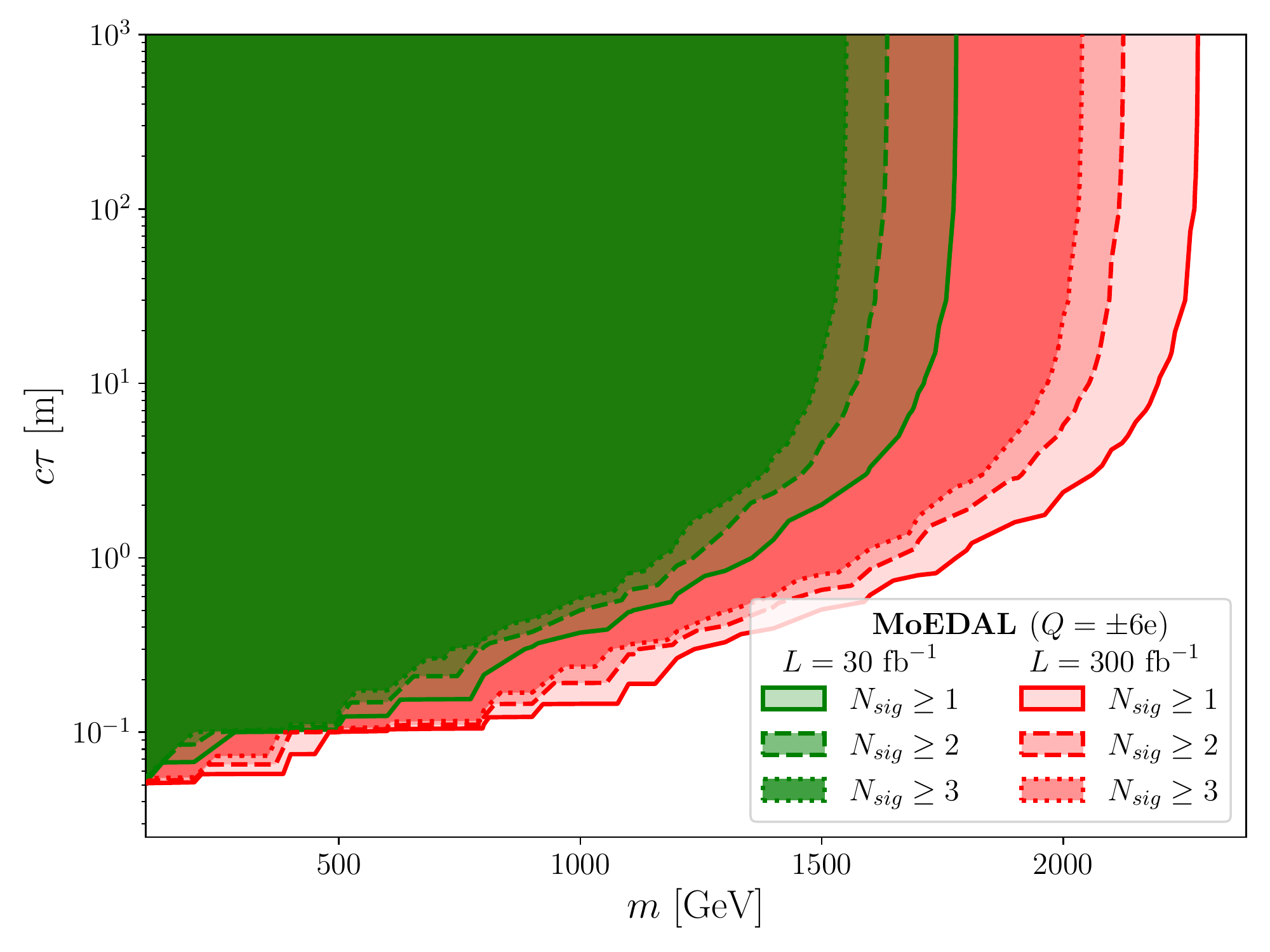}
      \includegraphics[width=0.4\textwidth]{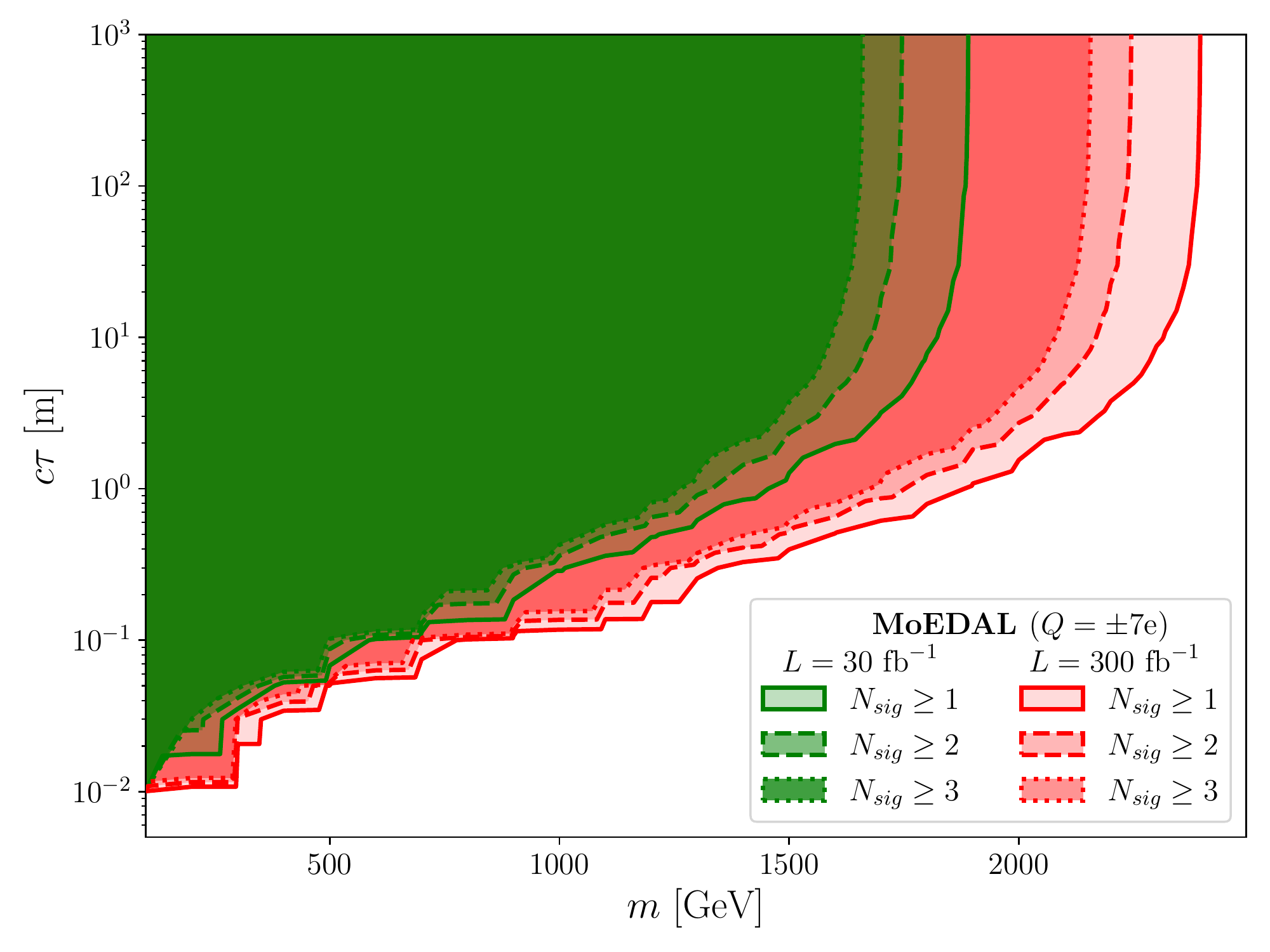}  \hspace{5mm}
      \includegraphics[width=0.4\textwidth]{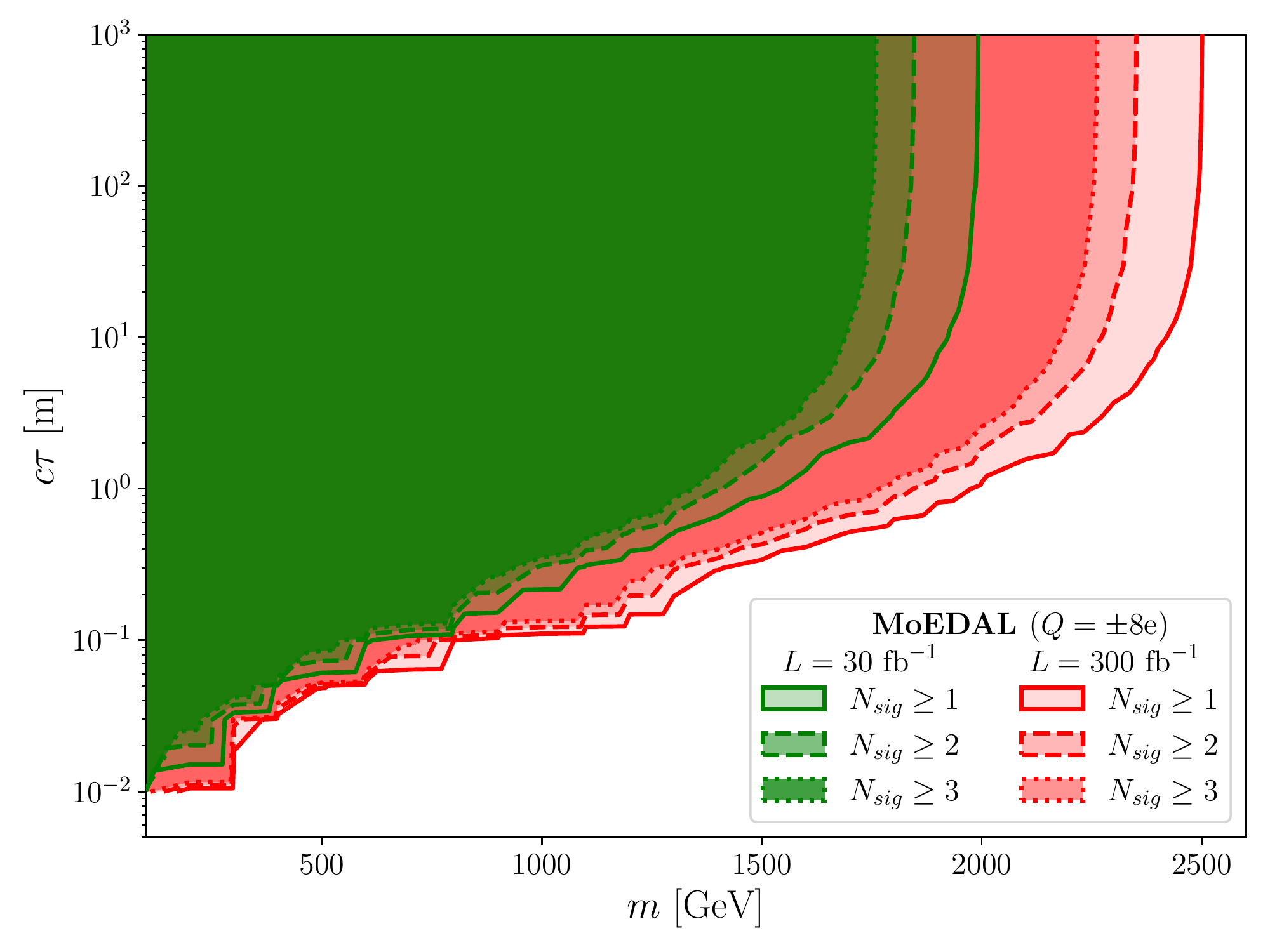}  
\caption{\small Model-independent detection reach at MoEDAL in the ($m$, $c
  \tau$) parameter plane for colourless fermions. The
  solid dashed and dotted contours correspond to $N_{\rm sig} = 1$, 2
  and 3, respectively. Green and red contours correspond to Run-3 
  $(L=30$ $\mathrm{fb}{}^{-1})$ and HL-LHC $(L=300$ $\mathrm{fb}^{-1})$ luminosities respectively.
  }
\label{fig:lim_fHighQ}
\end{figure}

\begin{figure}[t!]
\centering
      \includegraphics[width=0.4\textwidth]{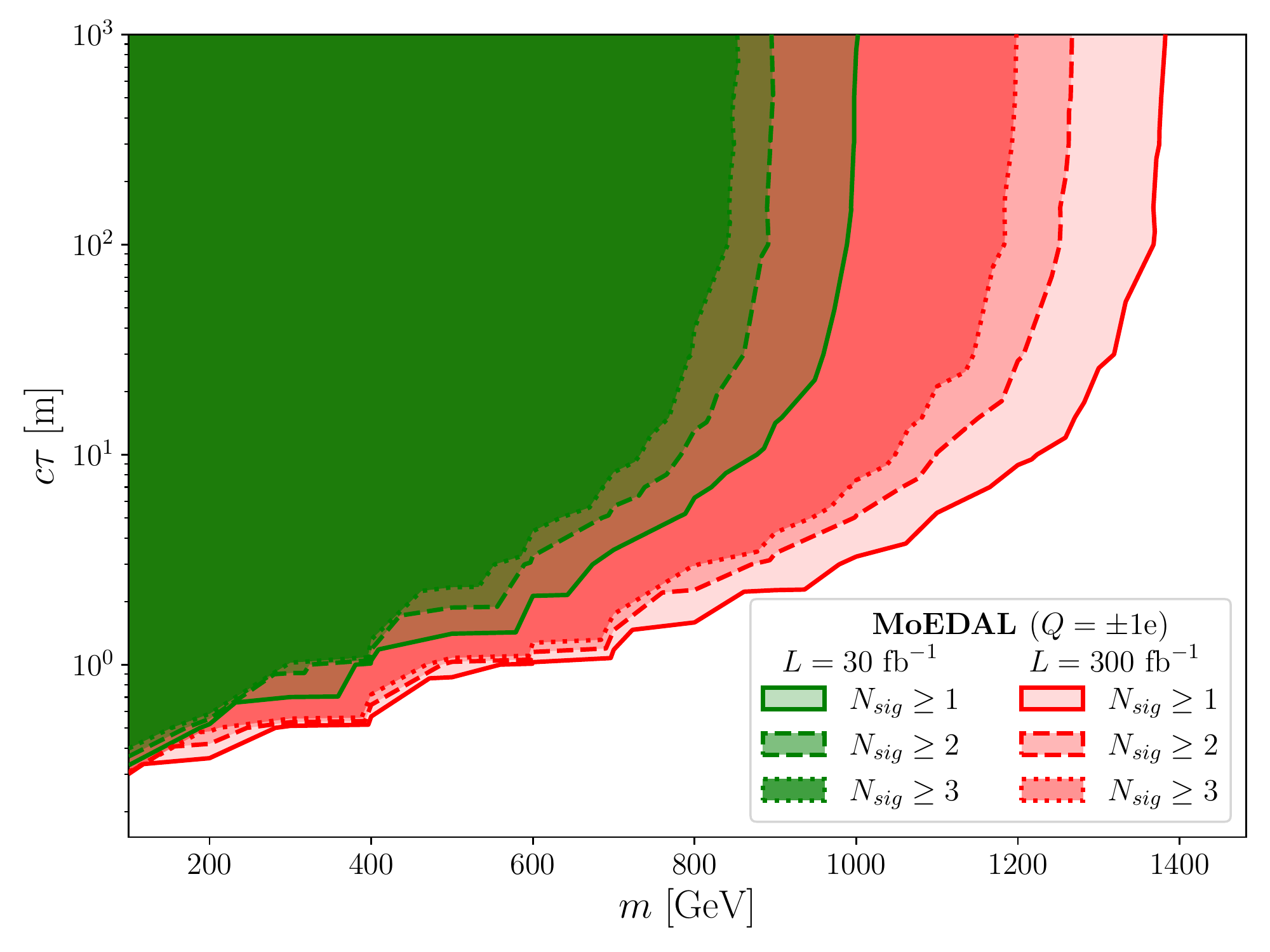} \hspace{5mm}
      \includegraphics[width=0.4\textwidth]{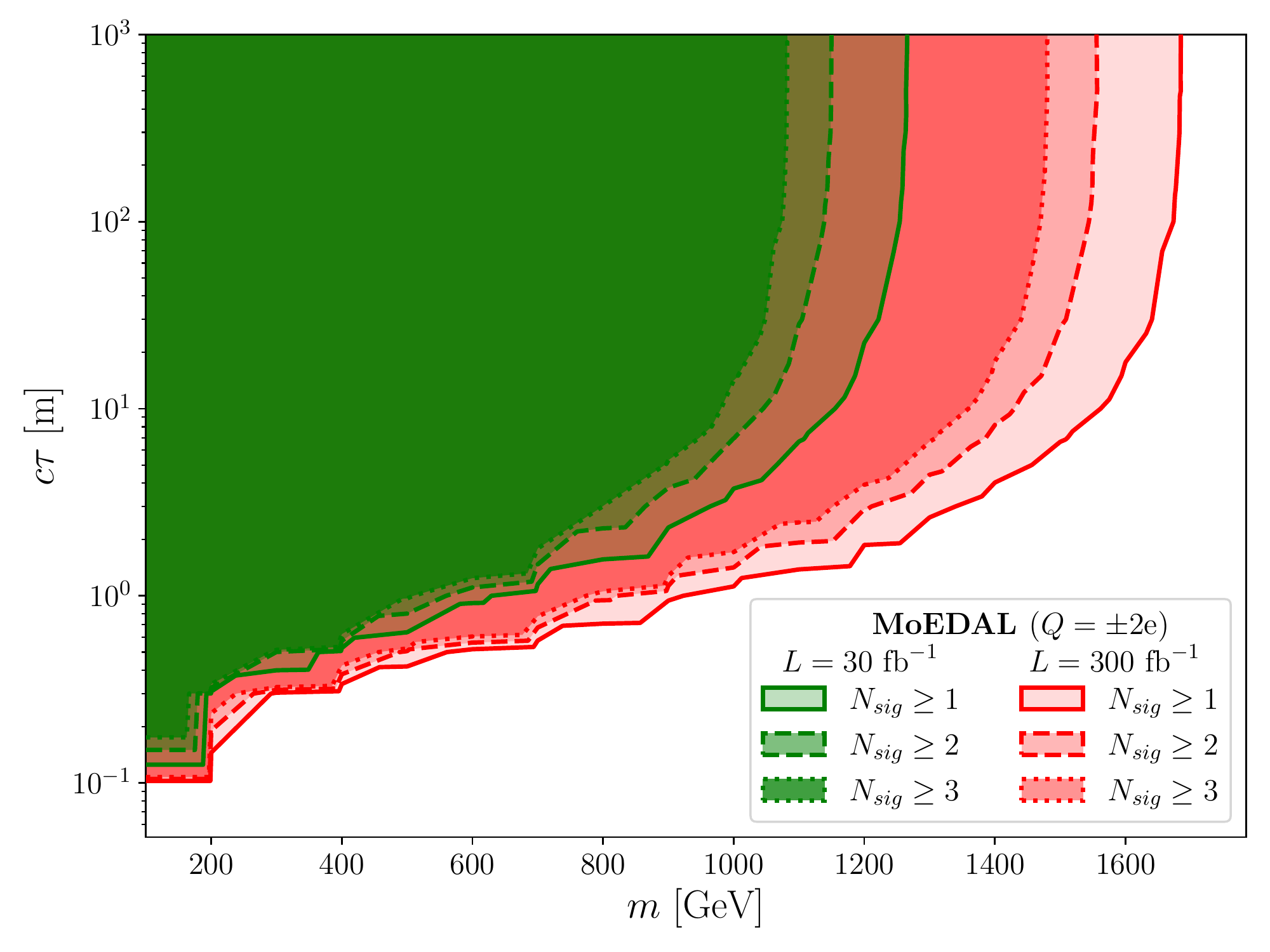}
      \includegraphics[width=0.4\textwidth]{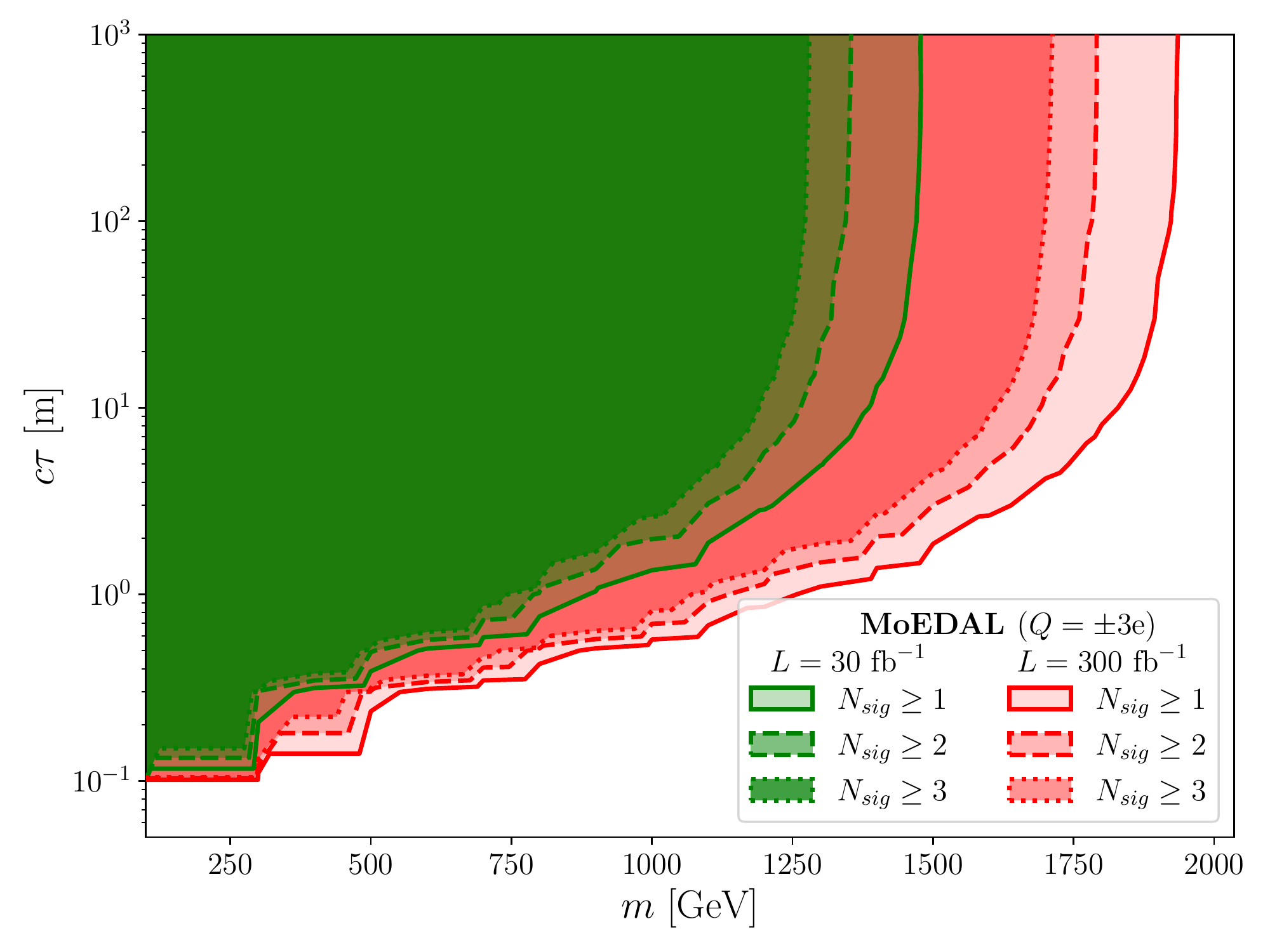}  \hspace{5mm}
      \includegraphics[width=0.4\textwidth]{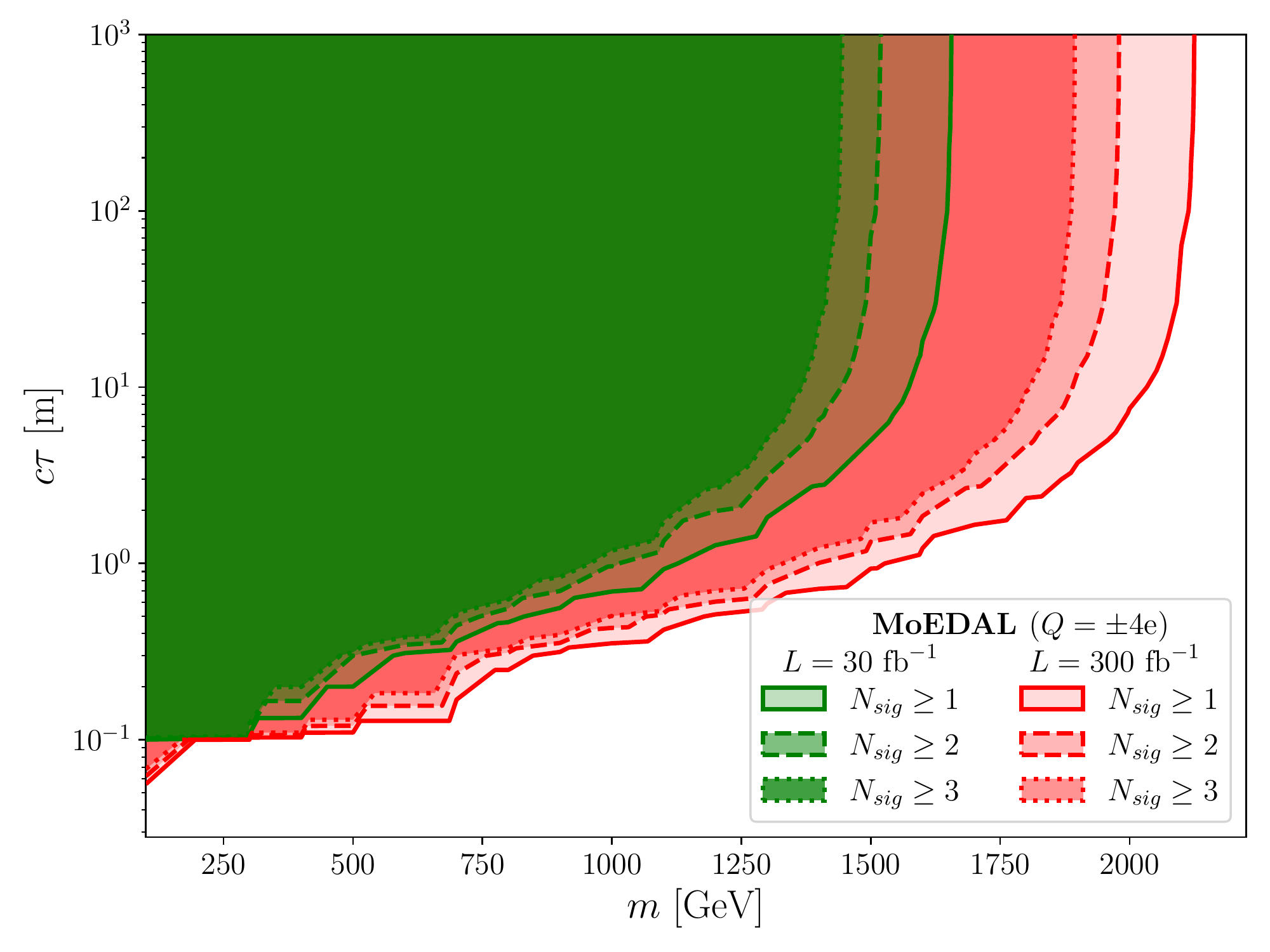}
      \includegraphics[width=0.4\textwidth]{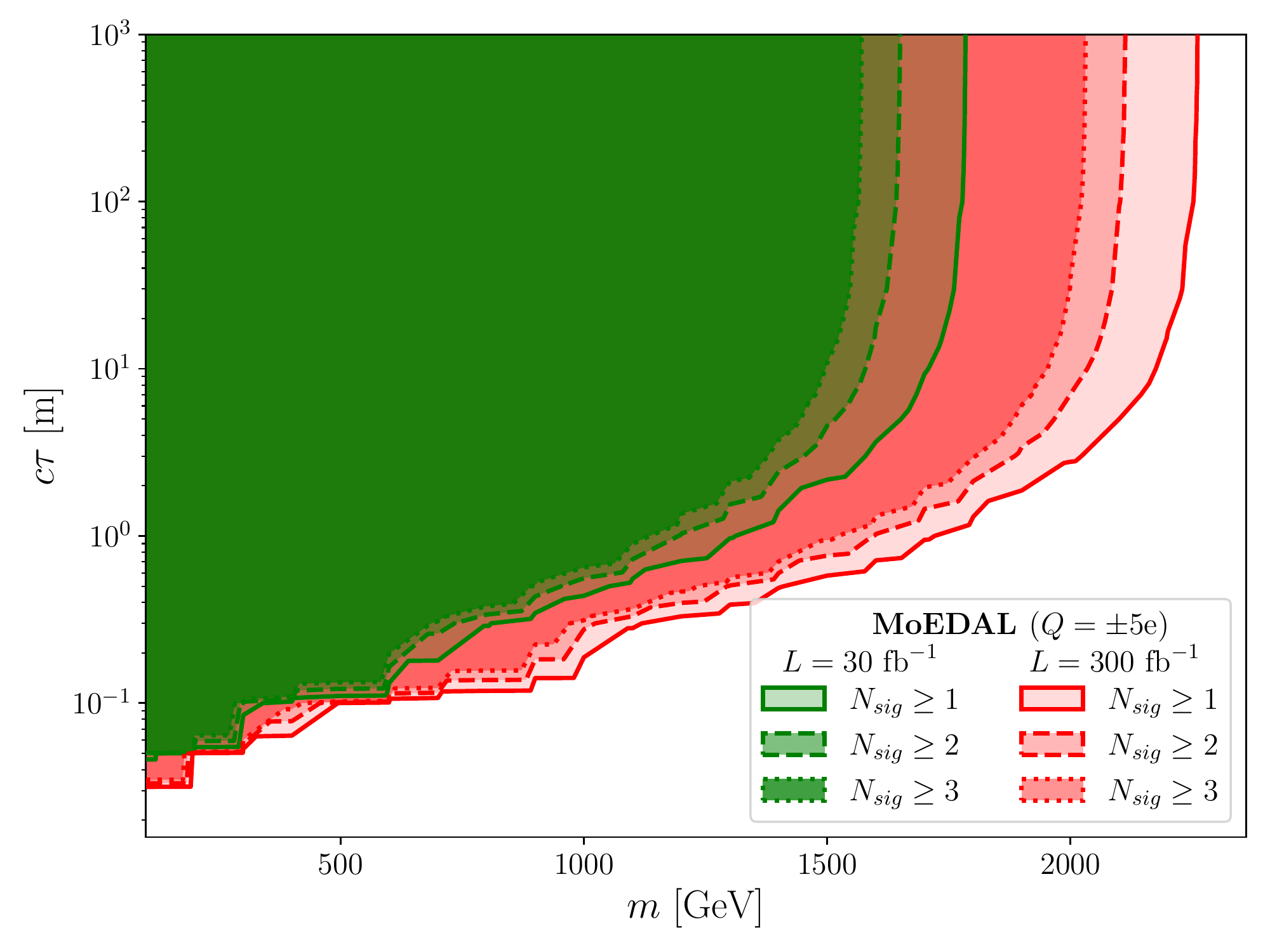}  \hspace{5mm}
      \includegraphics[width=0.4\textwidth]{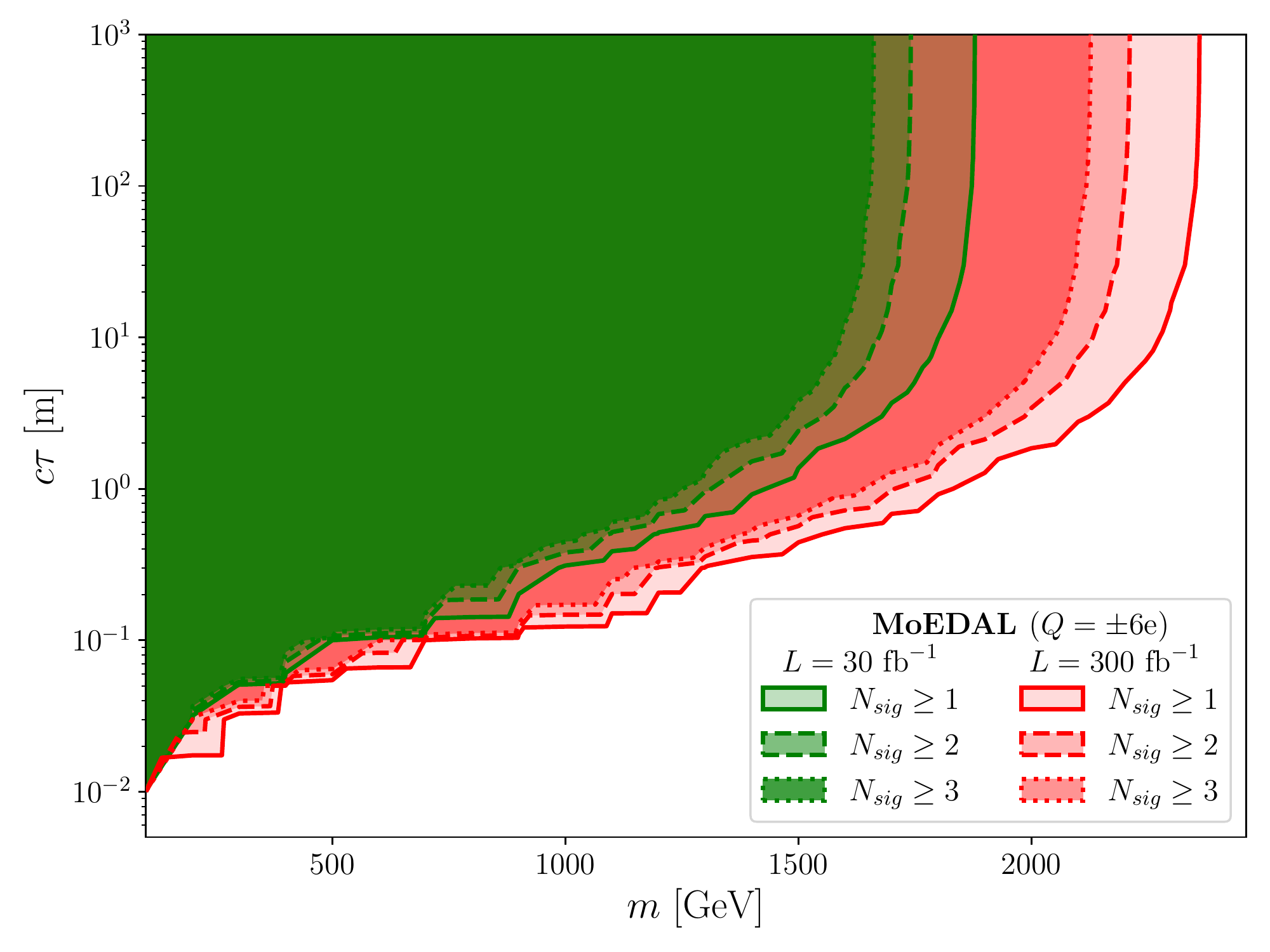}
      \includegraphics[width=0.4\textwidth]{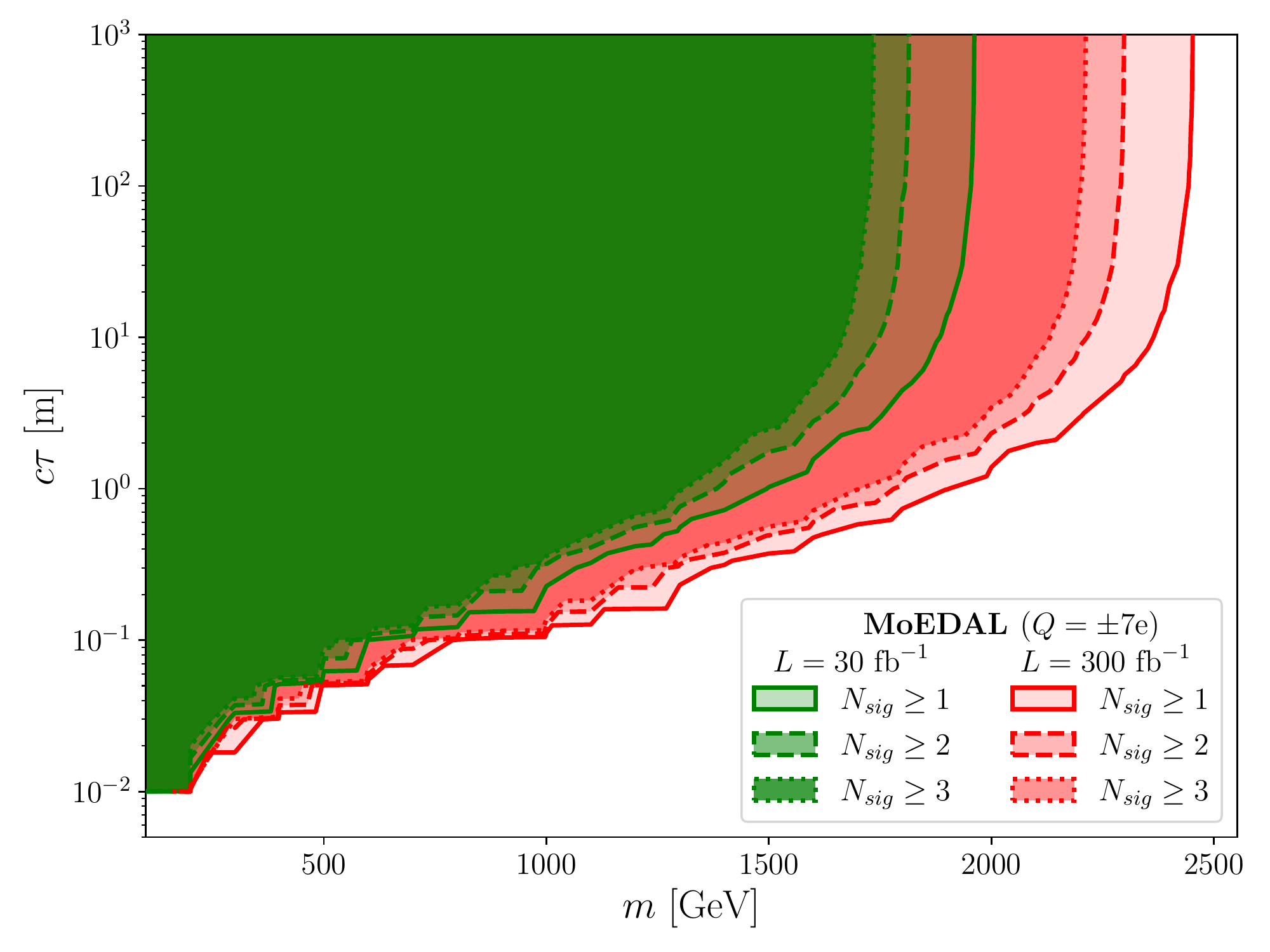}  \hspace{5mm}
      \includegraphics[width=0.4\textwidth]{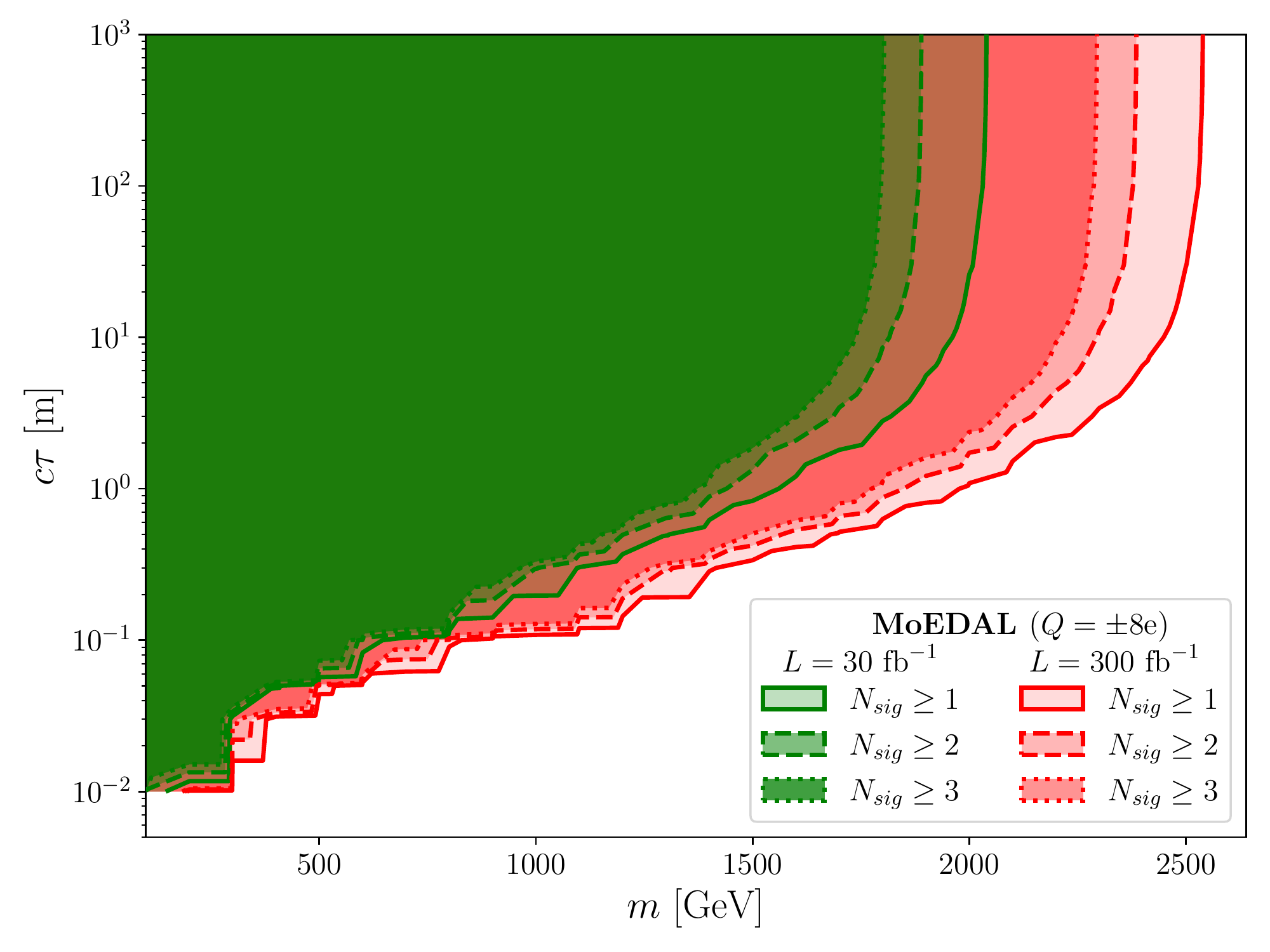}  
\caption{\small Model-independent detection reach at MoEDAL in the ($m$, $c
  \tau$) parameter plane for coloured scalars. The
  solid dashed and dotted contours correspond to $N_{\rm sig} = 1$, 2
  and 3, respectively. Green and red contours correspond to Run-3 
  $(L=30$ $\mathrm{fb}{}^{-1})$ and HL-LHC $(L=300$ $\mathrm{fb}^{-1})$ luminosities respectively.
  }
\label{fig:lim_csHighQ}
\end{figure}

\begin{figure}[t!]
\centering
      \includegraphics[width=0.4\textwidth]{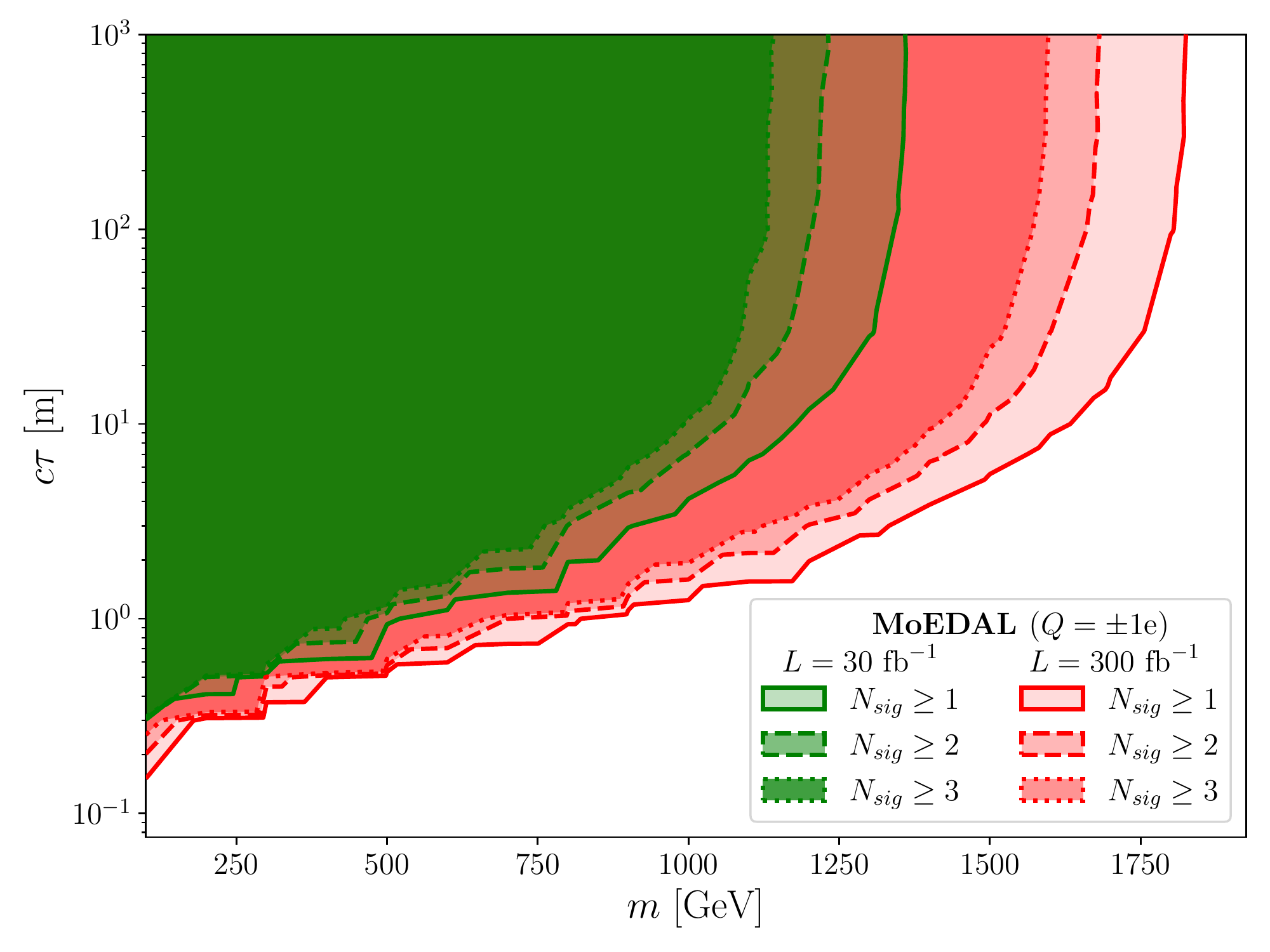} \hspace{5mm}
      \includegraphics[width=0.4\textwidth]{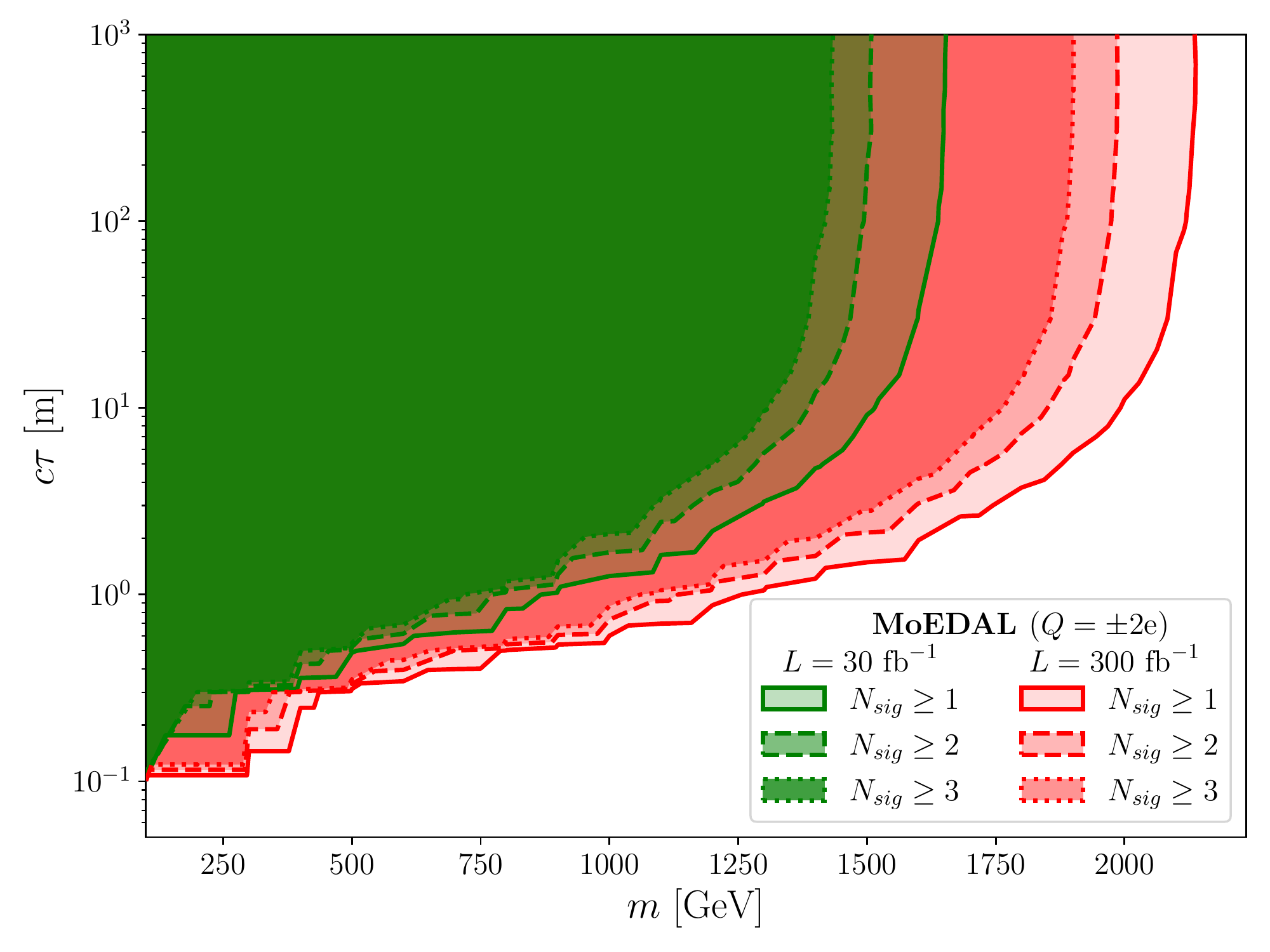}
      \includegraphics[width=0.4\textwidth]{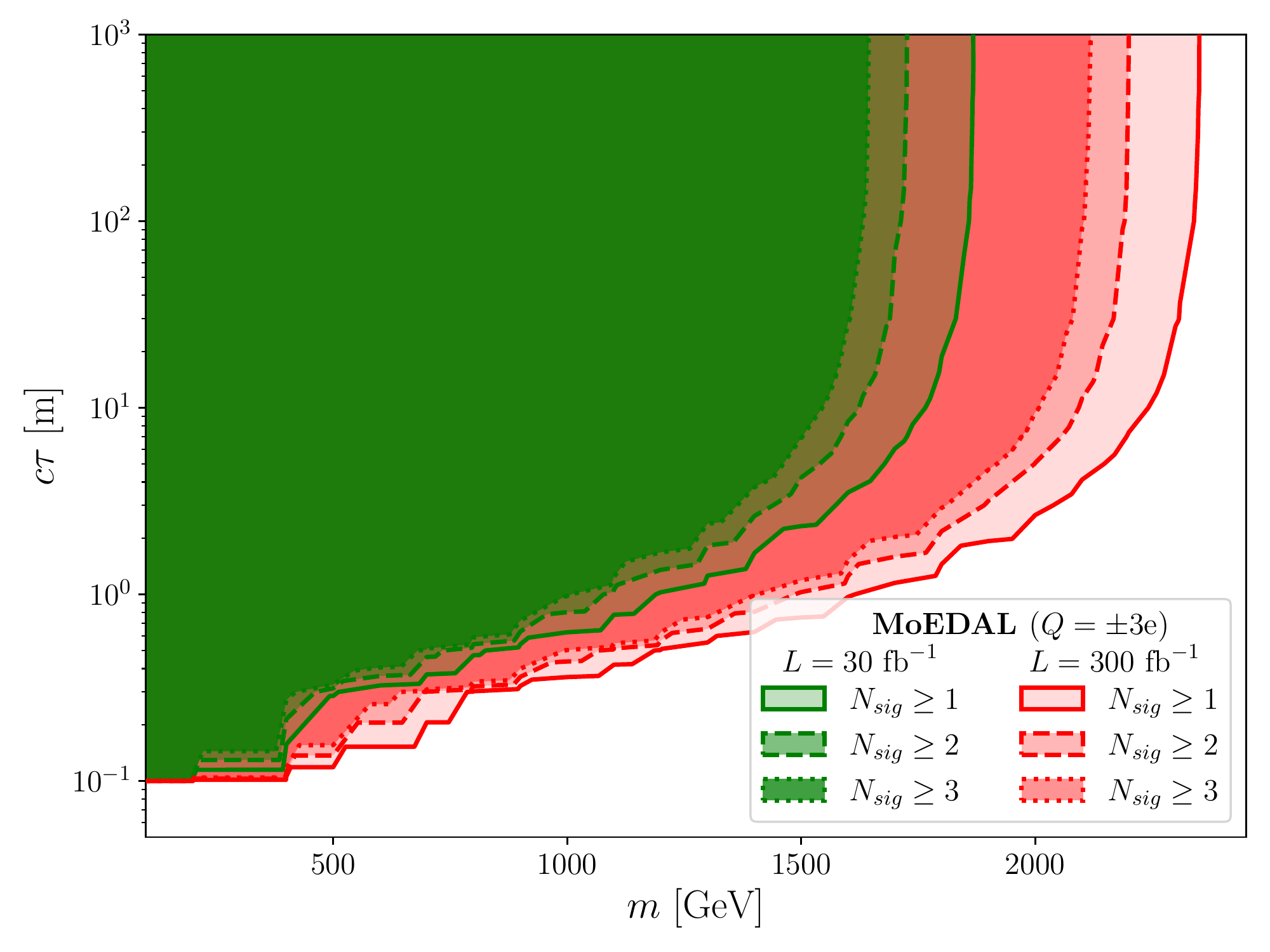}  \hspace{5mm}
      \includegraphics[width=0.4\textwidth]{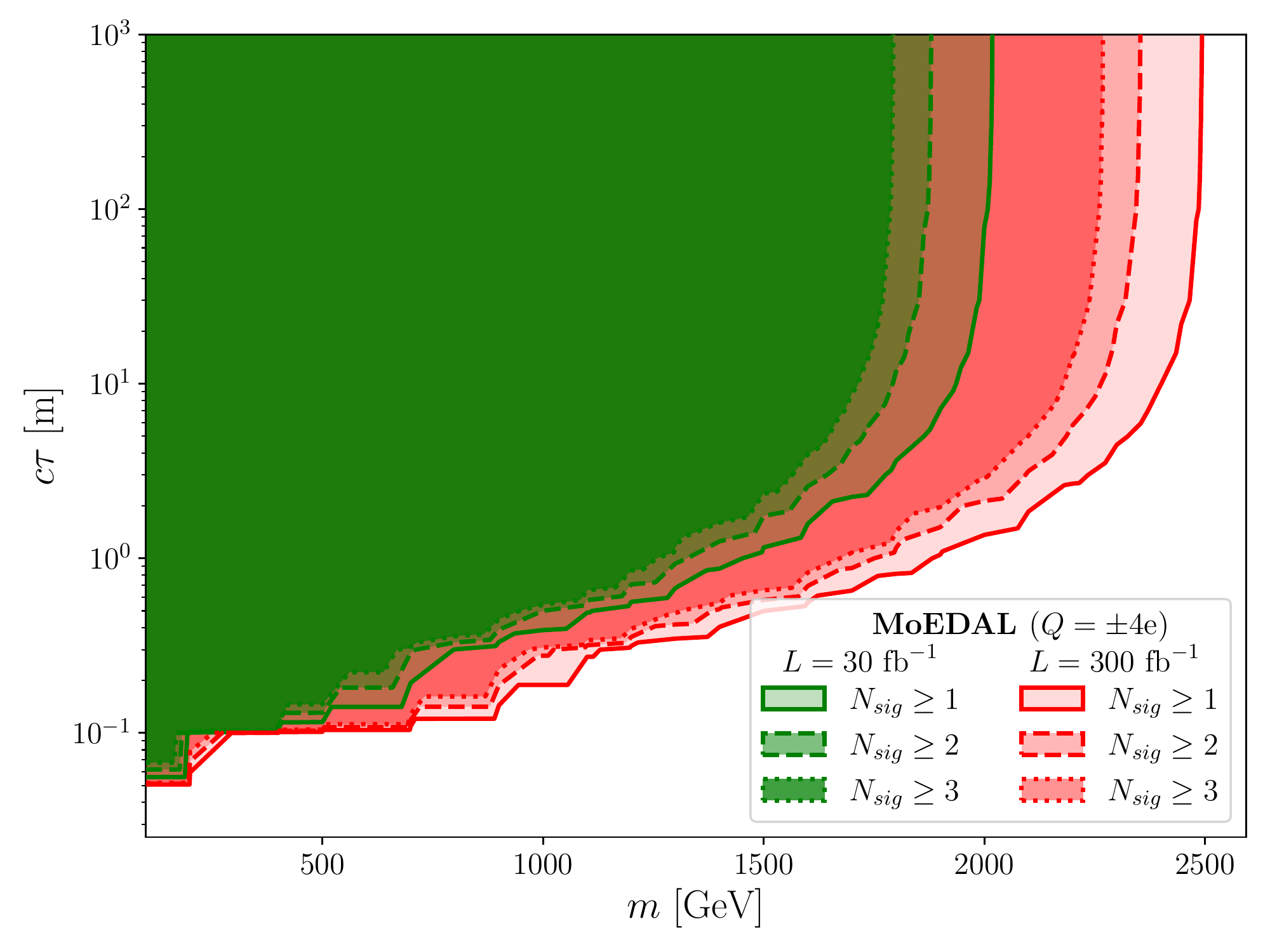}
      \includegraphics[width=0.4\textwidth]{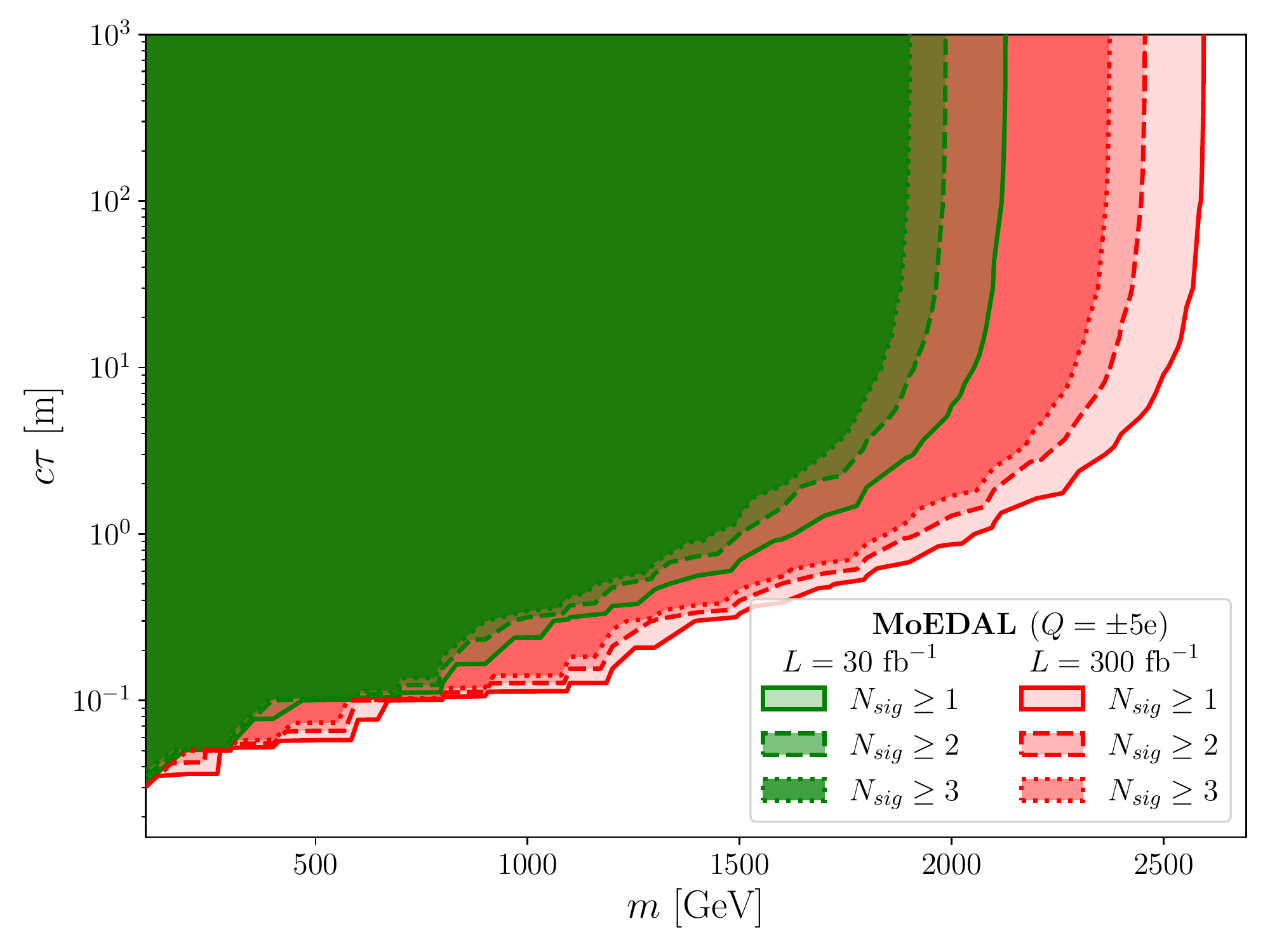}  \hspace{5mm}
      \includegraphics[width=0.4\textwidth]{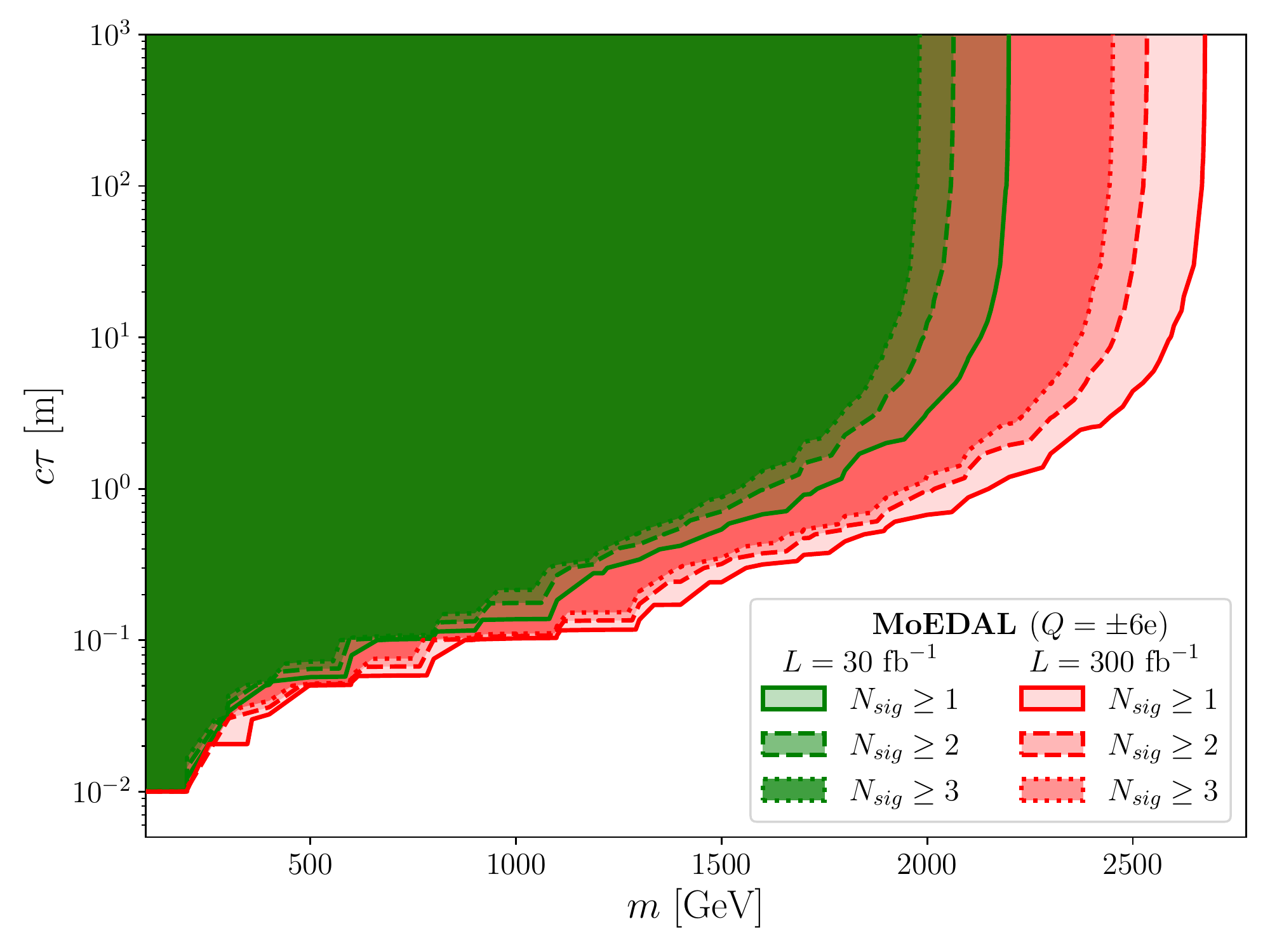}
      \includegraphics[width=0.4\textwidth]{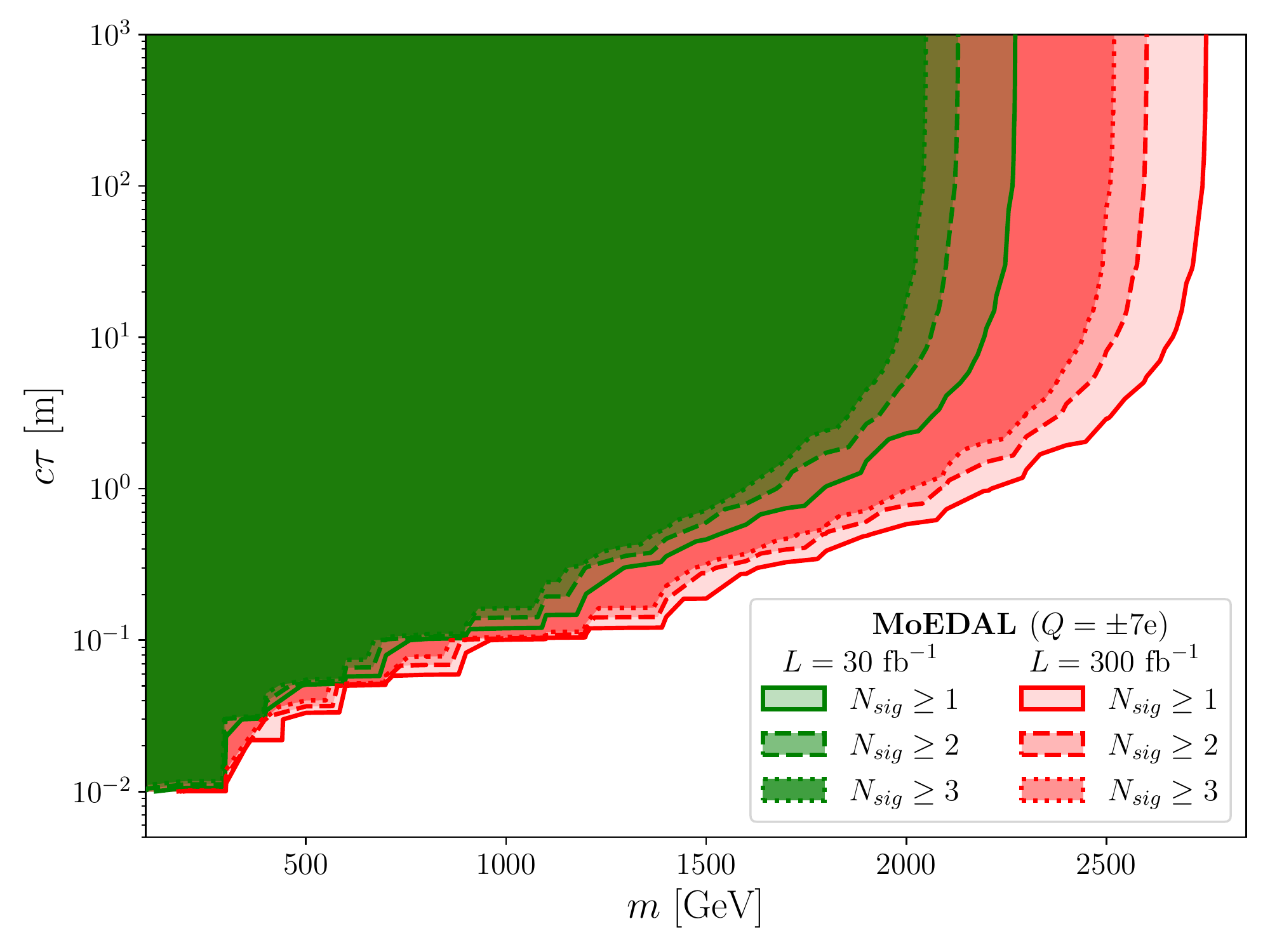}  \hspace{5mm}
      \includegraphics[width=0.4\textwidth]{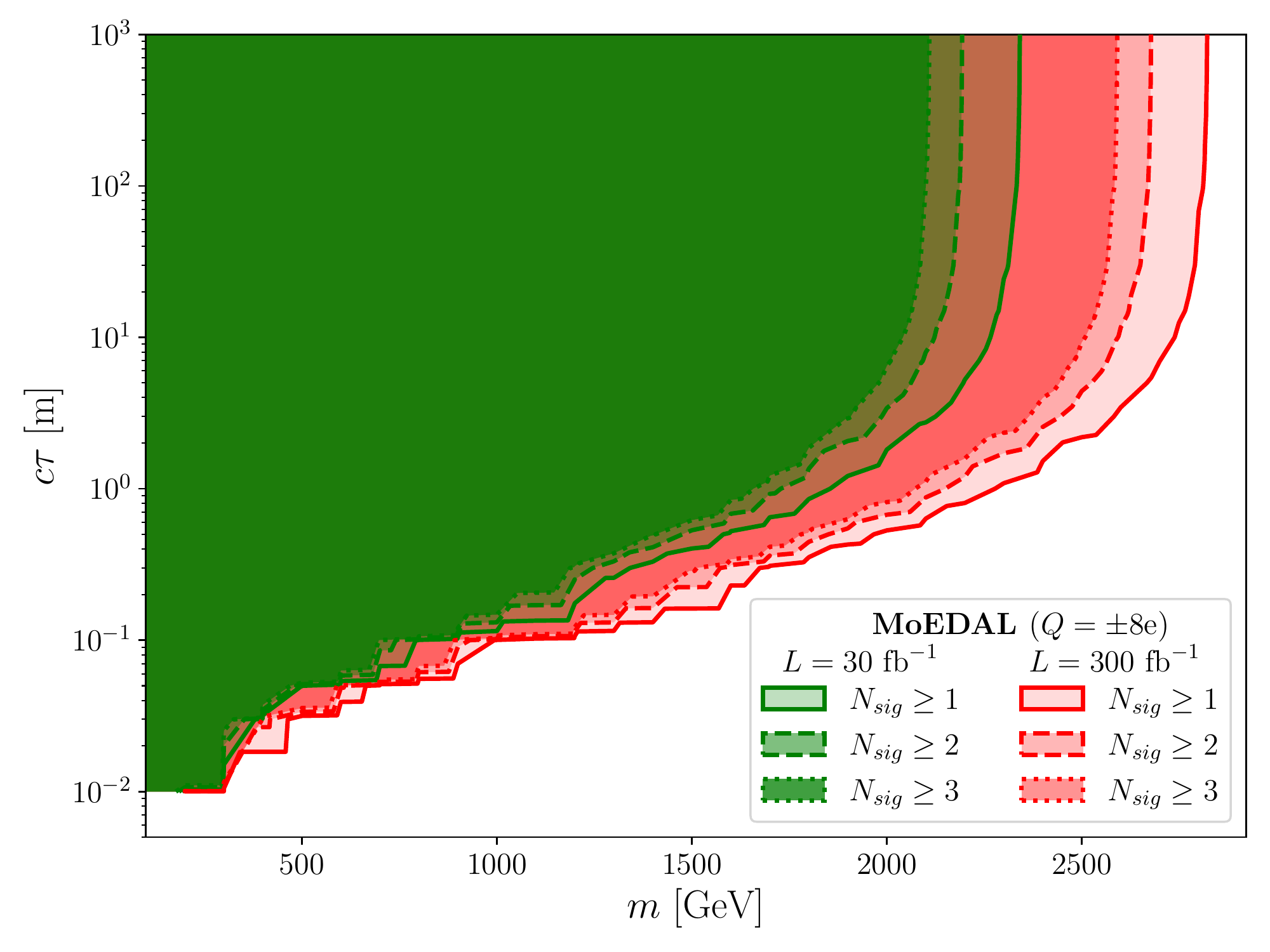}  
\caption{\small Model-independent detection reach at MoEDAL in the ($m$, $c
  \tau$) parameter plane for coloured fermions. The
  solid dashed and dotted contours correspond to $N_{\rm sig} = 1$, 2
  and 3, respectively. Green and red contours correspond to Run-3 
  $(L=30$ $\mathrm{fb}{}^{-1})$ and HL-LHC $(L=300$ $\mathrm{fb}^{-1})$ luminosities respectively.
  }
\label{fig:lim_cfHighQ}
\end{figure}

We show the model-independent mass reach ($N_{\rm sig} \ge 1$, 2 and 3)
for the long-lived multi-charged particles 
by the MoEDAL experiment 
expected at Run-3 ($L = 30$ fb$^{-1}$) 
and HL-LHC ($L = 300$ fb$^{-1}$)
in the mass versus lifetime planes ($m$, $c \tau$).
Figs.~\ref{fig:lim_sHighQ},
\ref{fig:lim_fHighQ},
\ref{fig:lim_csHighQ} and
\ref{fig:lim_cfHighQ}
correspond to colour-singlet scalars,
colour-singlet fermions, 
colour-triplet scalars 
and colour-triplet fermions, respectively.

\bibliography{ref}
\bibliographystyle{utphys}

\end{document}